\documentclass[sn-mathphys,Numbered]{sn-jnl}

\usepackage{graphicx}%
\usepackage{multirow}%
\usepackage{amsmath,amssymb,amsfonts}%
\usepackage{amsthm}%
\usepackage{mathrsfs}%
\usepackage[title]{appendix}%
\usepackage{xcolor}%
\usepackage{textcomp}%
\usepackage{manyfoot}%
\usepackage{booktabs}%
\usepackage{algorithm}%
\usepackage{algorithmicx}%
\usepackage{algpseudocode}%
\usepackage{listings}%
\usepackage{tabularx}%
\usepackage{float}%
\usepackage{lineno}

\begin{document}

\title[Article Title]{Heterogeneous Vulnerability of Zero-Carbon Power Grids under Climate-Technological Changes}


\author*[1]{\fnm{M. Vivienne } \sur{Liu}}\email{ml2589@cornell.edu}
\author[2]{\fnm{Vivek} \sur{Srikrishnan}}\email{vs498@cornell.edu}
\author[2]{\fnm{Kenji} \sur{Doering}}\email{kmd266@cornell.edu}
\author[3]{\fnm{Elnaz} \sur{Kabir}}\email{ekabir@tamu.edu}
\author[2]{\fnm{Scott} \sur{Steinschneider}}\email{ss3378@cornell.edu}

\author[1,2]{\fnm{C. Lindsay} \sur{Andersion}}\email{cla28@cornell.edu}

\affil*[1]{\orgdiv{Systems Engineering}, \orgname{Cornell University}, \orgaddress{\city{Ithaca}, \postcode{14853}, \state{NY}, \country{USA}}}

\affil[2]{\orgdiv{Department of Biological and Environmental Engineering}, \orgname{Cornell University}, \orgaddress{ \city{Ithaca}, \postcode{14853}, \state{NY}, \country{USA}}}

\affil[3]{\orgdiv{Department of Engineering Technology \& Industrial Distribution}, \orgname{Texas A\&M University}, \orgaddress{ \city{College Station}, \postcode{77843}, \state{TX}, \country{USA}}}



\abstract{The transition to decarbonized energy systems has become a priority globally to mitigate carbon emissions and, therefore, climate change. However, the vulnerabilities of zero-carbon power grids under climatic and technological changes have not been thoroughly examined. In this study, we focus on modeling the zero-carbon grid using a dataset that captures diverse future climatic-technological scenarios, with New York State as a case study. By accurately representing the topology and operational constraints of the power grid, we identify spatiotemporal heterogeneity in vulnerabilities arising from the interplay of renewable resource availability, high load, and severe transmission line congestion. Our findings reveal a need for 61-105\% more firm, zero-emission capacity to ensure system reliability. Merely increasing wind and solar capacity is ineffective in improving reliability due to transmission congestion and spatiotemporal variations in vulnerabilities. This underscores the importance of considering spatiotemporal dynamics and operational constraints when making decisions regarding additional investments in renewable resources.}


\maketitle

\section{Introduction}\label{introduction} 

In light of the Intergovernmental Panel on Climate Change's recommendation~\cite{bashmakov2022climate} to achieve net-zero emissions by 2050, countries worldwide are formulating and implementing clean energy transition policies. Despite variations across countries~\cite{sun2020review}, these policies aim to electrify the transportation and heating sectors while simultaneously transitioning from fossil fuel-based electricity to a mix of renewable resources such as wind, solar, hydro, geothermal, and various forms of storage. 

Changes in climate and technology are likely to have profound impacts on both the supply and demand sides of energy systems~\cite{yalew2020impacts}. Unlike fully dispatchable fossil fuel generators, the output of variable renewable resources such as wind, hydro, and solar depend on precipitation, temperature, wind speed, and solar irradiation, rendering them more sensitive to climate variability and change~\cite{crook2011climate,schaeffer2012energy,yalew2020impacts}. Consequently, the availability of diverse renewable energy sources is subject to a range of stressors, from short-term intense fluctuations—typical of solar energy—to milder, long-term variations, as observed in large-scale hydroelectric systems

On the demand side, the electric load profile is heavily influenced by the same weather variables, as well as by human behaviors. Furthermore, the planned electrification of heating and transportation is expected to impact energy demand profiles, both in terms of magnitude~\cite{tarroja2018translating} and shape (e.g., from summer peak to winter peak~\cite{NYISOphaseI2019, NYtrends2022}). As such, there is a strong dependence between the supply and demand sides of energy systems, driven by latent weather variables~\cite{doering2018summer, ElnazK2023} and climatic-technological changes. Identifying and analyzing the vulnerabilities that can arise from these complex interactions, which have been largely overlooked in previous studies, are essential to ensure zero-carbon system reliability~\cite{perera2020quantifying, tarroja2018translating}.

Integrated Assessment Models (IAMs) such as EnergyPLAN~\cite{lund2021energyplan} and REMix~\cite{scholz2012renewable} have been used to model the interaction between electricity supply and demand and to evaluate the potential impacts of different strategies for energy decarbonization on global and regional scales~\cite{keppo2021exploring}. The common objective of these studies is to determine the least cost approach to reach emission reduction targets under alternative long-term climate scenarios~\cite{luderer2022impact, keppo2021exploring}. Although IAMs incorporate a wide range of models for various sectors, including energy supply, demand, transportation, land use, and agriculture, they often oversimplify the energy system by employing low spatial resolution models for renewable resource modeling~\cite{brazil_dranka2018planning, Portuguese_fernandes2014renewable, Macedonia_cosic2012100}. Furthermore, most analyses do not adequately represent the power grid topology and operational constraints~\cite{scholz2017application}. As a result, the proposed strategies may not reliably achieve their targets~\cite{liu2023spatiotemporal}. 

In this study, we investigate the reliability of power systems in the context of the transformation towards a low-carbon energy sector over an ensemble of 22-year scenarios. Our framework centers on the power system, integrating the impacts of both climatic and technological changes on the supply side, power transmission structure, and demand side. We use the New York State (NYS) power system to illustrate the potential vulnerabilities of a zero-carbon power system and discuss various technology options to improve the reliability of the system given the vulnerabilities identified.

\section{Main Text}
\subsection{NYS Power Grid and CLCPA}\label{NYSandCLCPA} 

In order to achieve the goals of the Climate Leadership and Community Protection Act (CLCPA), the NYS Climate Action Council developed a scoping plan that outlines the specific details of the plan to decarbonize NYS~\cite{scopingplan}. In this section, we introduce the NYS power grid and the CLCPA.

\begin{figure}[H]%
\centering
\includegraphics[trim=2cm 5cm 2cm 2cm, clip,width=0.9\textwidth]{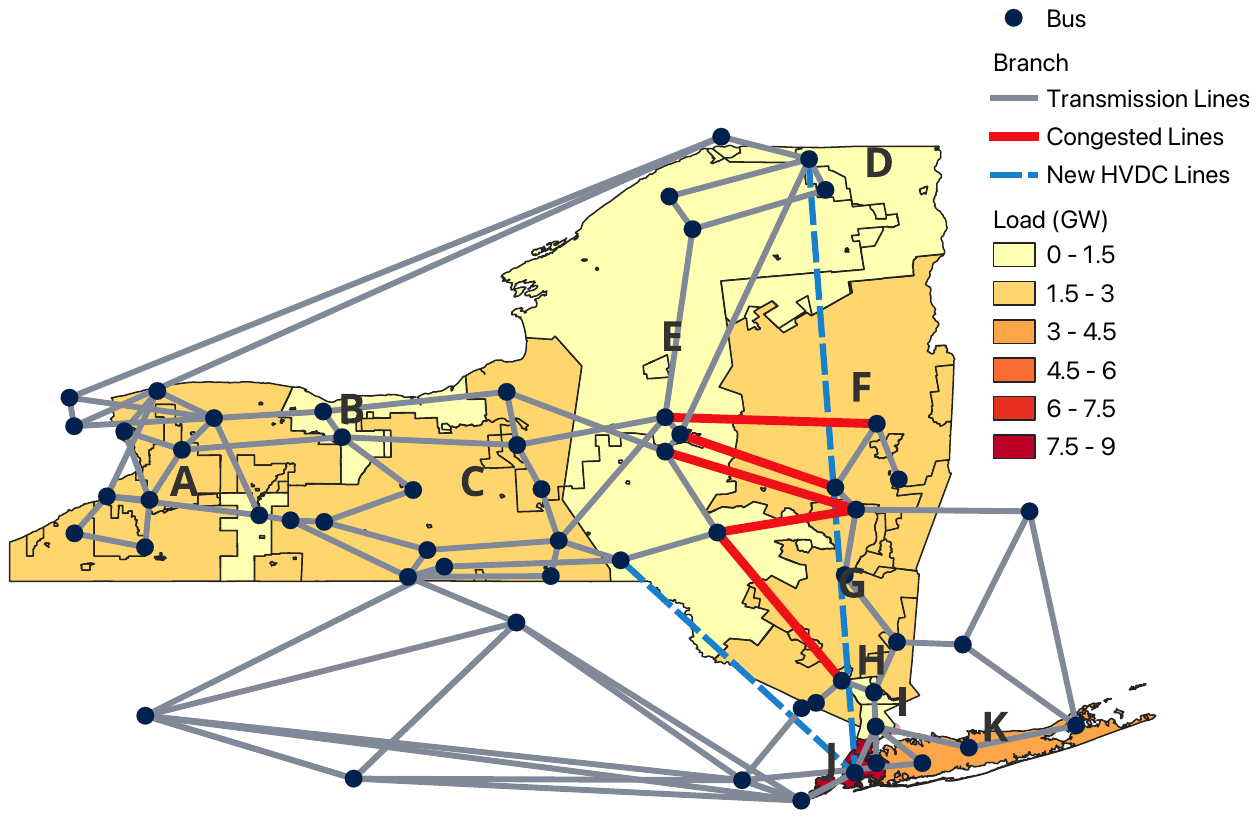}
\caption[Representation of the NYS grid]%
{\textbf{Representation of the NYS grid} - The NYS grid has 11 load zones indexed from A-K. Zones A-E are upstate zones and F-K are downstate zones. In the NYS representation, there are 57 buses shown as black circles in the figure. The 94 transmission lines are denoted by gray lines with red highlights to denote the Total East interface that transfers power from upstate to downstate and is subject to frequent congestion. Dashed lines represent the new HVDC lines that are designed to alleviate the congestion shown in red. One line is from Hydro-Quebec to NYC (D $\rightarrow$ J) and the other is from upstate NY to NYC (E $\rightarrow$ J), and are scheduled to be online in 2026 and 2027, respectively. The average load for each load zone (after electrification over 22 years) is indicated by the background color, where NYC and Long Island have the majority of the load due to high population density. }\label{fig: NYoverview}
\end{figure}

The NYS power grid has 11 load zones (Fig.~\ref{fig: NYoverview}) with spatially imbalanced supply and demand patterns, resulting in congested transmission lines that limit the transportation of energy from upstate (zones A-E) to downstate (zones F-K). NYS also relies heavily on hydropower, which is subject to seasonal and inter-annual fluctuations that are highly sensitive to climate change~\cite{cherry2005impacts}. To decarbonize the NYS grid, the CLCPA stands as one of the world's most ambitious climate-energy initiatives, targeting 70\% renewable energy generation by 2030, zero-emission electricity by 2040, and a minimum 85\% reduction in greenhouse gas emissions from 1990 levels by 2050. To reach these targets, onshore and offshore wind, utility solar, behind-the-meter (BTM) solar, transmission upgrades, new High Voltage Direct Current (HVDC) transmission lines, and electrification of multiple sectors are planned to be integrated into the current NYS grid in the next few decades~\cite{scopingplan} (see Supplementary Note 1 for more details about the NYS grid and the CLCPA). The NYS power grid and the ambitious CLCPA present an interesting case study for identifying vulnerabilities in future energy systems. 

The analysis in this study is based on scenarios proposed in~\cite{scopingplan}, which outline the detailed procedures and milestones defined in the CLCPA, coupled with an extensive dataset with high spatiotemporal resolution to characterize weather and load dynamics for 22 years.
While we primarily focus on NYS, the findings of this study provide valuable insights for stakeholders and policymakers on the general vulnerabilities of future carbon-free and renewable-dependent power grids under ambitious climate-energy policies, making them highly relevant and applicable to other regions.

\subsection{Zero-carbon energy system modeling}\label{framework}
\begin{figure}[h]%
\centering
\includegraphics[trim=0cm 0cm 0cm 0cm, clip,width=0.9\textwidth]{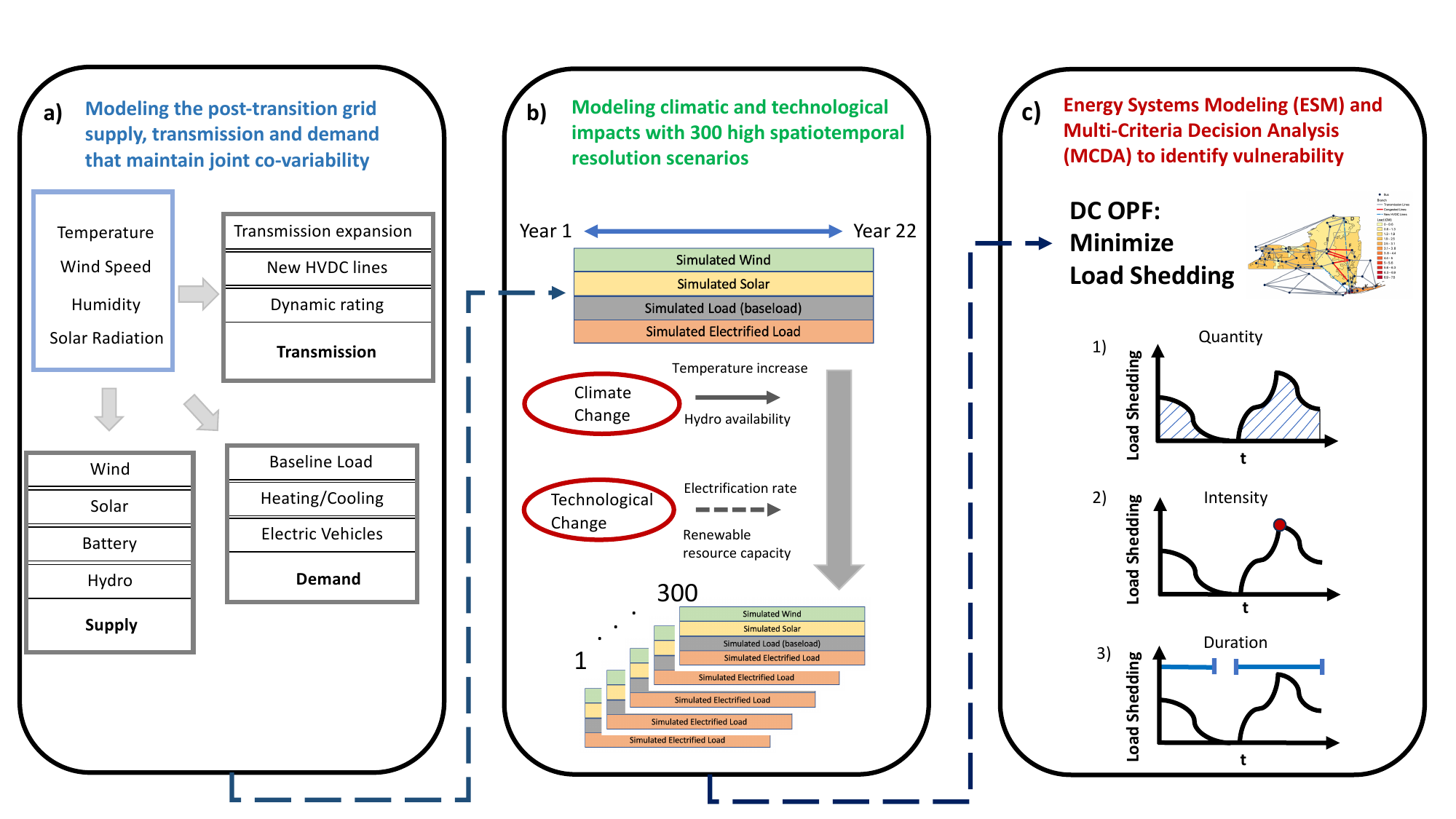}
\caption[Overview of the Framework]%
{\textbf{Overview of the Framework} - (a) Power grid demand, supply, and transmission are influenced by weather variables to maintain joint co-variability. (b) Climatic and technological factors are sampled to capture a wide range of 300 alternative SOWs for the 22-year horizon. Time series data is aggregated for illustration, but high spatial resolution is preserved in the model (see Methods). (c) The 300 scenarios, each with 22 years of hourly spatiotemporal trajectories, are input to the DC-OPF model, which seeks to minimize the total expected load shedding for a year. Simulated system performance is evaluated by three metrics to identify potential vulnerabilities, summarized in Table~\ref{tab: metricdef}. }\label{fig: Framework}
\end{figure}

\begin{figure}[h]%
\centering
\includegraphics[width=0.9\textwidth]{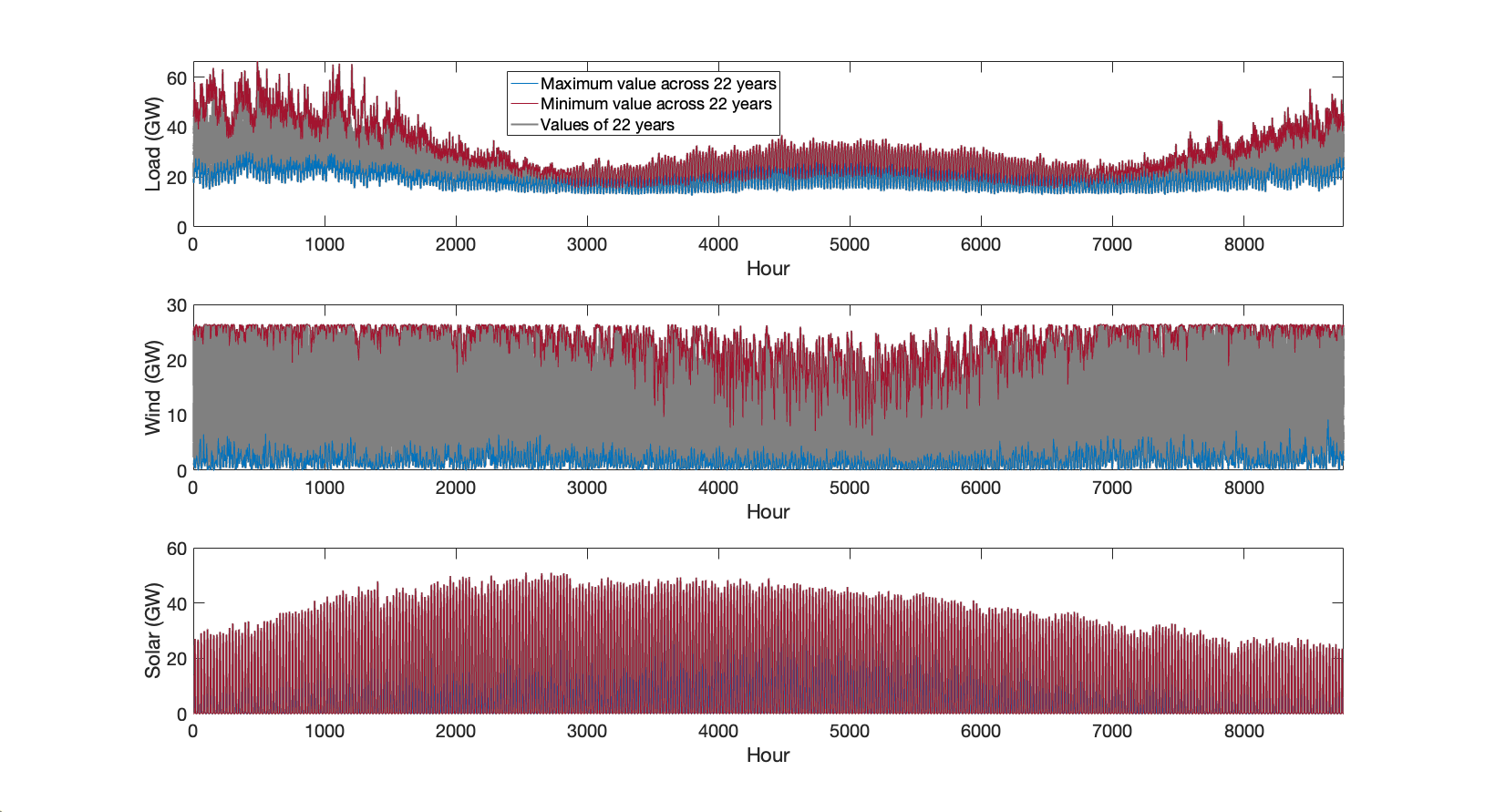}
\caption[Aggregated Seasonal Load, Wind and Solar Patterns for 22 years]%
{\textbf{Aggregated Seasonal Load, Wind and Solar Patterns for 22 years} - The simulated Load (upper), Wind (middle), and Solar (lower) for every year are shown in gray lines. The red and blue lines in each panel denote the maximum and minimum values for each hour over the 22 years, respectively. The load profiles have winter peaks after electrification, which is aligned with the expectation in~\cite{NYISOphaseI2019, NYtrends2022}. Wind power availability tends to be low over summer while solar power has the highest availability in spring with a little reduction in summer as temperature increases (solar panel efficiency decreases as temperature increases).}\label{fig: Loadwindsolar}
\end{figure}

Our approach integrates the interactive effects of weather and climatic-technological factors on the supply, transmission, and demand components of the energy systems, as shown in Fig.~\ref{fig: Framework}. We employ a coherent set of input weather variables to all modules in the energy system including wind output, solar output, hydro output, dynamic transmission line rating, baseline load, and electrification of buildings and transportation (Fig.~\ref{fig: Framework}(a); see Methods for details). Specifically, we use 22-year reanalysis weather data from MERRA2~\cite{MERRA2} at 120 locations to capture spatiotemporal co-variability, seasonal patterns, and climatic variability between load and renewable outputs (Fig.~\ref{fig: Loadwindsolar}). 

Given the lack of consensus on the impact of climate change on wind speeds and solar radiation~\cite{yalew2020impacts}, we focus on the potential effects of temperature increase. Proposed technical mechanisms to mitigate climate change, such as the integration of renewable resources, the expansion of transmission lines, and the electrification of buildings and transportation, are represented based on information from the CLCPA Scoping Plan~\cite{scopingplan} and the Reliability Needs Assessment (RNA) report~\cite{Assessment2020} from the New York Independent System Operator (NYISO). To account for deep uncertainties, which refers to an inability to specify a consensus probability distribution for the uncertainty of interest, we use the Latin Hypercube Sampling (LHS)~\cite{loh1996latin} method to generate 300 combinations (Methods) of a wide range of alternative States of the World (SOWs)~\cite{walker2010addressing}. The SOWs include scenarios of varying renewable integration levels, electrification levels, and temperature increases. Each sampled combination of deeply uncertain parameters is used to adjust the reanalysis weather data to represent the warming impacts on the time series output of each module along with renewable integration and electrification levels in the power system. The result is 300-by-22 years of time-series data with a high spatial resolution derived for evaluation of power grid reliability as illustrated in Fig.~\ref{fig: Framework}(b).

We employ the Optimal Power Flow (OPF) framework to determine the optimal dispatch of resources, aiming to minimize power shortages, also known as load shedding (Fig.~\ref{fig: Framework}(c)). A linearized DC-OPF formulation is used to reduce computational complexity while ensuring that constraints on power flow and nodal-level power balance are maintained (Methods). By adopting a Multi-Criteria Decision Analysis (MCDA) approach to quantify the vulnerabilities and identify different failure mechanisms, we can explore potential problems to improve systems reliability~\cite{baumann2019review}. In~\cite{scopingplan}, it is emphasized that a fully decarbonized grid is contingent on a firm, zero-carbon resource, which refers to emission-free dispatchable resources (see Supplemental Note 1 for more details). Throughout this paper, we adopt the term ``firm, zero-emission capacity (FZEC)" to refer to the additional resource capacity required to uphold power grid reliability. In Table~\ref{tab: metricdef}, we define three evaluation metrics to 1) identify the potential vulnerabilities caused by different underlying failure mechanisms and 2) determine the sufficient FZEC needed to improve overall system reliability. 

\begin{table}[h]
\centering
\caption {\label{tab: metricdef} Definition of the evaluation metrics}

\begin{tabular}{| m{7em} | m{4.5cm}| m{4.5cm} | }

\hline
 Metric & Definition  & Implication to Reliability \\   \hline
Load Shedding Quantity & the total amount of load shedding during a specific duration & the total amount of additional energy required to maintain system reliability during a specific horizon\\
 \hline
Load Shedding Hours (Duration) & the number of hours with energy deficiency within a fixed duration & the duration of FZEC needed to maintain system reliability\\
 \hline
Maximum Load shedding (Intensity) & maximum instantaneous quantity of load shedding during a specific duration & the FZEC needed to maintain system reliability\\
\hline

\end{tabular}
\end{table}

\subsection{Spatiotemporal heterogeneity of the future system vulnerability
}\label{baselineresult}

\begin{figure}[h]%
\centering
\includegraphics[width=0.99\textwidth]{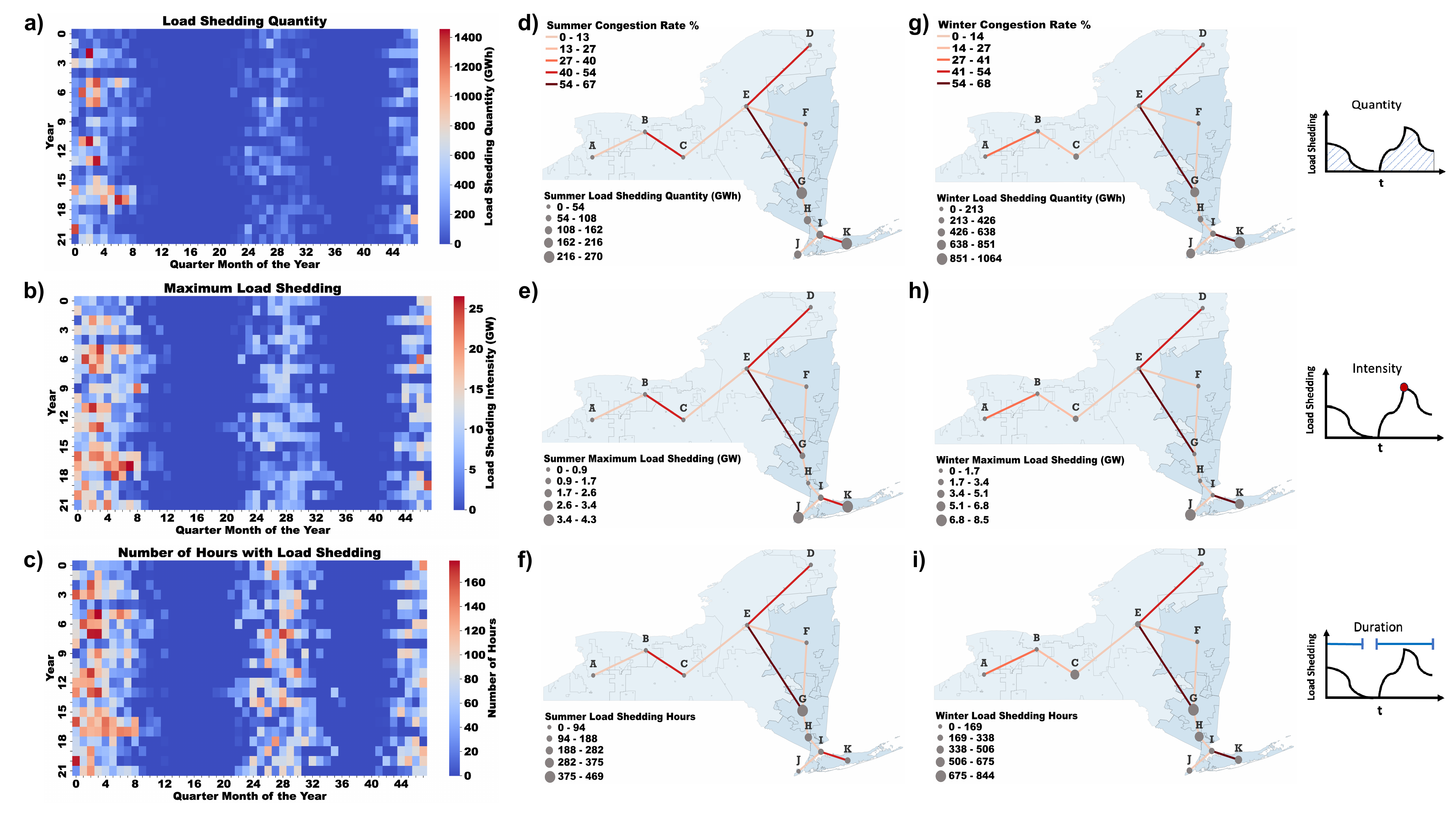}
\caption[Overview of baseline system vulnerability]%
{\textbf{Overview of baseline system vulnerability} - The figure highlights the spatial and temporal characteristics of NYS grid vulnerabilities over a 22-year horizon. The vulnerabilities are evaluated under three different metrics in panels \textbf{(a)-(c)}, which shows the quarter month (a quarter month is approximately one week, but adjusted for the duration of each month) of a year on the horizontal axis, and the index over the 22-year horizon on the vertical axis. Panel \textbf{(a)} shows the \emph{total quantity} of load shedding in a quarter month, panel \textbf{(b)} the \emph{maximum load shedding event} that occurs within each quarter month, and panel \textbf{(c)} shows the total \emph{number of hours} with load shedding in a quarter month. Pixel color in panels \textbf{(a)-(c)} indicates the severity of the vulnerabilities, ranging from minimal vulnerability (blue) and most severe vulnerability (red). Panels \textbf{(d)}-\textbf{(f)} consider the same vulnerability metrics for the \emph{summer season} within a spatial context. The size of the marker (dot) indicates the average intensity of each evaluation metric, while the color of the lines connecting each load zone denotes the average likelihood of congestion for each transmission interface connecting two load zones, with darker colors indicating a higher congestion rate. Similarly, panels \textbf{(g)}-\textbf{(i)} highlight the average spatial vulnerabilities for the \emph{winter season}.}\label{fig: baseresult}
\end{figure}

To provide a baseline assessment of the reliability of the system, we perform a detailed evaluation that spans a 22-year simulation period, focusing on the inherent complexities of the power system. The initial analysis, shown in Fig.~\ref{fig: baseresult}, excludes the deep uncertainties that arise from climatic and technological changes to provide a reliable baseline. These results consider the changes in generation technology and load profile described in the CLCPA Scoping Plan, without considering the impacts of climatic-technological changes on system performance.

The results in Fig.~\ref{fig: baseresult} demonstrate that the vulnerability of the future power grid exhibits spatiotemporal heterogeneity under the three evaluation metrics. Panels \textbf{(a)-(c)} highlight that the future system is more vulnerable in winter than in summer, while spring and fall exhibit much less load shedding overall. It is important to understand that the vulnerability patterns observed in winter and summer stem from different underlying mechanisms.

During winter, in load-centered zones J and K, load shedding intensity is the primary vulnerability, averaging 8549 MW and 6638 MW maximum load shedding over 22 years, which means that approximately 59\% and 89\% of load cannot be served during these severe energy deficits due to transmission congestion, respectively. Conversely, the generation-centered area (upstate zones A-E) exhibits a lower overall energy deficit. For example, zone D does not experience energy shortages throughout the 22-year simulation period, benefiting from its abundant and relatively stable hydropower supply. The load shedding primarily happens in zone C, due to relatively more congested interfaces (e.g., interface B-C) during cold days with wind droughts (wind drought refers to an extended period of very low wind speeds). This vulnerability can be attributed to two key factors:
\begin{itemize}
    \item The operational constraints on power flow pose frequent congestion (e.g., interfaces E-G and E-F), thereby impeding the transmission of needed energy from upstate to downstate. A different analysis that compares the vulnerability with and without key operational constraints reveals that the simplified representation of energy systems often used in these analyses is prone to underestimate system vulnerabilities. However, certain types of vulnerability can also be overestimated (see Supplementary Note 7 for detailed analysis).
    \item The co-variability between load and renewable outputs (e.g., heating demand caused by extremely cold days coincides with low renewable availability caused by wind droughts) makes the load-centered area more vulnerable and sensitive to weather conditions.
\end{itemize}

In general, under winter conditions, the amount of load shedding is the result of both load shedding intensity (at the load center) and duration (across the whole state), generally driven by high heating demand co-occurring with low renewable availability. The primary conditions are spatially different in intensity and duration (see Supplementary Note 9 for details and time-series illustrations).

During summer, the overall state-level energy deficit is much lower compared to winter, which primarily manifests in load centers. The load shedding is caused by the limited wind availability in summer (Fig.~\ref{fig: Loadwindsolar}), which leads to increased transmission line congestion (the dynamic rating modeling could exacerbate the congestion; see Supplementary Note 11 for details). The load shedding is particularly persistent after sunset when load centers heavily rely on stored energy and imports from upstate (see Supplementary Note 9 for detailed analysis). The total load shedding quantity is primarily influenced by high temperatures and wind droughts, with energy demand being particularly sensitive to extreme heat in densely populated areas. Additionally, hydro droughts can play a significant role, especially during the summer when hydropower contributes a larger portion of the supply due to reduced wind availability.

As described in Table~\ref{tab: metricdef}, the metrics measure vulnerability in a way that indicates the additional capacity needed to address power shortages, which in turn provides an estimate of the duration and capacity of the FZEC required. In addition, the spatial heterogeneity of vulnerability requires a spatially-disaggregated estimation of FZEC as load shedding is always accompanied by nearby upstream transmission line congestion, shown in all failure mechanisms. Congestion prevents the transfer of power from upstream zones to where it is demanded. As a result, although the maximum FZEC requirement identified without zonal analysis is 27 GW over the 22-year analysis period (Fig.~\ref{fig: baseresult}(b)), the actual need is likely as high as 37 GW when zonal requirements are included. Recall that in~\cite{scopingplan} 18-23 GW FZEC is estimated to ensure grid reliability. The more detailed study presented here indicates that the FZEC need is 61-105\% more than the scoping plan estimate. These findings underscore the importance of modeling energy systems with spatiotemporal co-variability and incorporating grid topology and operational constraints.

\subsection{Climatic and Technological Changes Exacerbate Vulnerabilities}

Building on the baseline analysis, the sensitivity of the findings can be analyzed under the impact of deeply uncertain climatic-technological factors (see Supplementary Note 5). These factors include temperature increase, capacity for renewable resources, and electrification rates for buildings and transportation. Using LHS, 300 combinations of these factors are analyzed to re-evaluate the 22-year analysis, enabling the understanding of system vulnerabilities and performance under a broad range of conditions.

\begin{figure}[H]%
\centering
\includegraphics[trim=0cm 0cm 0cm 0cm, clip,width=0.9\textwidth]{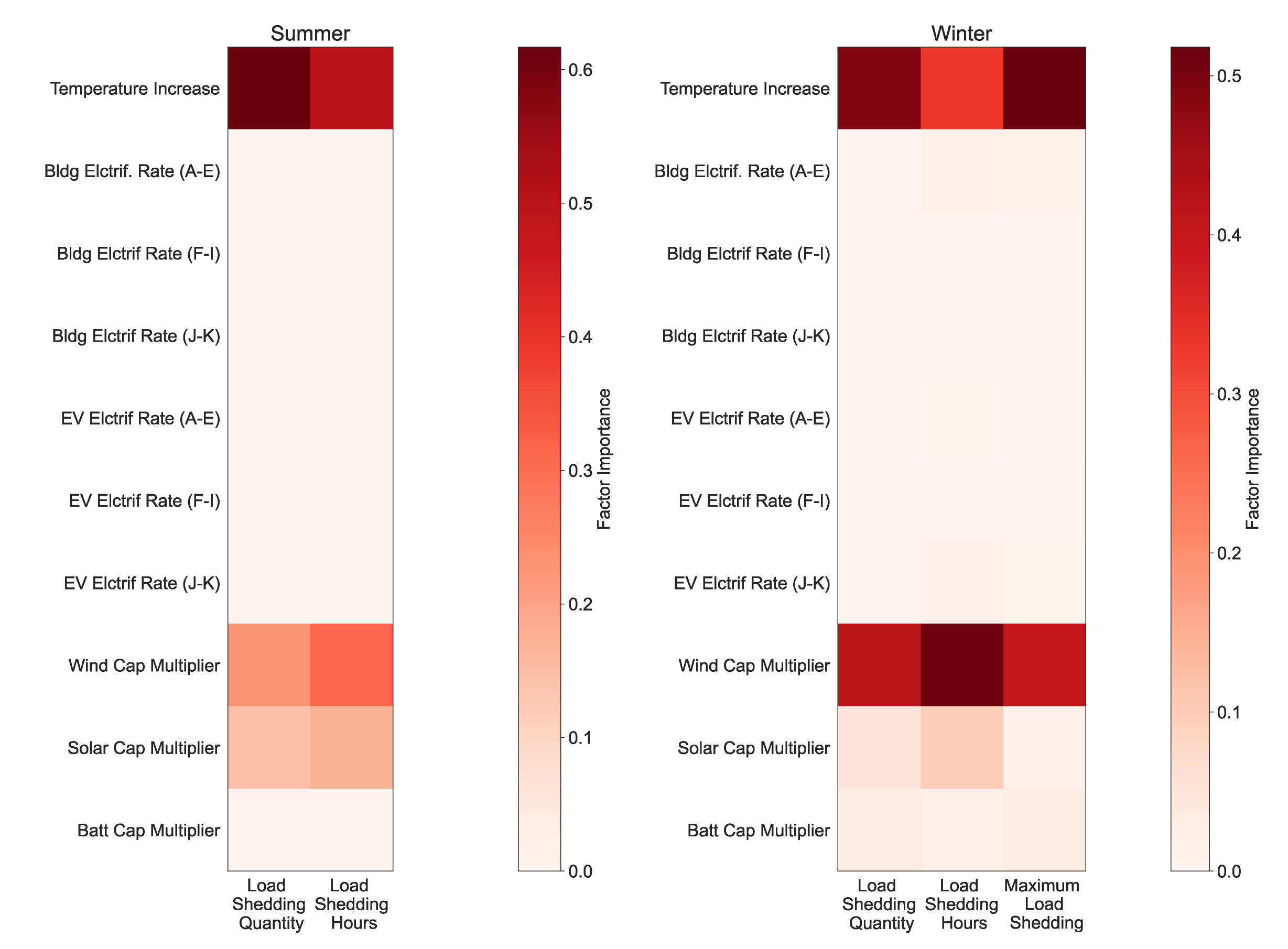}
\caption[Ranking of the climatic-technological factors for summer and winter]%
{\textbf{Ranking of the climatic-technological factors for summer and winter} - The importance score for each deeply uncertain parameter is derived from the Gradient Boosted Tree for winter and summer. The summer vulnerability is evaluated in terms of load shedding quantity and the hours of load shedding. (Maximum load shedding is not included in the figure because there are only a few data points that exceed the threshold. Therefore, the dataset is strongly biased for the Gradient Boosted Tree to classify.) The temperature increase is the most important driver for both load shedding quantity and load shedding hours. Winter operations are evaluated by all three metrics, showing temperature increase and wind capacity multiplier as the most important drivers of vulnerability.}\label{fig: Factorranking}
\end{figure}

\begin{figure}[h]%
\centering
\includegraphics[width=0.99\textwidth]{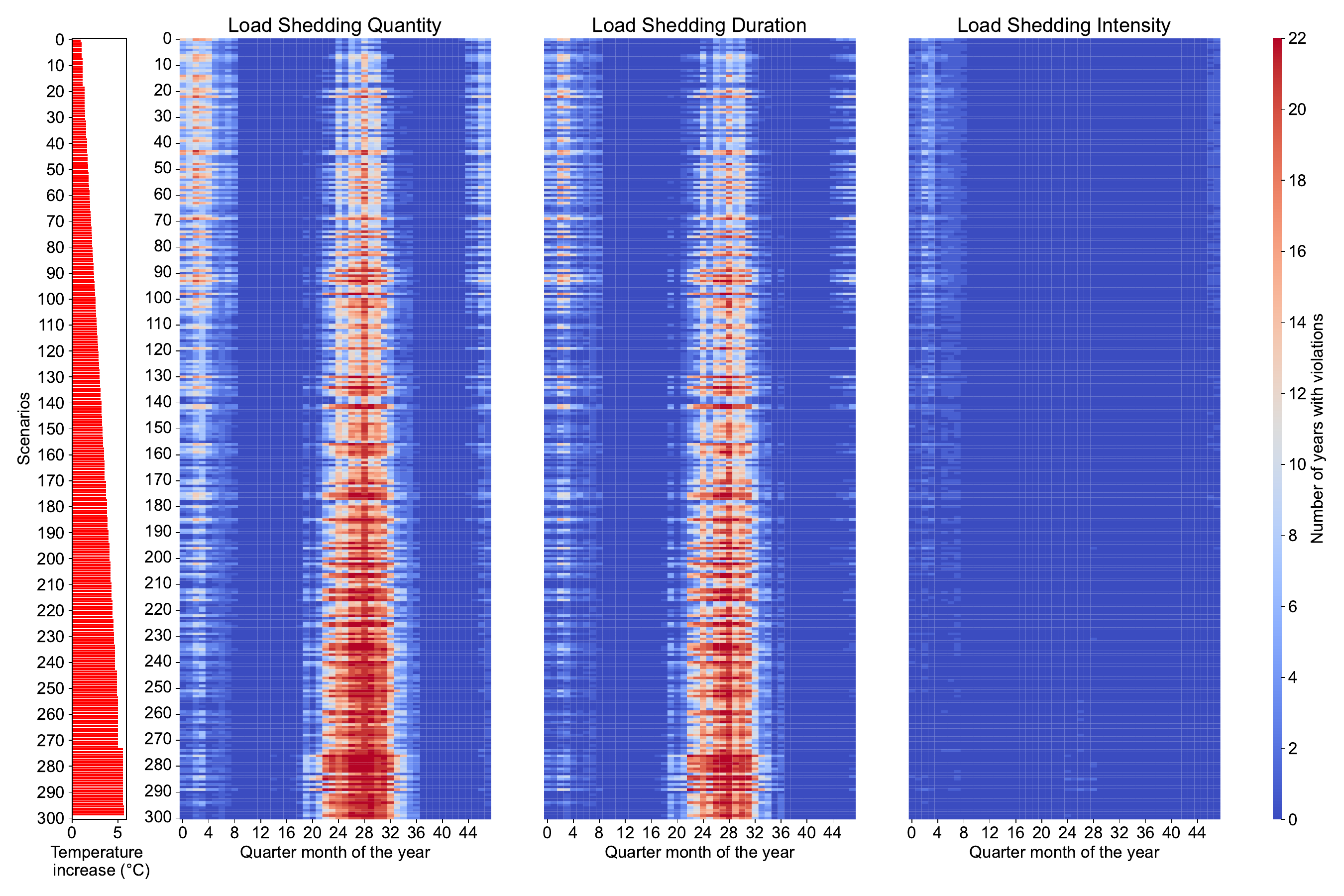}
\caption[Vulnerabilities under climatic-technological changes]%
{\textbf{Vulnerabilities under climatic-technological changes} - The scenarios are sorted in ascending order of temperature increase. a) Temperature increases across the 300 scenarios sorted in ascending order. b) - d) Vulnerabilities evaluated for every quarter monthly over the 22-year for each scenario. To quantify the vulnerability over 22 years, a threshold is chosen based on the information from~\cite{scopingplan} for each criterion. The pixels in the heatmaps represent the number of years with violations.}\label{fig: DU_result}
\end{figure}

To explore which factors are most significant for the vulnerability of the system, we first set a threshold based on information from~\cite{scopingplan} for each criterion to indicate violations as a binary variable. Then, we implement the Gradient Boosted Tree (GBT)~\cite{drucker1995boosting} method and use the importance scores of the inputs to rank the significance of the climatic-technological factors (see Methods), as shown in Fig.~\ref{fig: Factorranking}. It is obvious that temperature increase has crucial impacts on the energy system for both summer and winter. Therefore, as we count the number of violations for each evaluation metric over the 22-year horizon, we sort the scenarios in ascending order of temperature increase as shown in Fig.~\ref{fig: DU_result}. The sorted scenarios exhibit an increase in both violations of load shedding quantity and duration in summer while decreasing all types of vulnerability in winter. This shift highlights the changing dynamics of the system under climate change. Specifically, in winter, the overall vulnerability decreases as temperature increases, resulting from less extreme cold and reduced heating load, which is consistent with previous European studies~\cite{bloomfield2021quantifying}. It is worth noting that our analysis likely underestimates winter vulnerability due to the simplified representation of temperature increase as a step change across the year, and so we neglect increases in cold-weather extremes associated with climate change~\cite{wcd-3-1311-2022}.

In summer, the vulnerability increase is most notable in shedding event duration (Fig.~\ref{fig: DU_result}), while maximum load shedding is less significant. This results from the coinciding increase in summer cooling load with longer daylight hours and increased solar resources, resulting in smaller but more prolonged shortage events during daytime hours (see Supplementary Note 10 for a comparison between intensive temperature increase and no temperature increase). After sunset, cooling demand decreases, alleviating the need for additional generation. Consequently, maximum load shedding exceeding the threshold in summer is observed only in scenarios with temperature increases near $5^\circ$ C, combined with underbuilt wind and/or solar capacity. 

Finally, it is important to note that load shedding is almost always associated with nearby transmission line congestion that limits the effective use of available renewable resources. Therefore, increasing wind, solar, and battery capacity or decreasing electrification rates cannot fully compensate for the shortages caused by temperature increases. This supports our assertion that analyses of decarbonization strategies that neglect spatiotemporal co-variability and operational constraints are prone to underestimating system vulnerability.

\section{Discussion and Conclusion} \label{conclusion}

This study shows that the spatiotemporal heterogeneity of the NYS power system leads to substantial vulnerabilities under the proposed decarbonization plan outlined in \cite{scopingplan}. Possible approaches to address these vulnerabilities include increased deployment of long-duration battery storage, and/or development of complementary flexible resources, such as green hydrogen. In general, total curtailed renewable energy in the state is sufficient to cover the total load shedding quantity (assuming a 75\% renewable conversion efficiency), except in 17 scenarios where the planned wind and solar capacities are installed at less than 80\% of planned. Consequently, seasonal storage solutions such as long-duration batteries and hydrogen exhibit great promise for addressing these challenges.

\begin{figure}[h]%
\centering
\includegraphics[width=0.9\textwidth]{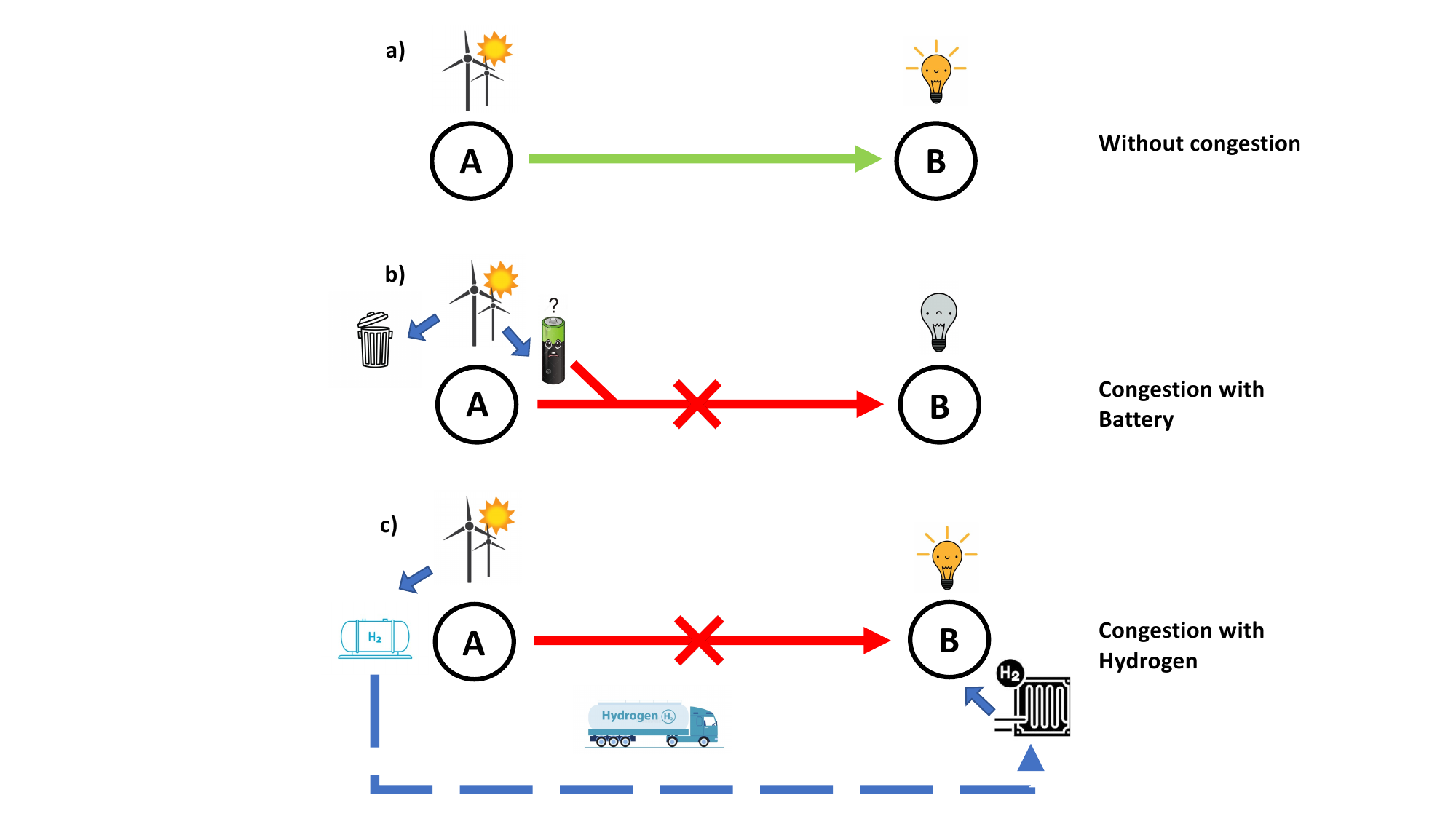}
\caption[Comparison between long-duration battery and hydrogen]%
{\textbf{Comparison between long-duration battery and hydrogen} - This figure illustrates the potential advantages of hydrogen compared to long-duration batteries. In Panel \textbf{(a)}, the scenario without congestion is depicted, where Node A, with abundant renewable resources, can transmit the required power to Node B, which lacks its own generators. In Panel \textbf{(b)}, the challenge of allocating long-duration batteries is presented. If the battery is placed at Node A, it can store excess renewable energy. However, during congested hours, the stored energy cannot be delivered to Node B. On the other hand, if the battery is located at Node B, only a limited amount of energy can be stored during non-congested hours. Consequently, a significant portion of renewable energy will be curtailed when congestion occurs. Panel \textbf{(c)} highlights the potential role of hydrogen during congested hours. Hydrogen can be stored and transported without utilizing transmission lines, providing a decoupling effect between the electrolysis of renewable energy and congestion situations. This decoupling enables the more efficient utilization of excess renewable energy, as hydrogen can be stored and delivered even during periods of congestion. }\label{fig: hydrogenvsbatt}
\end{figure}

While the topic of green hydrogen development is outside the scope of this paper, the concept involves producing hydrogen through electrolysis powered by the surplus renewable generation that cannot be directly consumed or stored. While the technology for hydrogen storage and transport is not yet fully mature, hydrogen offers the advantage of efficient use of curtailed renewable energy. This is particularly beneficial when renewable curtailment coincides with transmission line congestion, requiring excess renewable energy to be consumed locally. By locating the electrolyzer in areas with abundant renewable resources, hydrogen can be produced and then transported to high-demand regions that have limited \emph{local} renewable resources, without exacerbating transmission line congestion.  This offers a potential advantage over long-duration batteries, which are still subject to frequent transmission line congestion (Fig.~\ref{fig: hydrogenvsbatt}). In future work, a comprehensive analysis that compares hydrogen and long-duration batteries would be valuable to assess the cost, efficiency, and respective contributions to system reliability.

The interaction between climatic and technological impacts poses significant challenges for future carbon-free power systems. Exploring alternative scenarios and determining cost-effective pathways to decarbonize energy systems is crucial for combating climate change while ensuring reliability. While this study focuses on NYS, its ambitious climate policy and spatial heterogeneity allow us to identify broader insights applicable to other regions undergoing energy-system transitions.

Our findings show that identifying spatiotemporal heterogeneity in grid vulnerabilities requires consideration of grid topology, system constraints, and stressor co-variability over longer-term simulations. In the case study, vulnerability worsens during winter with heating and transportation electrification but shifts back to summer with a severe temperature increase. We also investigate the critical role of transmission congestion, coupled with various driving factors like extreme heat or wind drought, on load shedding intensity, duration, and renewable energy curtailment.

Our discussion provides valuable insights for policymakers and decision-makers, highlighting the importance of deploying FZEC of various technologies effectively to enhance system reliability. This information can aid in making informed decisions and shaping effective policies for a more resilient and reliable power grid in the face of climate-induced challenges.

\section{Methods} \label{methods}
\subsection{Modeling the zero-carbon energy system}
Like most zero-carbon energy systems, the envisioned NYS grid will rely heavily on newly installed wind and solar power, with existing hydro and nuclear sources, as the primary sources of energy. It is assumed that all fossil fuel generators will be retired by 2040. Our baseline zero-carbon configuration is based on the plan proposed by the Climate Action Council of NYS, which aims to achieve an 85\% reduction in economy-wide greenhouse gas emissions by 2050 compared to 1990 levels~\cite{CLCPA}. Detailed information regarding zonal capacity for wind, solar, and storage, as well as electrification assumptions, can be found in Supplementary Note 1.
 
To represent the zero-carbon configuration, we have adopted the grid topology and parameters from the baseline representation developed in~\cite{liu2022open} using 2019 data. These representations have been modified to reface the changes that are planned for the transition. The modeling details for each component of the energy system are provided below.

\subsubsection{Wind and Solar}
The capacity of wind and solar sites in the zero-carbon energy system is initially determined based on the information in~\cite{scopingplan}. Specifically, the system incorporates 11.6 GW of land-based wind, 14.7 GW of offshore wind, 13.6 GW of behind-the-meter (BTM) solar, and 51.2 GW of utility solar. The capacity allocations for different types of renewable resources are summarized in Table S2, and these zonal capacities are further distributed to bus-level (wind and utility solar on PV buses only; BTM solar on any type of buses) based on the potential wind and solar sites identified by the National Renewable Energy Laboratory (NREL). In this allocation, utility solar and wind generators are modeled as semi-dispatchable units, allowing them to generate power at any level below their forecasted limits. Conversely, BTM solar is modeled as a negative load to offset local energy needs.

To determine the bus-level renewable outputs for wind and solar resources, the Wind Integration National Dataset (WIND) Toolkit (WTK)~\cite{WTK} and Solar Integration National Dataset (SIND)~\cite{SIND} provided by NREL are employed. These toolkits enable the conversion of weather data spanning from 1998 to 2019 into unit power output values. As the WIND and SIND datasets only cover a limited time span, we use the reanalysis MERRA2~\cite{MERRA2} dataset to regenerate the wind and solar outputs for 22 years at different locations. By doing so, we preserve the spatial-temporal co-variability between the renewable outputs and other components in the zero-carbon grid that also take the MERRA2 dataset as inputs. 
 
The MERRA2 dataset offers a spatial resolution of 4 km × 4 km grids, and the nearest corresponding point is matched for each wind and solar site. To mitigate computational complexity, a subset of representative wind and solar sites is selected using the methodology outlined in~\cite{doering2022evaluating}. It is important to note that instead of computing exact power outputs, unit power output values are derived. This approach allows for flexibility in alternative capacity allocation during subsequent analyses.
 
For wind resources, the MERRA2 data is first bias corrected to the NREL WTK data, which is a more accurate dataset but has a limited time span. Then, the bias-corrected MERRA 2 data is used to generate the hourly unit power output as described in~\cite{doering2022evaluating}. On the other hand, the unit power output for solar resources is determined based on the bias-corrected temperature and solar radiation data from MERRA2. It is important to emphasize that other spatial and temporal modules, such as load, hydro, and dynamic transmission line ratings, are all modeled using the same MERRA2 data to maintain spatial-temporal correlation throughout the analysis.

\subsubsection{Hydro}
Given the abundance of hydropower resources in upstate NY, it is crucial to model the variability of hydropower availability for different times of the year under different climatic scenarios. In addition, water regulation practices mandate hydropower generation levels for each quarter month, particularly for larger hydropower plants, which depend on the availability of water resources. It is unrealistic to disregard available hydropower when there is excess wind and solar energy or to overuse non-existent hydropower when wind and solar have low outputs. As a result, distinct strategies are adopted for small and large hydro plants.
 
For the two major hydro plants in NYS, one located in zone A and the other in zone D, historical quarter monthly time series data for precipitation and average temperature are used as suggested by~\cite{semmendinger2022establishing}. These data are employed to predict the net basin supplies into the Great Lakes and ice conditions on the St. Lawrence River using an LSTM model~\cite{hochreiter1997long}. The predictions are then converted into quarter-monthly available hydropower. The power outputs serve as a basis for regulating the hydropower generation by enforcing the constraint that the aggregated quarter monthly dispatch from hydropower must match the available hydropower (see formulation in Supplementary Note 6). Implementing this constraint eliminates the curtailment of hydropower and prevents the unrealistic overuse of hydropower in real-world scenarios. 

On the other hand, smaller hydro plants rely heavily on the available river stream flow and are thus modeled as non-dispatchable generators, represented by the negative load. Average monthly capacity factors are calculated to account for the seasonal variation in stream flows, assuming a perfect correlation of power outputs among smaller hydro plants.

\subsubsection{Load}
The electricity demand profile in our analysis consists of three main components: baseline load, which represents the current load profile; the electrified load from residential and commercial buildings; and charging load from electric vehicles (EVs). To capture the relationship between the weather data and each load component, we leverage artificial neural networks (ANN). We present a comparison between the baseline load profile and the post-electrification load profile in Figure S11. This comparison illustrates the shift of peak demand from summer to winter due to widespread electrification.

\bmhead{Baseline load} To capture the intricate relationship between various factors influencing the load profiles, we leverage the zonal hourly load data from NYISO spanning from 2002 to 2019. Using this dataset, we train an artificial neural network (ANN) for each zone. The ANNs input temperature, load of the previous day, hour of the day, and day of the week, to generate the 24-hour load profile for a given day, as suggested by~\cite{satish2004effect, doering2022evaluating}. By incorporating the MERRA2 temperature data aggregated for each zone, the trained ANN models generate accurate hourly baseline zonal load profiles. These profiles are then allocated to individual buses to maintain the ratio of loads in the original model outlined in~\cite{liu2022open}.
 
\bmhead{Electrified building load}  We employ two advanced tools developed by NREL, namely ResStock~\cite{reyna2021us} and ComStock~\cite{parker2023comstock}, to accurately model energy consumption and potential energy savings associated with building upgrades in the residential and commercial sectors across the United States.
 
The ResStock tool enables us to simulate energy usage for five distinct types of residential buildings, including mobile homes, single-family attached and detached houses, and multi-family buildings with 2-4 or 5-plus units. ComStock simulates energy use for 14 different types of commercial buildings, including small office, medium office, large office, retail, strip mall, warehouse, primary school, secondary school, full-service restaurant, quick-service restaurant, small hotel, large hotel, hospital, and outpatient.
 
By leveraging these tools and the end-use load profile data~\cite{pigman2022end}, we estimate the energy savings potential achieved through different building upgrades. It is important to note that the energy savings are positive for fossil fuel usage, indicating a reduction, while they are negative for electricity, signifying increased electricity demand resulting from electrification. These state-level electricity savings, obtained for each building type, are available on an hourly basis for the Typical Meteorological Year, version 3 (TMY3)~\cite{wilcox2008users}.
 
To establish the connection between weather data and building load, we fit an ANN model for each building type. These models enable the prediction of electrified load based on input weather variables. By using the same MERRA2 data, our approach preserves the spatiotemporal co-variability between the electrified load and other weather-dependent modules, ensuring accurate predictions of the electrified load for different counties and years.
 
To properly scale the predicted electrified load, we consider the distribution of different building types within each county. Subsequently, the county-level loads are aggregated to the nearest bus in the power grid, ensuring alignment with the spatial representation of the power system (a comprehensive overview of our framework is provided in Supplementary Note 2). 
 
\bmhead{EV load} To accurately model the electric vehicle (EV) load, we employ the EVI-Pro Lite~\cite{wood2022evi} tool developed by NREL. The tool allows us to simulate the EV load profiles considering various factors that can influence their shape such as charging access, charging preferences, and the distribution of public and private charging stations.
 
In our study, we make certain assumptions regarding EV charging behavior because the transmission system is not sensitive to the assumptions of the EV charging load (primarily because the EV load is relatively small). We assume home charging is the preferred method, with Level 2 (L2) charging stations dominating the charging infrastructure. Additionally, since our primary focus is identifying vulnerabilities in the planned zero-carbon energy system, we assume minimal charging delays, namely no demand response in EV charging.

The simulation of EV load is conducted at the county level, taking into account the existing light-duty vehicles in each county. This allows us to capture the geographical distribution of EVs and their corresponding impact on the electricity grid. Further details regarding the selected parameters and a sample EV load profile can be found in Supplementary Note 3. 

\subsubsection{Battery}
As suggested in~\cite{scopingplan}, 19.8 GW of 8-hour duration lithium batteries is included in the model with a round trip efficiency of 85\%. The zonal capacity can be found in Table S1. The pumped hydro facility in zone E is modeled as a 12-hour battery with 1170 MW of capacity.
 
\subsubsection{Transmission Lines}
Transmission line constraints play a crucial role in the reliable operation of the power grid, particularly at the interfaces between load zones. In our study, we incorporate these constraints by including the transmission line upgrades outlined in the RNA report~\cite{Assessment2020}. In addition, we go beyond static ratings and introduce dynamic line ratings to account for the potential influence of climate change, particularly in relation to extreme heat events. This analysis also includes the new HVDC lines stated for construction in 2027.

\bmhead{Dynamic rating of transmission lines} The dynamic rating of transmission lines is modeled by the temperature-ampacity relationship described in~\cite{bartos2016impacts}. The ampacity is a function of ambient temperature, solar radiation, wind speed, and material parameters of the transmission cables. As the different cable models have a very small effect on the primary results, typical cable models identified in~\cite{bartos2016impacts} are used with MERRA2 data to derive the dynamic transmission line rating. Detailed model summary and parameter choice can be found in Supplementary Note 4.
 
\bmhead{New high voltage direct current transmission lines} The spatial imbalance between supply and demand in the NYS grid leads to significant congestion in the transmission interfaces connecting upstate and downstate regions, as discussed in~\cite{liu2023spatiotemporal}. To address this issue, NYS has contracted two new HVDC transmission lines: Clean Path New York and Champlain-Hudson. The Clean Path New York line, starting from the Fraser Substation in Delaware County, has a capacity of 1300 MW. The Champlain-Hudson line originates from Quebec and delivers hydropower with a capacity of 1250 MW. In our modeling, these HVDC lines are represented as sets of dummy generators to mimic the controllable power flow. They have the same output magnitude but opposite signs, enabling flexible control over the power flow on these lines as required. By simulating the operation and impact of these planned HVDC lines, we understand their impact on transmission capacity and facilitation of the smooth transfer of electricity across the grid.

\subsection{Alternative Scenarios Design}
The alternative SOWs design in our study aims to include a wide range of potential scenarios that acknowledge uncertainties arising from both climatic and technological factors.
 
To consider the impact of climate factors, we focus on temperature increase and design the scenarios as described in~\cite{semmendinger2022establishing}. To obtain projections for climate change, we use the CMIP6 climate model projection database~\cite{eyring2016overview}, which includes data from 46 distinct General Circulation Models (GCMs) and four different emission scenarios (SSP1-2.6, SSP2-4.5, SSP3-7.0, and SSP5-8.5). These projections are downscaled to the Great Lakes basin using the delta change method~\cite{maraun2016bias}. Monthly adjustments are made to historical precipitation data from 1952-2019, incorporating change factors derived from each GCM under each emission scenario. The projections reveal a warmer and wetter future, with temperature increases ranging from +1°C to +5.7°C  and precipitation changes between -3\% and +14\%  compared to historical averages. In total, 159 scenarios are developed. Note that not all the GCMs have the four emission scenarios and two GCMs do not provide future projections for the period of interest. These 159 scenarios represent different combinations of average temperature increase and corresponding changes to hydropower availability at a quarter-monthly time resolution.
 
Technological factors are represented by the electrification rate of buildings and electric vehicles (EVs) in NYS. We assume electrification rates range from 0.7 (representing a slower-than-expected rate) to 1.05 (indicating a slight over-electrification, potentially due to newly constructed buildings or vehicles). The chosen range is somewhat arbitrary, as the primary aim is to conduct a sensitivity analysis on the electrification rate's impact on overall system reliability. However, our findings reveal that the electrification rate is not a significant driving factor, as discussed in the main text. Additionally, we consider underbuilt/overbuilt scenarios for wind, solar, and storage units by scaling their capacities by factors ranging from 0.6 to 1.4.
 
Combining the climatic and technological factors leads to a six-dimensional sampling space. To generate a comprehensive set of combinations, we employ the Latin Hypercube sampling (LHS) technique, which ensures an even sampling from each dimension of the uncertainty space. This method outperforms random sampling approaches by avoiding clustered samples and reducing sampling variance~\cite{mckay2000comparison}. By generating 300 samples, our distributed sample set provides a better approximation of the underlying distribution within the high-dimensional sampling space than random sampling. Given that the goal of this study is to perform a sensitivity analysis to identify the failure mechanisms under different combinations of future uncertainties, 300 samples are considered sufficient to balance representation with computational requirements. 

\subsection{Enhanced DC-OPF Formulation}
The method employed in this study involves an enhanced formulation of the traditional DC-OPF (Optimal Power Flow) problem. The DC-OPF formulation is a linearized approximation that aims to optimize generator dispatch to meet demand at the lowest total cost. Since we aim to assess system reliability, we introduce a slack variable representing load shedding at each bus and minimize overall load shedding in the system.
 
In addition to the standard DC-OPF constraints, we incorporate additional constraints to account for HVDC lines, quarter-monthly hydro requirements, and battery state transitions. Furthermore, time-varying estimations for the maximum hourly outputs given the MERRA2 weather dataset are assigned to capture the upper limits of wind and solar generators, as well as upper and lower limits for transmission line capacities.
 
Given that the problem is solved at an hourly time-step for a full year length and this process is repeated for 22 years and 300 scenarios, we treat the renewable outputs calculated by historical weather data (with temperature adjustments for different climatic scenarios) as perfect forecasts to manage computation complexity. The full formulation of the problem can be found in Supplementary Note 6. We acknowledge that the uncertainties of renewable resources pose increased risks and vulnerabilities for the energy system. Future works focusing on designing accurate forecasting methodology and advanced control algorithms to manage the operation of the zero-carbon grid are extremely valuable but are out of the scope of this study.

\subsection{Gradient Boosting Tree and the Importance Score}
Gradient Boosting Trees combine multiple decision trees to make accurate predictions by iteratively building a series of weak learners, each of which corrects the errors of the previous model, resulting in a strong and highly predictive model capable of handling complex datasets. In this study, the GBT is used to map the climatic-technological factors (as features) to the violation indicators (as labels). The violation threshold is defined based on the information from as follows: the scoping plan identified 17722 MW of FZEC and claimed that 25GW of 100-hour long duration battery can ensure system reliability. We use 17722 MW as the threshold for Maximum Load Shedding, 100 hours as the threshold for load shedding hours, and 208 GWh as the threshold for load shedding quantity of a week, assuming 1/12th of the energy will be used for a quarter month from the 100-hour 25GW battery. 

The GBT method is able to capture the non-linear relationships in the feature space and uncover the most significant input factor~\cite{reed2022addressing}. The importance score quantifies the contribution of each feature in the model's decision-making process and is therefore calculated to help identify the most influential variables for the target variable.

\bmhead{Acknowledgments}

This work was supported by the National Institute for Agriculture and Food (NIFA) and National Science Foundation (NSF) INFEWS Project No. NYC-121574 (grant no. 2019-67019-30122).


\bmhead{Competing interests}
The authors declare no competing interests.


\bibliography{sn-bibliography}


\begin{thebibliography}{48}
\ifx \bisbn   \undefined \def \bisbn  #1{ISBN #1}\fi
\ifx \binits  \undefined \def \binits#1{#1}\fi
\ifx \bauthor  \undefined \def \bauthor#1{#1}\fi
\ifx \batitle  \undefined \def \batitle#1{#1}\fi
\ifx \bjtitle  \undefined \def \bjtitle#1{#1}\fi
\ifx \bvolume  \undefined \def \bvolume#1{\textbf{#1}}\fi
\ifx \byear  \undefined \def \byear#1{#1}\fi
\ifx \bissue  \undefined \def \bissue#1{#1}\fi
\ifx \bfpage  \undefined \def \bfpage#1{#1}\fi
\ifx \blpage  \undefined \def \blpage #1{#1}\fi
\ifx \burl  \undefined \def \burl#1{\textsf{#1}}\fi
\ifx \doiurl  \undefined \def \doiurl#1{\url{https://doi.org/#1}}\fi
\ifx \betal  \undefined \def \betal{\textit{et al.}}\fi
\ifx \binstitute  \undefined \def \binstitute#1{#1}\fi
\ifx \binstitutionaled  \undefined \def \binstitutionaled#1{#1}\fi
\ifx \bctitle  \undefined \def \bctitle#1{#1}\fi
\ifx \beditor  \undefined \def \beditor#1{#1}\fi
\ifx \bpublisher  \undefined \def \bpublisher#1{#1}\fi
\ifx \bbtitle  \undefined \def \bbtitle#1{#1}\fi
\ifx \bedition  \undefined \def \bedition#1{#1}\fi
\ifx \bseriesno  \undefined \def \bseriesno#1{#1}\fi
\ifx \blocation  \undefined \def \blocation#1{#1}\fi
\ifx \bsertitle  \undefined \def \bsertitle#1{#1}\fi
\ifx \bsnm \undefined \def \bsnm#1{#1}\fi
\ifx \bsuffix \undefined \def \bsuffix#1{#1}\fi
\ifx \bparticle \undefined \def \bparticle#1{#1}\fi
\ifx \barticle \undefined \def \barticle#1{#1}\fi
\bibcommenthead
\ifx \bconfdate \undefined \def \bconfdate #1{#1}\fi
\ifx \botherref \undefined \def \botherref #1{#1}\fi
\ifx \url \undefined \def \url#1{\textsf{#1}}\fi
\ifx \bchapter \undefined \def \bchapter#1{#1}\fi
\ifx \bbook \undefined \def \bbook#1{#1}\fi
\ifx \bcomment \undefined \def \bcomment#1{#1}\fi
\ifx \oauthor \undefined \def \oauthor#1{#1}\fi
\ifx \citeauthoryear \undefined \def \citeauthoryear#1{#1}\fi
\ifx \endbibitem  \undefined \def \endbibitem {}\fi
\ifx \bconflocation  \undefined \def \bconflocation#1{#1}\fi
\ifx \arxivurl  \undefined \def \arxivurl#1{\textsf{#1}}\fi
\csname PreBibitemsHook\endcsname

\bibitem[\protect\citeauthoryear{Bashmakov et~al.}{2022}]{bashmakov2022climate}
\begin{botherref}
\oauthor{\bsnm{Bashmakov}, \binits{I.}},
\oauthor{\bsnm{Nilsson}, \binits{L.}},
\oauthor{\bsnm{Acquaye}, \binits{A.}},
\oauthor{\bsnm{Bataille}, \binits{C.}},
\oauthor{\bsnm{Cullen}, \binits{J.}},
\oauthor{\bsnm{Fischedick}, \binits{M.}},
\oauthor{\bsnm{Geng}, \binits{Y.}},
\oauthor{\bsnm{Tanaka}, \binits{K.}}, et al.:
Climate change 2022: Mitigation of climate change. contribution of working group iii to the sixth assessment report of the intergovernmental panel on climate change, chapter 11
(2022)
\end{botherref}
\endbibitem

\bibitem[\protect\citeauthoryear{Sun et~al.}{2020}]{sun2020review}
\begin{barticle}
\bauthor{\bsnm{Sun}, \binits{K.}},
\bauthor{\bsnm{Xiao}, \binits{H.}},
\bauthor{\bsnm{Liu}, \binits{S.}},
\bauthor{\bsnm{You}, \binits{S.}},
\bauthor{\bsnm{Yang}, \binits{F.}},
\bauthor{\bsnm{Dong}, \binits{Y.}},
\bauthor{\bsnm{Wang}, \binits{W.}},
\bauthor{\bsnm{Liu}, \binits{Y.}}:
\batitle{A review of clean electricity policies—from countries to utilities}.
\bjtitle{Sustainability}
\bvolume{12}(\bissue{19}),
\bfpage{7946}
(\byear{2020})
\end{barticle}
\endbibitem

\bibitem[\protect\citeauthoryear{Yalew et~al.}{2020}]{yalew2020impacts}
\begin{barticle}
\bauthor{\bsnm{Yalew}, \binits{S.G.}},
\bauthor{\bsnm{Vliet}, \binits{M.T.}},
\bauthor{\bsnm{Gernaat}, \binits{D.E.}},
\bauthor{\bsnm{Ludwig}, \binits{F.}},
\bauthor{\bsnm{Miara}, \binits{A.}},
\bauthor{\bsnm{Park}, \binits{C.}},
\bauthor{\bsnm{Byers}, \binits{E.}},
\bauthor{\bsnm{De~Cian}, \binits{E.}},
\bauthor{\bsnm{Piontek}, \binits{F.}},
\bauthor{\bsnm{Iyer}, \binits{G.}}, \betal:
\batitle{Impacts of climate change on energy systems in global and regional scenarios}.
\bjtitle{Nature Energy}
\bvolume{5}(\bissue{10}),
\bfpage{794}--\blpage{802}
(\byear{2020})
\end{barticle}
\endbibitem

\bibitem[\protect\citeauthoryear{Crook et~al.}{2011}]{crook2011climate}
\begin{barticle}
\bauthor{\bsnm{Crook}, \binits{J.A.}},
\bauthor{\bsnm{Jones}, \binits{L.A.}},
\bauthor{\bsnm{Forster}, \binits{P.M.}},
\bauthor{\bsnm{Crook}, \binits{R.}}:
\batitle{Climate change impacts on future photovoltaic and concentrated solar power energy output}.
\bjtitle{Energy \& Environmental Science}
\bvolume{4}(\bissue{9}),
\bfpage{3101}--\blpage{3109}
(\byear{2011})
\end{barticle}
\endbibitem

\bibitem[\protect\citeauthoryear{Schaeffer et~al.}{2012}]{schaeffer2012energy}
\begin{barticle}
\bauthor{\bsnm{Schaeffer}, \binits{R.}},
\bauthor{\bsnm{Szklo}, \binits{A.S.}},
\bauthor{\bsnm{Lucena}, \binits{A.F.P.}},
\bauthor{\bsnm{Borba}, \binits{B.S.M.C.}},
\bauthor{\bsnm{Nogueira}, \binits{L.P.P.}},
\bauthor{\bsnm{Fleming}, \binits{F.P.}},
\bauthor{\bsnm{Troccoli}, \binits{A.}},
\bauthor{\bsnm{Harrison}, \binits{M.}},
\bauthor{\bsnm{Boulahya}, \binits{M.S.}}:
\batitle{Energy sector vulnerability to climate change: A review}.
\bjtitle{Energy}
\bvolume{38}(\bissue{1}),
\bfpage{1}--\blpage{12}
(\byear{2012})
\end{barticle}
\endbibitem

\bibitem[\protect\citeauthoryear{Tarroja et~al.}{2018}]{tarroja2018translating}
\begin{barticle}
\bauthor{\bsnm{Tarroja}, \binits{B.}},
\bauthor{\bsnm{Chiang}, \binits{F.}},
\bauthor{\bsnm{AghaKouchak}, \binits{A.}},
\bauthor{\bsnm{Samuelsen}, \binits{S.}},
\bauthor{\bsnm{Raghavan}, \binits{S.V.}},
\bauthor{\bsnm{Wei}, \binits{M.}},
\bauthor{\bsnm{Sun}, \binits{K.}},
\bauthor{\bsnm{Hong}, \binits{T.}}:
\batitle{Translating climate change and heating system electrification impacts on building energy use to future greenhouse gas emissions and electric grid capacity requirements in california}.
\bjtitle{Applied energy}
\bvolume{225},
\bfpage{522}--\blpage{534}
(\byear{2018})
\end{barticle}
\endbibitem

\bibitem[\protect\citeauthoryear{}{2019}]{NYISOphaseI2019}
\begin{botherref}
{N}ew {Y}ork {ISO} {C}limate {C}hange {I}mpact {S}tudy - {Phase I}: {L}ong-term {L}oad {I}mpact.
Technical report,
Itron Incorporation
(2019)
\end{botherref}
\endbibitem

\bibitem[\protect\citeauthoryear{}{2022}]{NYtrends2022}
\begin{botherref}
{P}ower {T}rends 2022: {T}he {P}ath to a {R}eliable, {G}reener {G}rid for {N}ew {Y}ork.
Technical report,
THE NEW YORK ISO ANNUAL GRID \& MARKETS REPORT
(2022)
\end{botherref}
\endbibitem

\bibitem[\protect\citeauthoryear{Doering and Steinschneider}{2018}]{doering2018summer}
\begin{barticle}
\bauthor{\bsnm{Doering}, \binits{K.}},
\bauthor{\bsnm{Steinschneider}, \binits{S.}}:
\batitle{Summer covariability of surface climate for renewable energy across the contiguous united states: Role of the north atlantic subtropical high}.
\bjtitle{Journal of Applied Meteorology and Climatology}
\bvolume{57}(\bissue{12}),
\bfpage{2749}--\blpage{2768}
(\byear{2018})
\end{barticle}
\endbibitem

\bibitem[\protect\citeauthoryear{Kabir et~al.}{2023}]{ElnazK2023}
\begin{botherref}
\oauthor{\bsnm{Kabir}, \binits{E.}},
\oauthor{\bsnm{Srikrishnan}, \binits{V.}},
\oauthor{\bsnm{Liu}, \binits{M.V.}},
\oauthor{\bsnm{Steinschneider}, \binits{S.}},
\oauthor{\bsnm{Anderson}, \binits{C.L.}}:
Quantifying the multi-scale and multi-resource impacts of large-scale adoption of renewable energy sources.
arXiv preprint arXiv:2307.11076
(2023)
\end{botherref}
\endbibitem

\bibitem[\protect\citeauthoryear{Perera et~al.}{2020}]{perera2020quantifying}
\begin{barticle}
\bauthor{\bsnm{Perera}, \binits{A.}},
\bauthor{\bsnm{Nik}, \binits{V.M.}},
\bauthor{\bsnm{Chen}, \binits{D.}},
\bauthor{\bsnm{Scartezzini}, \binits{J.-L.}},
\bauthor{\bsnm{Hong}, \binits{T.}}:
\batitle{Quantifying the impacts of climate change and extreme climate events on energy systems}.
\bjtitle{Nature Energy}
\bvolume{5}(\bissue{2}),
\bfpage{150}--\blpage{159}
(\byear{2020})
\end{barticle}
\endbibitem

\bibitem[\protect\citeauthoryear{Lund et~al.}{2021}]{lund2021energyplan}
\begin{barticle}
\bauthor{\bsnm{Lund}, \binits{H.}},
\bauthor{\bsnm{Thellufsen}, \binits{J.Z.}},
\bauthor{\bsnm{{\O}stergaard}, \binits{P.A.}},
\bauthor{\bsnm{Sorkn{\ae}s}, \binits{P.}},
\bauthor{\bsnm{Skov}, \binits{I.R.}},
\bauthor{\bsnm{Mathiesen}, \binits{B.V.}}:
\batitle{Energyplan--advanced analysis of smart energy systems}.
\bjtitle{Smart Energy}
\bvolume{1},
\bfpage{100007}
(\byear{2021})
\end{barticle}
\endbibitem

\bibitem[\protect\citeauthoryear{Scholz}{2012}]{scholz2012renewable}
\begin{botherref}
\oauthor{\bsnm{Scholz}, \binits{Y.}}:
Renewable energy based electricity supply at low costs: development of the remix model and application for europe
(2012)
\end{botherref}
\endbibitem

\bibitem[\protect\citeauthoryear{Keppo et~al.}{2021}]{keppo2021exploring}
\begin{barticle}
\bauthor{\bsnm{Keppo}, \binits{I.}},
\bauthor{\bsnm{Butnar}, \binits{I.}},
\bauthor{\bsnm{Bauer}, \binits{N.}},
\bauthor{\bsnm{Caspani}, \binits{M.}},
\bauthor{\bsnm{Edelenbosch}, \binits{O.}},
\bauthor{\bsnm{Emmerling}, \binits{J.}},
\bauthor{\bsnm{Fragkos}, \binits{P.}},
\bauthor{\bsnm{Guivarch}, \binits{C.}},
\bauthor{\bsnm{Harmsen}, \binits{M.}},
\bauthor{\bsnm{Lefevre}, \binits{J.}}, \betal:
\batitle{Exploring the possibility space: taking stock of the diverse capabilities and gaps in integrated assessment models}.
\bjtitle{Environmental Research Letters}
\bvolume{16}(\bissue{5}),
\bfpage{053006}
(\byear{2021})
\end{barticle}
\endbibitem

\bibitem[\protect\citeauthoryear{Luderer et~al.}{2022}]{luderer2022impact}
\begin{barticle}
\bauthor{\bsnm{Luderer}, \binits{G.}},
\bauthor{\bsnm{Madeddu}, \binits{S.}},
\bauthor{\bsnm{Merfort}, \binits{L.}},
\bauthor{\bsnm{Ueckerdt}, \binits{F.}},
\bauthor{\bsnm{Pehl}, \binits{M.}},
\bauthor{\bsnm{Pietzcker}, \binits{R.}},
\bauthor{\bsnm{Rottoli}, \binits{M.}},
\bauthor{\bsnm{Schreyer}, \binits{F.}},
\bauthor{\bsnm{Bauer}, \binits{N.}},
\bauthor{\bsnm{Baumstark}, \binits{L.}}, \betal:
\batitle{Impact of declining renewable energy costs on electrification in low-emission scenarios}.
\bjtitle{Nature Energy}
\bvolume{7}(\bissue{1}),
\bfpage{32}--\blpage{42}
(\byear{2022})
\end{barticle}
\endbibitem

\bibitem[\protect\citeauthoryear{Dranka and Ferreira}{2018}]{brazil_dranka2018planning}
\begin{barticle}
\bauthor{\bsnm{Dranka}, \binits{G.G.}},
\bauthor{\bsnm{Ferreira}, \binits{P.}}:
\batitle{Planning for a renewable future in the brazilian power system}.
\bjtitle{Energy}
\bvolume{164},
\bfpage{496}--\blpage{511}
(\byear{2018})
\end{barticle}
\endbibitem

\bibitem[\protect\citeauthoryear{Fernandes and Ferreira}{2014}]{Portuguese_fernandes2014renewable}
\begin{barticle}
\bauthor{\bsnm{Fernandes}, \binits{L.}},
\bauthor{\bsnm{Ferreira}, \binits{P.}}:
\batitle{Renewable energy scenarios in the portuguese electricity system}.
\bjtitle{Energy}
\bvolume{69},
\bfpage{51}--\blpage{57}
(\byear{2014})
\end{barticle}
\endbibitem

\bibitem[\protect\citeauthoryear{{\'C}osi{\'c} et~al.}{2012}]{Macedonia_cosic2012100}
\begin{barticle}
\bauthor{\bsnm{{\'C}osi{\'c}}, \binits{B.}},
\bauthor{\bsnm{Kraja{\v{c}}i{\'c}}, \binits{G.}},
\bauthor{\bsnm{Dui{\'c}}, \binits{N.}}:
\batitle{A 100\% renewable energy system in the year 2050: The case of macedonia}.
\bjtitle{Energy}
\bvolume{48}(\bissue{1}),
\bfpage{80}--\blpage{87}
(\byear{2012})
\end{barticle}
\endbibitem

\bibitem[\protect\citeauthoryear{Scholz et~al.}{2017}]{scholz2017application}
\begin{barticle}
\bauthor{\bsnm{Scholz}, \binits{Y.}},
\bauthor{\bsnm{Gils}, \binits{H.C.}},
\bauthor{\bsnm{Pietzcker}, \binits{R.C.}}:
\batitle{Application of a high-detail energy system model to derive power sector characteristics at high wind and solar shares}.
\bjtitle{Energy Economics}
\bvolume{64},
\bfpage{568}--\blpage{582}
(\byear{2017})
\end{barticle}
\endbibitem

\bibitem[\protect\citeauthoryear{Liu et~al.}{2023}]{liu2023spatiotemporal}
\begin{bchapter}
\bauthor{\bsnm{Liu}, \binits{M.}},
\bauthor{\bsnm{Doering}, \binits{K.}},
\bauthor{\bsnm{Gupta}, \binits{A.}},
\bauthor{\bsnm{Anderson}, \binits{C.L.}}:
\bctitle{A spatiotemporal analysis of new york state grid transition under the clcpa energy strategy}.
(\byear{2023})
\end{bchapter}
\endbibitem

\bibitem[\protect\citeauthoryear{{New York State Climate Action Council}}{2022}]{scopingplan}
\begin{botherref}
\oauthor{\bsnm{{New York State Climate Action Council}}}:
{{N}ew {Y}ork {S}tate {C}limate {A}ction {C}ouncil {S}coping {P}lan}.
\url{https://www.climate.ny.gov/ScopingPlan}
(2022)
\end{botherref}
\endbibitem

\bibitem[\protect\citeauthoryear{Cherry et~al.}{2005}]{cherry2005impacts}
\begin{barticle}
\bauthor{\bsnm{Cherry}, \binits{J.}},
\bauthor{\bsnm{Cullen}, \binits{H.}},
\bauthor{\bsnm{Visbeck}, \binits{M.}},
\bauthor{\bsnm{Small}, \binits{A.}},
\bauthor{\bsnm{Uvo}, \binits{C.}}:
\batitle{Impacts of the north atlantic oscillation on scandinavian hydropower production and energy markets}.
\bjtitle{Water resources management}
\bvolume{19},
\bfpage{673}--\blpage{691}
(\byear{2005})
\end{barticle}
\endbibitem

\bibitem[\protect\citeauthoryear{Molod et~al.}{2015}]{MERRA2}
\begin{barticle}
\bauthor{\bsnm{Molod}, \binits{A.}},
\bauthor{\bsnm{Takacs}, \binits{L.}},
\bauthor{\bsnm{Suarez}, \binits{M.}},
\bauthor{\bsnm{Bacmeister}, \binits{J.}}:
\batitle{Development of the geos-5 atmospheric general circulation model: Evolution from merra to merra2}.
\bjtitle{Geoscientific Model Development}
\bvolume{8}(\bissue{5}),
\bfpage{1339}--\blpage{1356}
(\byear{2015})
\end{barticle}
\endbibitem

\bibitem[\protect\citeauthoryear{{New York Independent System Operator}}{2020}]{Assessment2020}
\begin{botherref}
\oauthor{\bsnm{{New York Independent System Operator}}}:
{2020 {R}eliability {N}eeds {A}ssessment}.
Technical report
(2020)
\end{botherref}
\endbibitem

\bibitem[\protect\citeauthoryear{Loh}{1996}]{loh1996latin}
\begin{barticle}
\bauthor{\bsnm{Loh}, \binits{W.-L.}}:
\batitle{On latin hypercube sampling}.
\bjtitle{The annals of statistics}
\bvolume{24}(\bissue{5}),
\bfpage{2058}--\blpage{2080}
(\byear{1996})
\end{barticle}
\endbibitem

\bibitem[\protect\citeauthoryear{Walker et~al.}{2010}]{walker2010addressing}
\begin{barticle}
\bauthor{\bsnm{Walker}, \binits{W.E.}},
\bauthor{\bsnm{Marchau}, \binits{V.A.}},
\bauthor{\bsnm{Swanson}, \binits{D.}}:
\batitle{Addressing deep uncertainty using adaptive policies: Introduction to section 2}.
\bjtitle{Technological forecasting and social change}
\bvolume{77}(\bissue{6}),
\bfpage{917}--\blpage{923}
(\byear{2010})
\end{barticle}
\endbibitem

\bibitem[\protect\citeauthoryear{Baumann et~al.}{2019}]{baumann2019review}
\begin{barticle}
\bauthor{\bsnm{Baumann}, \binits{M.}},
\bauthor{\bsnm{Weil}, \binits{M.}},
\bauthor{\bsnm{Peters}, \binits{J.F.}},
\bauthor{\bsnm{Chibeles-Martins}, \binits{N.}},
\bauthor{\bsnm{Moniz}, \binits{A.B.}}:
\batitle{A review of multi-criteria decision making approaches for evaluating energy storage systems for grid applications}.
\bjtitle{Renewable and Sustainable Energy Reviews}
\bvolume{107},
\bfpage{516}--\blpage{534}
(\byear{2019})
\end{barticle}
\endbibitem

\bibitem[\protect\citeauthoryear{Drucker and Cortes}{1995}]{drucker1995boosting}
\begin{botherref}
\oauthor{\bsnm{Drucker}, \binits{H.}},
\oauthor{\bsnm{Cortes}, \binits{C.}}:
Boosting decision trees.
Advances in neural information processing systems
\textbf{8}
(1995)
\end{botherref}
\endbibitem

\bibitem[\protect\citeauthoryear{Bloomfield et~al.}{2021}]{bloomfield2021quantifying}
\begin{barticle}
\bauthor{\bsnm{Bloomfield}, \binits{H.}},
\bauthor{\bsnm{Brayshaw}, \binits{D.}},
\bauthor{\bsnm{Troccoli}, \binits{A.}},
\bauthor{\bsnm{Goodess}, \binits{C.}},
\bauthor{\bsnm{De~Felice}, \binits{M.}},
\bauthor{\bsnm{Dubus}, \binits{L.}},
\bauthor{\bsnm{Bett}, \binits{P.}},
\bauthor{\bsnm{Saint-Drenan}, \binits{Y.-M.}}:
\batitle{Quantifying the sensitivity of european power systems to energy scenarios and climate change projections}.
\bjtitle{Renewable Energy}
\bvolume{164},
\bfpage{1062}--\blpage{1075}
(\byear{2021})
\end{barticle}
\endbibitem

\bibitem[\protect\citeauthoryear{Faranda et~al.}{2022}]{wcd-3-1311-2022}
\begin{barticle}
\bauthor{\bsnm{Faranda}, \binits{D.}},
\bauthor{\bsnm{Bourdin}, \binits{S.}},
\bauthor{\bsnm{Ginesta}, \binits{M.}},
\bauthor{\bsnm{Krouma}, \binits{M.}},
\bauthor{\bsnm{Noyelle}, \binits{R.}},
\bauthor{\bsnm{Pons}, \binits{F.}},
\bauthor{\bsnm{Yiou}, \binits{P.}},
\bauthor{\bsnm{Messori}, \binits{G.}}:
\batitle{A climate-change attribution retrospective of some impactful weather extremes of 2021}.
\bjtitle{Weather and Climate Dynamics}
\bvolume{3}(\bissue{4}),
\bfpage{1311}--\blpage{1340}
(\byear{2022})
\doiurl{10.5194/wcd-3-1311-2022}
\end{barticle}
\endbibitem

\bibitem[\protect\citeauthoryear{}{}]{CLCPA}
\begin{botherref}
{C}limate {L}eadership and {C}ommunity {P}rotection {A}ct {(CLCPA)}.
\url{https://climate.ny.gov/-/media/Project/Climate/Files/Draft-Scoping-Plan.pdf}.
(accessed November 15, 2022)
\end{botherref}
\endbibitem

\bibitem[\protect\citeauthoryear{Liu et~al.}{2022}]{liu2022open}
\begin{botherref}
\oauthor{\bsnm{Liu}, \binits{M.V.}},
\oauthor{\bsnm{Yuan}, \binits{B.}},
\oauthor{\bsnm{Wang}, \binits{Z.}},
\oauthor{\bsnm{Sward}, \binits{J.A.}},
\oauthor{\bsnm{Zhang}, \binits{K.M.}},
\oauthor{\bsnm{Anderson}, \binits{C.L.}}:
An open source representation for the nys electric grid to support power grid and market transition studies.
IEEE Transactions on Power Systems
(2022)
\end{botherref}
\endbibitem

\bibitem[\protect\citeauthoryear{}{}]{WTK}
\begin{botherref}
{N}ational {R}enewable {E}nergy {L}ab ({NREL}) {W}ind {I}ntegration {N}ational {D}ataset (WIND) {T}oolkit (WTK).
\url{https://data.nrel.gov/submissions/54}.
(accessed Jun 28, 2021)
\end{botherref}
\endbibitem

\bibitem[\protect\citeauthoryear{Bloom et~al.}{2016}]{SIND}
\begin{botherref}
\oauthor{\bsnm{Bloom}, \binits{A.}},
\oauthor{\bsnm{Townsend}, \binits{A.}},
\oauthor{\bsnm{Palchak}, \binits{D.}},
\oauthor{\bsnm{Novacheck}, \binits{J.}},
\oauthor{\bsnm{King}, \binits{J.}},
\oauthor{\bsnm{Barrows}, \binits{C.}},
\oauthor{\bsnm{Ibanez}, \binits{E.}},
\oauthor{\bsnm{O'Connell}, \binits{M.}},
\oauthor{\bsnm{Jordan}, \binits{G.}},
\oauthor{\bsnm{Roberts}, \binits{B.}}, et al.:
{Eastern Renewable Generation Integration Study}.
Technical report,
National Renewable Energy Lab.(NREL), Golden, CO ({United States})
(2016)
\end{botherref}
\endbibitem

\bibitem[\protect\citeauthoryear{Doering et~al.}{2022}]{doering2022evaluating}
\begin{barticle}
\bauthor{\bsnm{Doering}, \binits{K.}},
\bauthor{\bsnm{Anderson}, \binits{C.L.}},
\bauthor{\bsnm{Steinschneider}, \binits{S.}}:
\batitle{Evaluating the intensity, duration and frequency of flexible energy resources needed in a zero-emission, hydropower reliant power system}.
\bjtitle{Oxford Open Energy}
\bvolume{1},
\bfpage{003}
(\byear{2022})
\end{barticle}
\endbibitem

\bibitem[\protect\citeauthoryear{Semmendinger et~al.}{2022}]{semmendinger2022establishing}
\begin{barticle}
\bauthor{\bsnm{Semmendinger}, \binits{K.}},
\bauthor{\bsnm{Lee}, \binits{D.}},
\bauthor{\bsnm{Fry}, \binits{L.}},
\bauthor{\bsnm{Steinschneider}, \binits{S.}}:
\batitle{Establishing opportunities and limitations of forecast use in the operational management of highly constrained multiobjective water systems}.
\bjtitle{Journal of Water Resources Planning and Management}
\bvolume{148}(\bissue{8}),
\bfpage{04022044}
(\byear{2022})
\end{barticle}
\endbibitem

\bibitem[\protect\citeauthoryear{Hochreiter and Schmidhuber}{1997}]{hochreiter1997long}
\begin{barticle}
\bauthor{\bsnm{Hochreiter}, \binits{S.}},
\bauthor{\bsnm{Schmidhuber}, \binits{J.}}:
\batitle{Long short-term memory}.
\bjtitle{Neural computation}
\bvolume{9}(\bissue{8}),
\bfpage{1735}--\blpage{1780}
(\byear{1997})
\end{barticle}
\endbibitem

\bibitem[\protect\citeauthoryear{Satish et~al.}{2004}]{satish2004effect}
\begin{barticle}
\bauthor{\bsnm{Satish}, \binits{B.}},
\bauthor{\bsnm{Swarup}, \binits{K.}},
\bauthor{\bsnm{Srinivas}, \binits{S.}},
\bauthor{\bsnm{Rao}, \binits{A.H.}}:
\batitle{Effect of temperature on short term load forecasting using an integrated ann}.
\bjtitle{Electric Power Systems Research}
\bvolume{72}(\bissue{1}),
\bfpage{95}--\blpage{101}
(\byear{2004})
\end{barticle}
\endbibitem

\bibitem[\protect\citeauthoryear{Reyna et~al.}{2021}]{reyna2021us}
\begin{botherref}
\oauthor{\bsnm{Reyna}, \binits{J.}},
\oauthor{\bsnm{Wilson}, \binits{E.}},
\oauthor{\bsnm{Parker}, \binits{A.}},
\oauthor{\bsnm{Satre-Meloy}, \binits{A.}},
\oauthor{\bsnm{Egerter}, \binits{A.}},
\oauthor{\bsnm{Bianchi}, \binits{C.}},
\oauthor{\bsnm{Praprost}, \binits{M.}},
\oauthor{\bsnm{Speake}, \binits{A.}},
\oauthor{\bsnm{Liu}, \binits{L.}},
\oauthor{\bsnm{Horsey}, \binits{R.}}, et al.:
Us building stock characterization study: A national typology for decarbonizing us buildings.
Technical report,
National Renewable Energy Lab.(NREL), Golden, CO (United States)
(2021)
\end{botherref}
\endbibitem

\bibitem[\protect\citeauthoryear{Parker et~al.}{2023}]{parker2023comstock}
\begin{botherref}
\oauthor{\bsnm{Parker}, \binits{A.}},
\oauthor{\bsnm{Horsey}, \binits{H.}},
\oauthor{\bsnm{Dahlhausen}, \binits{M.}},
\oauthor{\bsnm{Praprost}, \binits{M.}},
\oauthor{\bsnm{CaraDonna}, \binits{C.}},
\oauthor{\bsnm{LeBar}, \binits{A.}},
\oauthor{\bsnm{Klun}, \binits{L.}}:
Comstock reference documentation: Version 1.
Technical report,
National Renewable Energy Lab.(NREL), Golden, CO (United States)
(2023)
\end{botherref}
\endbibitem

\bibitem[\protect\citeauthoryear{Pigman et~al.}{2022}]{pigman2022end}
\begin{botherref}
\oauthor{\bsnm{Pigman}, \binits{M.}},
\oauthor{\bsnm{Frick}, \binits{N.M.}},
\oauthor{\bsnm{Wilson}, \binits{E.}},
\oauthor{\bsnm{Parker}, \binits{A.}},
\oauthor{\bsnm{Present}, \binits{E.}}:
End-use load profiles for the us building stock: Practical guidance on accessing and using the data.
Technical report,
Lawrence Berkeley National Lab.(LBNL), Berkeley, CA (United States)
(2022)
\end{botherref}
\endbibitem

\bibitem[\protect\citeauthoryear{Wilcox and Marion}{2008}]{wilcox2008users}
\begin{botherref}
\oauthor{\bsnm{Wilcox}, \binits{S.}},
\oauthor{\bsnm{Marion}, \binits{W.}}:
Users manual for tmy3 data sets
(2008)
\end{botherref}
\endbibitem

\bibitem[\protect\citeauthoryear{Wood}{2022}]{wood2022evi}
\begin{botherref}
\oauthor{\bsnm{Wood}, \binits{E.}}:
Evi-pro lite api.
Technical report,
Livewire Data Platform; NREL; Pacific Northwest National Lab.(PNNL~…
(2022)
\end{botherref}
\endbibitem

\bibitem[\protect\citeauthoryear{Bartos et~al.}{2016}]{bartos2016impacts}
\begin{barticle}
\bauthor{\bsnm{Bartos}, \binits{M.}},
\bauthor{\bsnm{Chester}, \binits{M.}},
\bauthor{\bsnm{Johnson}, \binits{N.}},
\bauthor{\bsnm{Gorman}, \binits{B.}},
\bauthor{\bsnm{Eisenberg}, \binits{D.}},
\bauthor{\bsnm{Linkov}, \binits{I.}},
\bauthor{\bsnm{Bates}, \binits{M.}}:
\batitle{Impacts of rising air temperatures on electric transmission ampacity and peak electricity load in the united states}.
\bjtitle{Environmental Research Letters}
\bvolume{11}(\bissue{11}),
\bfpage{114008}
(\byear{2016})
\end{barticle}
\endbibitem

\bibitem[\protect\citeauthoryear{Eyring et~al.}{2016}]{eyring2016overview}
\begin{barticle}
\bauthor{\bsnm{Eyring}, \binits{V.}},
\bauthor{\bsnm{Bony}, \binits{S.}},
\bauthor{\bsnm{Meehl}, \binits{G.A.}},
\bauthor{\bsnm{Senior}, \binits{C.A.}},
\bauthor{\bsnm{Stevens}, \binits{B.}},
\bauthor{\bsnm{Stouffer}, \binits{R.J.}},
\bauthor{\bsnm{Taylor}, \binits{K.E.}}:
\batitle{Overview of the coupled model intercomparison project phase 6 (cmip6) experimental design and organization}.
\bjtitle{Geoscientific Model Development}
\bvolume{9}(\bissue{5}),
\bfpage{1937}--\blpage{1958}
(\byear{2016})
\end{barticle}
\endbibitem

\bibitem[\protect\citeauthoryear{Maraun}{2016}]{maraun2016bias}
\begin{barticle}
\bauthor{\bsnm{Maraun}, \binits{D.}}:
\batitle{Bias correcting climate change simulations-a critical review}.
\bjtitle{Current Climate Change Reports}
\bvolume{2},
\bfpage{211}--\blpage{220}
(\byear{2016})
\end{barticle}
\endbibitem

\bibitem[\protect\citeauthoryear{McKay et~al.}{2000}]{mckay2000comparison}
\begin{barticle}
\bauthor{\bsnm{McKay}, \binits{M.D.}},
\bauthor{\bsnm{Beckman}, \binits{R.J.}},
\bauthor{\bsnm{Conover}, \binits{W.J.}}:
\batitle{A comparison of three methods for selecting values of input variables in the analysis of output from a computer code}.
\bjtitle{Technometrics}
\bvolume{42}(\bissue{1}),
\bfpage{55}--\blpage{61}
(\byear{2000})
\end{barticle}
\endbibitem

\bibitem[\protect\citeauthoryear{Reed et~al.}{2022}]{reed2022addressing}
\begin{botherref}
\oauthor{\bsnm{Reed}, \binits{P.}},
\oauthor{\bsnm{Hadjimichael}, \binits{A.}},
\oauthor{\bsnm{Malek}, \binits{K.}},
\oauthor{\bsnm{Karimi}, \binits{T.}},
\oauthor{\bsnm{Vernon}, \binits{C.R.}},
\oauthor{\bsnm{Srikrishnan}, \binits{V.}},
\oauthor{\bsnm{Gupta}, \binits{R.S.}},
\oauthor{\bsnm{Gold}, \binits{D.F.}},
\oauthor{\bsnm{Lee}, \binits{B.}},
\oauthor{\bsnm{Keller}, \binits{K.}}, et al.:
Addressing Uncertainty in Multisector Dynamics Research. Zenodo
(2022)
\end{botherref}
\endbibitem

\end{thebibliography}


\begin{thebibliography}{10}
\newcommand{\enquote}[1]{``#1''}

\bibitem{nys2tail}
{New York Independent System Operator}, \enquote{New transmission investments
  add a new chapter to the “tale of two grids”,}
  \url{https://www.nyiso.com/-/new-transmission-investments-add-a-new-chapter-to-the-tale-of-two-grids-}
  (2022). Accessed: 2023-05-23.

\bibitem{scopingplan}
{New York State Climate Action Council}, \enquote{{{N}ew {Y}ork {S}tate
  {C}limate {A}ction {C}ouncil {S}coping {P}lan},}
  \url{https://www.climate.ny.gov/ScopingPlan} (2022).

\bibitem{Assessment2020}
N.~Y. I.~S. Operator, \enquote{2020 {R}eliability {N}eeds {A}ssessment,} Tech.
  rep., New York Independent System Operator, Rensselaer, NY 12144 (2020).

\bibitem{sepulveda2018role}
N.~A. Sepulveda, J.~D. Jenkins, F.~J. De~Sisternes, and R.~K. Lester,
  \enquote{The role of firm low-carbon electricity resources in deep
  decarbonization of power generation,} {\protect\JournalTitle{Joule}}
  \textbf{2}, 2403--2420 (2018).

\bibitem{CHPE}
{New York State Climate Action Council}, \enquote{{{C}anada to {N}ew {Y}ork
  {C}ity hydropower pipeline breaks ground. Here’s what it means.}}
  \url{https://www.silive.com/news/2022/11/canada-to-new-york-city-hydropower-pipeline-breaks-ground-heres-what-it-means.html}
  (2022).

\bibitem{NYCP}
{New York State Climate Action Council}, \enquote{{{C}lean {P}ath {NY}
  {A}nnounces {P}ublic {I}nformation {O}pen {H}ouses in {B}ronx {C}ounty},
  howpublished =
  "\url{https://www.cleanpathny.com/cpny-announces-\%20openhouses-bronx}", year
  = {2022},} .

\bibitem{reyna2021us}
J.~Reyna, E.~Wilson, A.~Parker, A.~Satre-Meloy, A.~Egerter, C.~Bianchi,
  M.~Praprost, A.~Speake, L.~Liu, R.~Horsey \emph{et~al.}, \enquote{Us building
  stock characterization study: A national typology for decarbonizing us
  buildings,} Tech. rep., National Renewable Energy Lab.(NREL), Golden, CO
  (United States) (2021).

\bibitem{parker2023comstock}
A.~Parker, H.~Horsey, M.~Dahlhausen, M.~Praprost, C.~CaraDonna, A.~LeBar, and
  L.~Klun, \enquote{Comstock reference documentation: Version 1,} Tech. rep.,
  National Renewable Energy Lab.(NREL), Golden, CO (United States) (2023).

\bibitem{NYVehicle}
{NY Open Data}, \enquote{Vehicle, snowmobile, and boat registrations,}
  \url{https://data.ny.gov/Transportation/Vehicle-Snowmobile-and-Boat-Registrations/w4pv-hbkt}
  (2023). Accessed: 2023-03-08.

\bibitem{NYpopuden}
{Department of Health}, \enquote{Vital statistics of new york state 2018,}
  \url{https://www.health.ny.gov/statistics/vital_statistics/2018/table02.htm}
  (2020). Accessed: 2023-03-08.

\bibitem{wood2022evi}
E.~Wood, \enquote{Evi-pro lite api,} Tech. rep., Livewire Data Platform; NREL;
  Pacific Northwest National Lab.(PNNL~… (2022).

\bibitem{bartos2016impacts}
M.~Bartos, M.~Chester, N.~Johnson, B.~Gorman, D.~Eisenberg, I.~Linkov, and
  M.~Bates, \enquote{Impacts of rising air temperatures on electric
  transmission ampacity and peak electricity load in the united states,}
  {\protect\JournalTitle{Environmental Research Letters}} \textbf{11}, 114008
  (2016).

\bibitem{NYISOphaseI2019}
\enquote{{N}ew {Y}ork {ISO} {C}limate {C}hange {I}mpact {S}tudy - {Phase I}:
  {L}ong-term {L}oad {I}mpact,} Tech. rep., Itron Incorporation (2019).

\bibitem{Dyresult}
A.~Dhabi, \enquote{Innovation landscape brief: Dynamic line rating,,} Tech.
  Rep. ISBN 978-92-9260-182-9, International Renewable Energy Agency (2020).

\end{thebibliography}

\end{document}


\maketitle

\section{Detailed NYS grid features and the CLCPA plan}

\begin{figure}[H]%
\centering
\includegraphics[width=0.9\textwidth]{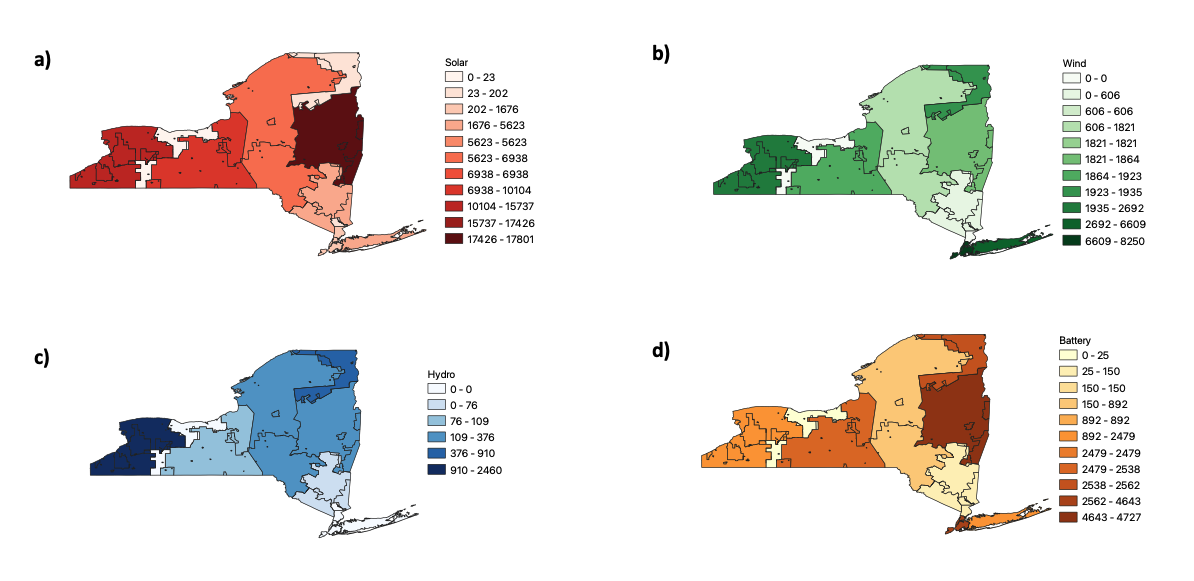}
\caption[Overview of zonal CLCPA renewable resources capacities]%
{\textbf{Overview of zonal CLCPA renewable resources capacities} }\label{NYScapacity}
\end{figure}

The upstate zones A-E had 92\% of zero-emission generation with a combination of hydro, nuclear, and wind in 2021, whereas downstate zones (F-K) had 8\% of zero-emission generation~\cite{nys2tail}. As the CLCPA mandates zero-emission by 2040, the Scoping Plan~\cite{scopingplan} outlined wind, solar, and storage unit allocation for different regions in NYS. Figure~\ref{NYScapacity} shows the planned capacity for different resources and the post-electrification load distribution. With the rich land availability in upstate and zone F, the majority of wind and solar are allocated to these regions. The darker color in zone J and K for the wind map are offshore winds. The majority of hydro is in zones A and D, contributing to approximately 80\% of the annual hydro generation. The zonal capacities for different resources are summarized in Table~\ref{tab:zonalcap2050}. 

\begin{figure}[H]%
\centering
\includegraphics[width=0.9\textwidth]{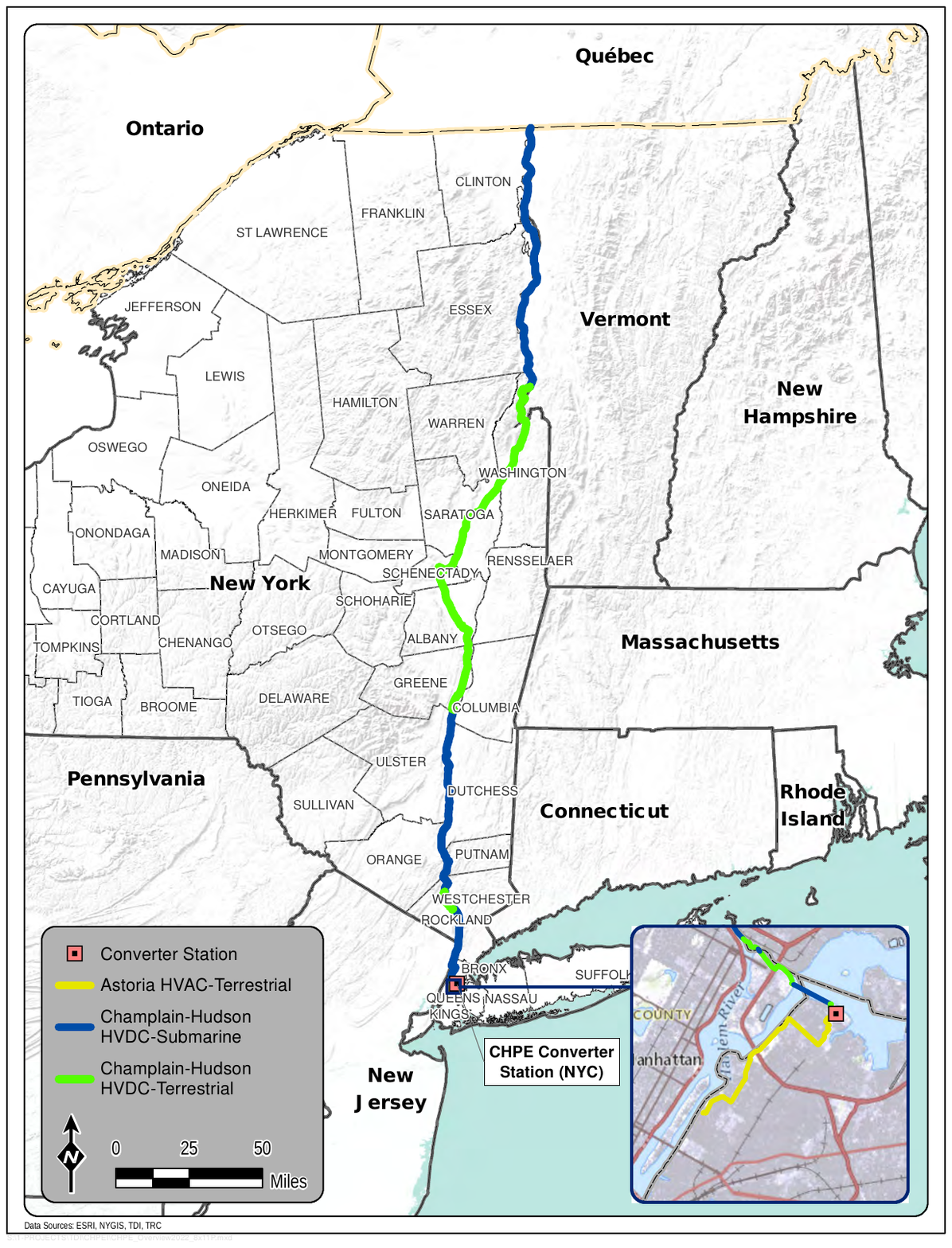}
\caption[Overview of the Champlain Hudson Power Express HVDC line]%
{\textbf{Overview of the Champlain Hudson Power Express HVDC line~\cite{CHPE}} }\label{CHPE}
\end{figure}

\begin{figure}[H]%
\centering
\includegraphics[width=0.9\textwidth]{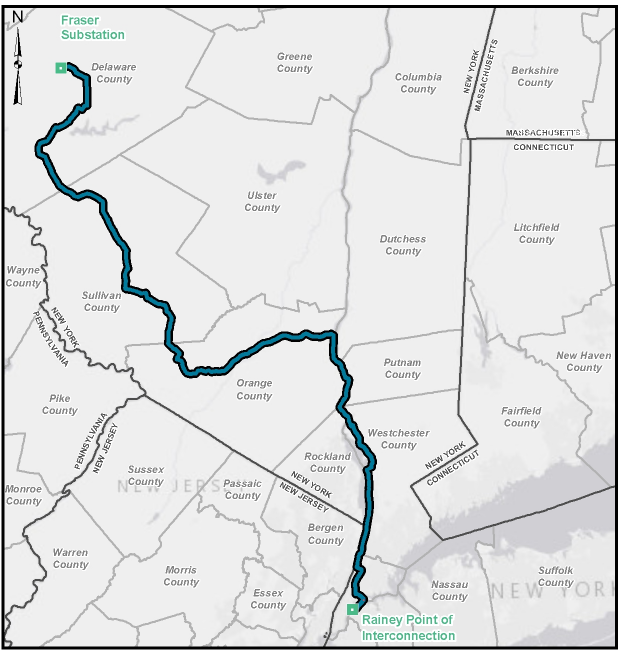}
\caption[Overview of the New York Clean Path HVDC line]%
{\textbf{Overview of the New York Clean Path HVDC line~\cite{NYCP}} }\label{CPNY}
\end{figure}

This plan further exacerbates the unbalance and presses more pressure on the transmission interfaces between load zones. To account for the transmission expansion plan from NYISO, we use the interface limits from the Reliability Needs Assessment (RNA) report to update transmission limits from~\cite{Assessment2020}. The upper and lower limits can be found in Table~\ref{tab:interface2050}. Additionally, there are two High Voltage Direct Current (HVDC) lines in the works: the Champlain Hudson Power Express, with a capacity of 1250 MW, and the New York Clean Path, with a capacity of 1300 MW, scheduled for deployment in 2026 and 2027, respectively. Figures~\ref{CHPE} and~\ref{CPNY} depict the routes of these new HVDC lines. 

The CLCPA emphasized that a fully decarbonized grid is contingent on a firm, zero-carbon resource with capacities ranging from 18 to 23 GW to ensure system reliability. This ``firm, zero-carbon capacity" encompasses emission-free dispatchable resources, such as fuel cells and long-duration batteries. There is no formal definition of firm, zero-carbon capacity. It refers to a combination of existing and new combustion-based resources (i.e., combustion turbines and combined cycle gas turbines) converted to use hydrogen as a zero-carbon fuel~\cite{sepulveda2018role}.

\begin{table}[H]
\centering
\caption {\label{tab:zonalcap2050} Zonal capacity of renewable resources}
\resizebox{\columnwidth}{!}{
\begin{tabular}{llllllllllll}
\hline
\hline
Renewable Resource  & A & B & C & D & E & F & G & H & I & J & K\\   \hline
Land-Based Wind & 2692 &	390	&1923 &	1935 &	1821 & 1864	& 606 & 303	& 0	& 0	& 121\\
Offshore Wind & 0&	0&	0&	0	&0	&0&	0	&0	&0	&8250&	6488\\
BTM Solar  & 1297&	402&	1098	&127&	1240	&2154	&2270	&202	&299&	1676	&2883\\
Utility Solar  & 14440&	1648	&9006&	0	&5698	&15647&	3353	&0	&0&	0	&1441\\
Hydro  &2460	&63.8&	109.4&	909.8&	376.3&	269.6	&75.8&	0&0&0	&0\\
Battery  & 2479&	10	&2538&	2562&	892&	4727&	150	&140	&140&	4263	&1924\\
\hline
\hline
\end{tabular}}
\end{table}

\begin{table}[H]
\centering
\caption {\label{tab:interface2050} Interface flow limits (MW)}
\begin{tabular}{lll}
\hline
\hline
 Interface & Lower Bound  & Upper Bound \\   \hline
 A-B   &  -2,200 & 2,200\\
 B-C   &  -1,600 & 1,500\\
 C-E   &  -5,650 & 5,650\\
 D-E   &  -1,600 & 2,650\\
 E-F   &  -3,925 & 3,925\\
 E-G   &  -1,600 & 2,300\\
 F-G   &  -5,400 & 5,400\\
 G-H   &  -7,375 & 7,375\\
 H-I   &  -8,450 & 8,450\\
 I-J   &  -4,350 & 4,350\\
 I-K   &  -515 & 1,293\\
 Total East   &  -3,400 & 5,600\\
 NY-NE   &  -1,700 & 1,300\\
 NY-IESO   &  -2,000 & 1,650\\
 NY-PJM   &  -900 & 500\\
\hline
\hline
\end{tabular}
\end{table}

\section{Building electrification module}

The data regarding electrification is sourced from the National Renewable Energy Laboratory (NREL), specifically the ResStock~\cite{reyna2021us} and ComStock~\cite{parker2023comstock} toolkits. These toolkits provide information on energy consumption and potential energy savings resulting from various upgrades for both residential and commercial buildings. The state-level data provides upgrades simulations regards to fossil fuel usage and electrified load. Additionally, the weather data used for building simulation corresponds to a typical meteorological year and is also provided.

Simulating building energy consumption requires significant computational resources. To overcome this challenge, we employ an emulator to project electricity savings (which are mostly negative, namely, electrified load) for two specific upgrades: "Whole-home electrification, high efficiency" (referred to as "upgrade 8") for residential buildings, and "DOAS HP Minisplits" (referred to as "upgrade 3") for commercial buildings. These upgrades primarily focus on electrifying the heating and cooling load, making them suitable for the focus of this study.

Firstly, we establish a function called $F_1(other fuel)$ using an Artificial Neural Network (ANN). This function maps the usage of other fuel types, such as natural gas and oil, to electricity savings at the state level. This choice is driven by the limitation that energy-saving data is only available at the aggregated state level. The assumption here is that the conversion from other energy sources to electricity remains consistent for each individual building type within the state. Consequently, county-level data on fossil fuel usage can be fed into the function to determine the electricity load for individual building types for each county. By leveraging the distribution of building types within each county, we can calculate the electrified load for every county.

\begin{figure}[H]%
\centering
\includegraphics[width=0.9\textwidth]{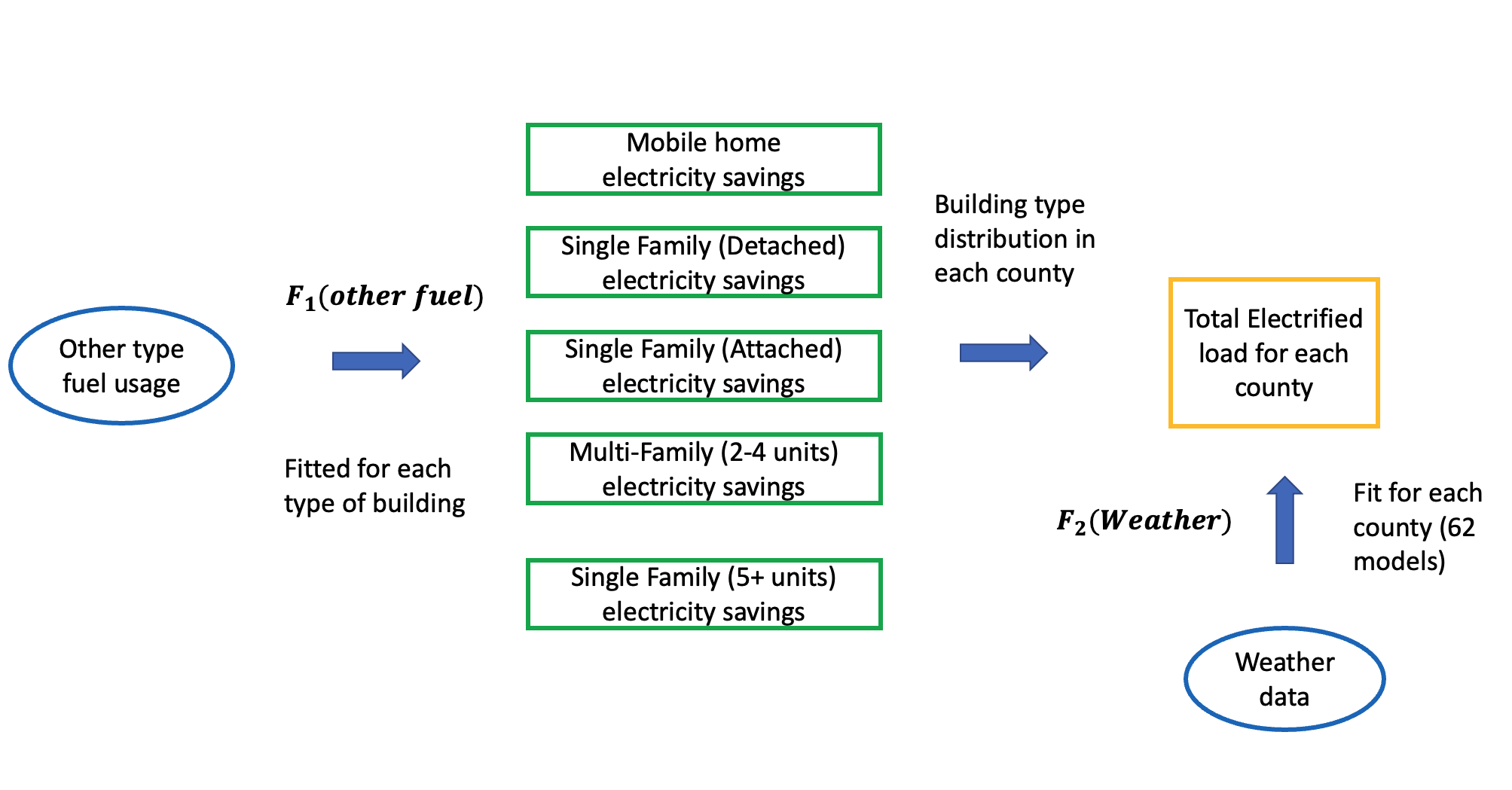}
\caption[Framework for modeling building electrification]%
{\textbf{Framework for modeling building electrification} }\label{fig: buildingeletrification}
\end{figure}

Next, we model the relationship between the weather data of each county and the electrified load. This enables the joint modeling of the co-variability among the electrified load, baseline load, and renewable outputs using latent weather data. For each county, we employ another ANN approximation called $F_2(Weather data)$ to predict the total electrified load. Subsequently, the county-level load is aggregated to the power grid model's buses. The process outlined in Figure~\ref{fig: buildingeletrification} illustrates the framework for residential buildings, while for commercial buildings, the process remains the same, with the building types being replaced by the 14 commercial building types.

\section{EV module}
\begin{figure}[H]%
\centering
\includegraphics[width=0.9\textwidth]{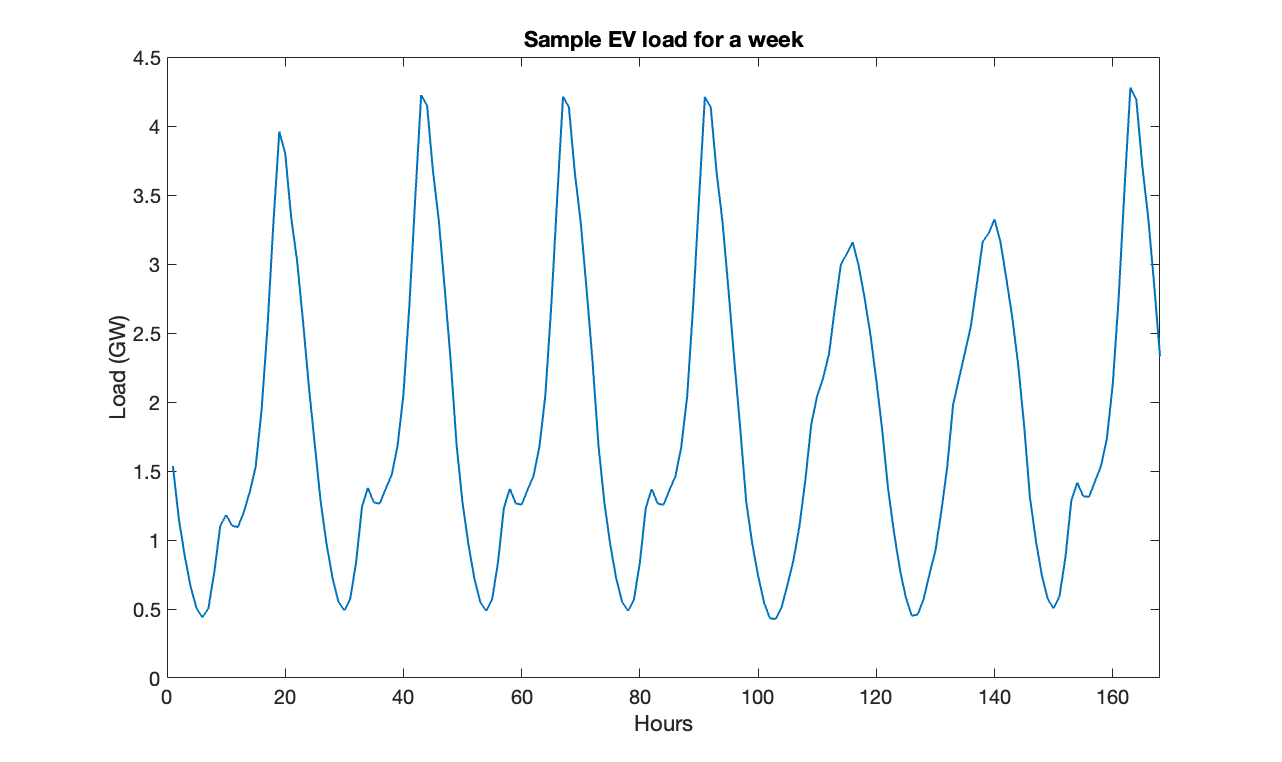}
\caption[Framework for modeling building electrification]%
{\textbf{Framework for modeling building electrification} }\label{fig: evprofile}
\end{figure}

The light-duty vehicle data is processed from the Vehicle, Snowmobile, and Boat Registrations~\cite{NYVehicle} dataset by filtering out the number of light-duty fleets for each county summarized in Table~\ref{tab: EV_county}. The median distance travel for each county is a required input for the EVI-Pro Lite model and is estimated based on the population density of 
each county~\cite{NYpopuden} using the population density to daily VMT from~\cite{wood2022evi}.The other parameters used for the EVI-Pro Lite model are summarized in Table~\ref{tab: EVIproparams}. A sample EV load profile for a week is shown in Fig~\ref{fig: evprofile}. 
\begin{table}[H]
\centering
\caption {\label{tab: EV_county} Light duty vehicle fleet size and median travel distance for each county }
\begin{tabular}{lll}
\hline
\hline
county &	fleet size	& median dvmt \\   \hline
Bronx&	26320&	14 \\ 
Kings&	60920	&14\\ 
New York&	25890	&14\\ 
Queens&	84524	&16\\ 
Richmond&	29686	&17\\ 
Albany&	15800&	22\\ 
Allegany	&1513&	31\\ 
Broome&	10253	&26\\ 
Cattaraugus&	2691	&31\\ 
Cayuga&	2818	&26\\ 
Chautauqua&	4679	&26\\ 
Chemung&	3998	&26\\ 
Chenango	&1847&	31\\ 
Clinton&	3632&	31\\ 
Columbia	&3298&	31\\ 
Cortland	&1597&	31\\ 
Delaware	&1904	&31\\ 
Dutchess	&15384&	26\\ 
Erie&	40570	&22\\ 
Essex	&1568	&31\\ 
Franklin&	1949	&31\\ 
Fulton&	2332&	26\\ 
Genesee&	2376	&26\\ 
Greene	&2107&	31\\ 
Hamilton	&276&	31\\ 
Herkimer	&2196	&31\\ 
Jefferson&	4378&	31\\ 
Lewis&	913	&31\\ 
Livingston	&2240	&26\\ 
Madison&	2691	&26\\ 
Monroe	&36962&	20\\ 
Montgomery&	2048	&26\\ 
Nassau&	96201&	17\\ 

\hline
\hline
\end{tabular}
\end{table}

\begin{table}[H]
\centering
\caption {\label{tab: DUparameters} Table \ref{tab: EV_county} continued }
\begin{tabular}{lll}
\hline
\hline
county &	fleet size	& mean dvmt \\   \hline
Niagara&	7823&	26\\ 
Oneida&	9393	&26\\ 
Onondaga	&21959&	22\\ 
Ontario	&5688&	26\\ 
Orange&	22821&	26\\ 
Orleans	&1138&	26\\ 
Oswego	&4344&	26\\ 
Otsego	&2421&	31\\ 
Putnam&	5739	&26\\ 
Rensselaer&	7317	&26\\ 
Rockland&	22377	&20\\ 
St Lawrence	&3544	&31\\ 
Saratoga&	14484&26\\ 
Schenectady	&9035&	22\\ 
Schoharie&	1235&	31\\ 
Schuyler	&845&	31\\ 
Seneca&	1305	&26\\ 
Steuben&	3783&	31\\ 
Suffolk	&94765	&20\\ 
Sullivan&	3885&	31\\ 
Tioga	&2234	&31\\ 
Tompkins	&3699&	26\\ 
Ulster&	8954	&26\\ 
Warren	&3476&	31\\ 
Washington	&2553	&31\\ 
Wayne&	3875	&26\\ 
Westchester	&57581&	18\\ 
Wyoming&	1407&	31\\ 
Yates	&922&	31\\

\hline
\hline
\end{tabular}
\end{table}

\begin{table}[H]
\centering
\caption {\label{tab: EVIproparams} EVI Lite Pro parameters }
\begin{tabular}{lll}
\hline
\hline
EVI pro params & Modeling choice \\ \hline
pev\_type & PHEV50 \\
pev\_dist	& EQUAL \\
class\_dist	& Sedan \\
home\_access\_dist	& HA100 \\
home\_power\_dist	& Equal \\
work\_power\_dist	& MostL2 \\
pref\_dist	& Home100 \\
res\_charging	& min\_delay \\
work\_charging & min\_delay \\

\hline
\hline
\end{tabular}
\end{table}

\section{Dynamic rating formulation and result} \label{sec: dynamicrating}

The carrying capacity of electric power cables decreases as ambient air temperatures rise. Bartos et al., \cite{bartos2016impacts} estimate the impacts of rising air temperatures on electric transmission ampacity across the United States. They estimate the climate-attributable capacity reductions of transmission lines by constructing a thermal balance model, which estimates the rated ampacity of transmission lines based on cable properties and meteorological forcings. In the following, we explain this approach in detail. 

Assuming steady-state conditions and no conduction, the energy balance formula is:
\begin{align}
& q_j + q_s = q_c + q_r, \label{reducedEnergyBalance}
\end{align}
where $q_j$ is the resistive heating of the conductor ($W/m^{-1}$), $q_s$ is the radiative heat transfer from the sun to the conductor ($W/m^{-1}$), and $q_c$ and $q_r$, respectively, denote the convective and radiative heat losses from the conductor to surroundings ($W/m^{-1}$). Equation \ref{reducedEnergyBalance} implies that the total heat gain from the electrical current flowing
through the conductor and from the solar radiation striking the top half of the surface of the conductor equals the total heat loss due to convection and radiation. The heat gain due to electrical current ($q_j$) is also a function of the transferred current and the resistance of the conductor.
\begin{align}
& q_j = I^2 . R(T_{cond}), \label{electricalLoadHeatGain}
\end{align}
where $I$ is the current transferred through the conductor, and $R(T_{cond})$ is the resistance of the conductor at the given conductor temperature $T_{cond}$. Then, rearranging the heat balance formula \ref{reducedEnergyBalance} yields the maximum allowable current (the rated ampacity):
\begin{align}
& I = \sqrt{\frac{q_c+q_r-q_s}{R(T_{cond})}}, \label{ratedAmpacity}
\end{align}
In the following, we explain how to compute each term of the equation \ref{ratedAmpacity}. The convective heat loss per unit length, $q_c$, is calculated as:
\begin{align}
& q_c = \Bar{h}.\pi.D.(T_{cond}-T_{amb}), \label{convectiveHeatLoss}
\end{align}
where $D$ and $T_{amb}$, respectively, denote the conductor diameter (m) and the ambient air temperature (K). $\Bar{h}$ is the average heat transfer coefficient (W/m2-K) which is a function of wind speed, cable diameter, and some constants.

\begin{align}
& \Bar{h} = 0.3+\frac{0.62.(V.D/\nu)^{1/2}.Pr^{1/3}}{\left( 1+(\frac{0.4}{Pr})^{2/3}\right)^{1/4}}\left( 1+(\frac{V.D/\nu}{282000})^{5/8}\right)^{4/5}.k/D, \label{convectiveHeatLoss}
\end{align}

$V$ is the wind speed, $\nu$ is the dynamic viscosity of air ($m^2/s$), $Pr$ is the Prandtl number, and $k$ is the thermal conductivity of air (W/m-K). Table \ref{Table1} specifies the $(\nu,k, Pr)$ parameters under several air temperature scenarios. 

The radiation heat loss, $q_r$, and the solar heat gain, $q_s$, per unit length is computed as:
\begin{align}
& q_r = \epsilon.\sigma.\pi.D.(T_{cond}^4-T_{amb}^4), \label{radiationHeatLoss} \\
& q_s = \delta.D.a_s, \label{sunHeatGain}
\end{align}
where $\epsilon$ is the emissivity of the conductor surface, $\sigma$ is the Stefan–Boltzmann constant ($5.670e-8 W/m^2-K^4$), $\delta$ is the incident solar radiation, and $a_s$ is the solar absorptivity of the conductor surface. For a stranded aluminum conductor, it is assumed that $\epsilon = 0.7$ and $a_s = 0.9$. $\delta$ is the fixed value of $1000 W/m^2$ to represent full sun conditions.

Finally, in the expanded form, the rated ampacity of an overhead conductor can be expressed in terms of meteorological variables and cable properties:

\begin{align}
& I = \sqrt{\frac{\Bar{h}.\pi.D.(T_{cond}-T_{amb}) + \epsilon.\sigma.\pi.D.(T_{cond}^4-T_{amb}^4) - \delta.D.a_s}{R(T_{cond})}}.\label{ExpandedRatedAmpacity}
\end{align}


\begin{table}[H]
\centering
\caption{Air properties as a function of ambient air temperature.}
\label{Table1}
\begin{tabular}{|l|l|l|l|}
\hline
\begin{tabular}[c]{@{}l@{}}Temperature \\ (K)\end{tabular} & \begin{tabular}[c]{@{}l@{}}Dynamic Viscosity \\ (m\textasciicircum{}2/s)e-6\end{tabular} & \begin{tabular}[c]{@{}l@{}}Thermal Conductivity\\ (W/m-K)e-3\end{tabular} & Prandtl Number \\ \hline
200                                                        & 7.59                                                                                     & 18.1                                                                      & 0.737          \\
250                                                        & 11.44                                                                                    & 22.3                                                                      & 0.720          \\
300                                                        & 15.89                                                                                    & 26.3                                                                      & 0.707          \\
350                                                        & 20.92                                                                                    & 30.0                                                                      & 0.700          \\
400                                                        & 26.41                                                                                    & 33.8                                                                      & 0.690          \\
450                                                        & 32.39                                                                                    & 37.3                                                                      & 0.686          \\
500                                                        & 38.79                                                                                    & 40.7                                                                      & 0.684          \\
550                                                        & 45.57                                                                                    & 43.9                                                                      & 0.683          \\
600                                                        & 52.69                                                                                    & 46.9                                                                      & 0.689          \\ \hline
\end{tabular}
\end{table}

As per Equation~\ref{ExpandedRatedAmpacity}, the ampacity of transmission lines depends on several factors, including ambient temperature, solar radiation, wind speed, and specific parameters related to the cable models and conductor voltage classes. In~\cite{bartos2016impacts}, it was found that the choice of cable models had minimal impact on the primary results. Therefore, we adopt the representative cable models proposed in this study, which are outlined in Table~\ref{tab: cablemodel}. To determine the ambient temperature, solar radiation, and wind speed for each transmission line, we consider the spatial MERRA2 data and choose the three closest data points. The minimum dynamic rating among these three data points is selected to represent the worst-case scenario.

\begin{table}[H]
\centering
\caption{Characteristic cable models for conductor voltage classes}
\label{tab: cablemodel}
\begin{tabular}{|c|l|c|llc|}
\hline
\multirow{2}{*}{Nominal voltage (kV)} & \multirow{2}{*}{Model cable} & \multirow{2}{*}{Diameter (cm)} & \multicolumn{3}{l|}{AC Resistance (Ohms/km)}                    \\ \cline{4-6} 
                                      &                              &                                & \multicolumn{1}{l|}{25 C}  & \multicolumn{1}{l|}{50 C}  & 75 C  \\ \hline
500                                   & 3 x 954 kcm ACSR Cardinal    & 3.04                           & \multicolumn{1}{l|}{0.061} & \multicolumn{1}{l|}{0.067} & 0.073 \\
345                                   & 2 x 954 kcm ACSR Cardinal    & 3.04                           & \multicolumn{1}{l|}{0.061} & \multicolumn{1}{l|}{0.067} & 0.073 \\
230                                   & 1 x 1351 kcm ACSR Martin     & 3.62                           & \multicolumn{1}{l|}{0.044} & \multicolumn{1}{l|}{0.048} & 0.052 \\
115                                   & 1 x 795 kcm ACSR Condor      & 2.77                           & \multicolumn{1}{l|}{0.073} & \multicolumn{1}{l|}{0.080} & 0.087 \\
69                                    & 1 x 336 kcm ACSR Linnet      & 1.83                           & \multicolumn{1}{l|}{0.170} & \multicolumn{1}{l|}{0.186} & 0.203 \\ \hline
\end{tabular}
\end{table}

As the static rating for the interfaces of the NYS system is provided by NYISO, we assume that the static rating is set with the standard industry protocol with wind speed equal to $0.61m/s$, solar radiation equal to $1000 W/m^2$, conductor temperature equal to $75 \degree C$ and ambient temperature equal to $75 \degree C$. Then if we denote the interface flow calculated by the above condition for each line in the interfaces as $\Bar{L_{normal}}$, the interface limits given by the NYISO as $L_{nyiso}$ and the dynamic rating calculated for each hour based on Equation~\ref{ExpandedRatedAmpacity} as $L_s$, then the dynamic limits can be calculated by $\dfrac{L_s}{L_{nyiso}}L_{normal}$

\section{Deeply uncertain parameters}

The deeply uncertain parameters for the climatic and technological factors are summarized in Table~\ref{tab: DUparameters}

\begin{table}[H]
\centering
\caption {\label{tab: DUparameters} Climatic and technological factors }
\begin{tabular}{lll}
\hline
\hline
 Parameter & Lower Bound  & Upper Bound \\   \hline
Temperature increase & 0.95 & 5.64 \\
  Building electrification rate & 0.7 & 1.05\\
  EV electrification rate  & 0.7 & 1.05\\
Wind capacity scaling factor & 0.6 & 1.4 \\
Solar capacity scaling factor & 0.6 & 1.4 \\
Battery capacity scaling factor & 0.6 & 1.4 \\
\hline
\hline
\end{tabular}
\end{table}

\section{Enhanced DCOPF formulation} \label{sec: dcopf}
\subsection*{Nomenclature}
\begin{mdframed}
\textbf{Sets and Indexes}
\item[$\mathcal{T}$] length of the planning horizon
\item[$\mathcal{Q}_q$] a set of time interval in a quarter month $q$
\item[$\mathcal{B}$]  a set of buses in the system
\item[$\mathcal{L}$]  a set of transmission lines in the system
\item[$\mathcal{C}$] a set of storage units
\item[$\mathcal{H}$] a set of large hydro generators
\item[$\mathcal{W}$] a set of wind generators
\item[$\mathcal{S}$] a set of solar generators
\item[$\mathcal{G}$] a set of nuclear generators
\item[$\mathcal{H}_{b}$]  a set of large hydro generators connected to bus $b$
\item[$\mathcal{W}_{b}$]  a set of wind generators connected to bus $b$
\item[$\mathcal{S}_{b}$]  a set of solar generators connected to bus $b$
\item[$\mathcal{G}_{b}$]  a set of nuclear generators connected to bus $b$
\item[$\mathcal{I}_b$]  a set of lines flow into bus $b$
\item[$\mathcal{O}_b$]  a set of lines flow out of bus $b$
\item[$\mathcal{C}_b$]  a set of storage units connected to bus $b$
\item[$\mathcal{IF}_i$]  a set of lines in zonal interface $i$
\item[$b \in \mathcal{B}$]  a bus in the system
\item[$t \in \mathcal{T}$]  a time interval
\item[$g \in \mathcal{H}\cup\mathcal{W}\cup\mathcal{S}\cup\mathcal{G}$]  a generator in the system
\item[$l \in \mathcal{L}$]  a transmission line in the system 
\item[$q \in \mathcal{Q}$]  a quarter month in the year \\

\textbf{Parameters} 
\item[$\overline{R}_g / \underline{R}_g$] upper/lower ramp rate limit of generator $g \in\mathcal{H}\cup\mathcal{G} $
\item[$\overline{P}_g / \underline{P}_g$]  generation upper/lower bound of generator $g \in\mathcal{H}\cup\mathcal{G} $
\item[$\overline{P}_{g,t} / \underline{P}_{g,t}$]  generation upper/lower bound of generator $g \in\mathcal{W}\cup\mathcal{S} $
\item[$\overline{L_{l,t}} / \underline{L_{l,t}}$] upper/lower bound of transmission line $l \in \mathcal{L}$ at time $t$
\item[$\overline{L_{IF_{i},t}} / \underline{L_{IF_{i},t}}$]  upper/lower bound of interface flow $i \in \mathcal{IF}$ at time $t$
\item[$D_{b,t}^{base}$]  baseline demand for bus $b \in \mathcal{B}$ in hour $t$ 
\item[$D_{b,t}^{bldg}$]  electrified building demand on bus $b\in \mathcal{B}$ in hour $t$ 
\item[$D_{b,t}^{ev}$]  EV charging demand on bus $b\in \mathcal{B}$ in hour $t$ 
\item[$D_{b,t}^{shydro}$] output of small hydro plants on bus $b\in \mathcal{B}$ at time $t$
\item[$D_{b,t}^{btm}$] output of behind the meter solar plants on bus $b\in \mathcal{B}$ at time $t$
\item[$\eta_{s}$]  round-trip efficiency of the storage unit at bus $s \in \mathcal{C}$ 
\item[$SOC_{s}$]  storage size for storage unit $s \in \mathcal{C}$ 
\item[$\Delta_{s}$]  charging/discharging capacity of storage unit $s \in \mathcal{C}$ 
\item[$H_{g,q}$] quarter monthly hydro power availability for hydro plant $g \in \mathcal{H}$ for quarter month $q$
\item[$B_l$] the susceptance of line $l$ \\

\textbf{Variables}  
\item[$p_{g,t}$] generation of generator $g$ in hour $t$ 
\item[$e_{l,t}$] power flow of branch $l$ in hour $t$ 
\item[$\theta_{b,t}$] phase angle of bus $b$ in hour $t$ 
\item[$\delta_{s,t}^+, \delta_{s,t}^-$] charge/discharge power of storage unit $s$ in hour $t$ 
\item[$soc_{s,t}$] amount of stored energy in the storage unit $s$ at hour $t$ 
\item[$\mu_{b,t}$] the amount of load shedding on bus $b$ at hour $t$

\end{mdframed}

\begin{equation}\label{eq: obj}
\begin{aligned}
Min  \sum_{t=1}^{\mathcal{T}}(\sum_{b \in \mathcal{B}}(\mu_{b,t})+\lambda\sum_{s \in \mathcal{C}_b}(\delta_{s,t}^+ + \delta_{s,t}^-))
\end{aligned}
\end{equation}

\begin{equation} \label{eq: eq2}
    \begin{aligned}  
\sum_{g \in \mathcal{H}\cup\mathcal{W}\cup\mathcal{S}\cup\mathcal{G}} p_{g,t} + \sum_{l \in \mathcal{I}_b}e_{l,t} + \sum_{s \in \mathcal{C}_{b}}\delta_{s,t}^- =\sum_{l \in \mathcal{O}_b}e_{l,t} + D_{b,t}^{base}+ D_{b,t}^{bldg}+ D_{b,t}^{ev} - D_{b,t}^{btm} - D_{b,t}^{shdyro}\\ + \sum_{s \in \mathcal{C}_{b}}\delta_{s,t}^+ 
\quad \forall t \in \mathcal{T},  b \in \mathcal{B}
    \end{aligned}
\end{equation}

\begin{gather}
\vspace{-20pt}
\underline{P}_g \leq p_{g,t}\leq  \overline{P}_g \quad \forall t \in \mathcal{T} ,  \quad g \in \mathcal{H}\cup\mathcal{G} \label{eq: eq3}\\
\underline{P}_{g,t} \leq p_{g,t}\leq  \overline{P}_{g,t} \quad\forall t \in \mathcal{T} ,  \quad g \in \mathcal{W}\cup\mathcal{S} \label{eq: eq4}\\
\underline{R}_{g} \leq p_{g,t}  - p_{g,t-1}  \leq \overline{R}_g \quad \forall t \in \mathcal{T} , \quad g \in \mathcal{H}\cup\mathcal{G} \label{eq: eq5}\\
\underline{L} \leq e_{l,t} \leq \overline{L}  \quad \forall t \in \mathcal{T} , \quad l \in \mathcal{L} \label{eq: eq6} \\
 -\pi \leq \theta_{b,t} \leq \pi \quad \forall t \in \mathcal{T} ,\quad b \in \mathcal{B} \label{eq: eq7} \\
e_{l,t} = B_l (\theta_{b,t} - \theta_{b^\prime,t})  \quad \forall t \in \mathcal{T} , \quad l \in \mathcal{L}  \label{eq: eq8}\\
soc_{s,t+1} = soc_{s,t} + \dfrac{1}{\sqrt{\eta_s}}\delta_{s,t}^+ - \sqrt{\eta_s}\delta_{s,t}^- \quad \forall t \in \mathcal{T} , \quad s \in \mathcal{C} \label{eq: eq9}  \\
0 \leq soc_{s,t} \leq \overline{SOC_{s}}  \quad \forall t \in \mathcal{T} ,  \quad s \in \mathcal{C}  \label{eq: eq10}\\
0 \leq \delta_{s,t}^- \leq \Delta_{s}  \quad \forall t \in \mathcal{T} , \quad s \in \mathcal{C}  \label{eq: eq11}\\
0 \leq \delta_{s,t}^+ \leq  \Delta_{s}  \quad \forall t \in \mathcal{T}, \quad  s \in \mathcal{C}   \label{eq: eq12}\\
\underline{L_{IF_i,t}} \leq \sum_{l \in IF_i}e_{l,t} \leq \overline{L_{IF_i,t}}  \quad \forall t \in \mathcal{T} , \quad i \in \mathcal{IF}  \label{eq: eq13} \\
\sum_{t \in \mathcal{Q}_q}p_{g,t} = H_{g,q}  \quad  \forall q \in \mathcal{Q}, \quad g \in \mathcal{H} \label{eq: eq14}\\
 0 \leq \mu_{b,t} \leq max(0,D_{b,t}^{base}+ D_{b,t}^{bldg}+ D_{b,t}^{ev} - D_{b,t}^{btm} - D_{b,t}^{shdyro} )\quad  \forall b \in \mathcal{B}, \quad t \in \mathcal{T} \label{eq: eq15}
\end{gather}

Equation~\ref{eq: obj} is the objective function for the enhanced DC-OPF problem. The first term represents load shedding for the whole system. The second term prevents the battery from charging and discharging simultaneously by assigning $\lambda = 0.01$, which is decided after the experiments to balance the optimality of the first term while limiting the simultaneous charging and discharging. Equation~\ref{eq: eq2} is the nodal load balancing constraint including the battery charging/discharging, the electrified load, and non-dispatchable resources modeled as negative load. Equation~\ref{eq: eq3} is the generation upper and lower limits constraint for dispatchable large hydro and nuclear generators. Equation~\ref{eq: eq4} is the generation upper and lower limits constraint for semi-dispatchable wind and solar generators. Noticing that the upper and lower bounds are modeled separately for dispatchable and semi-dispatchable generators. The upper limits for semi-dispatchable constraints are the predicted output calculated given the MERRA2 data. Equation~\ref{eq: eq5} is the ramping constraint for dispatchable generators. Equation~\ref{eq: eq6} is the transmission line limit constraint. Equation~\ref{eq: eq7} is the phase angle constraint. Equation~\ref{eq: eq8} describes the relationship between line flows with phase angle. Equation~\ref{eq: eq9}-\ref{eq: eq12} models the battery state transition, battery state of charge limits, and battery discharging and charging limits, respectively. Equation~\ref{eq: eq13} models the interface flow limits, noticing that the upper and lower limits are calculated by the method outlined in Section~\ref{sec: dynamicrating}. Equation~\ref{eq: eq14} enforces that the large hydro plants have to dispatch a certain amount of energy due to water regulations requirements.  Finally, Equation~\ref{eq: eq15} ensures that load shedding on each bus is non-negative and less than the available load that could be shed.

\section{Comparison for the impact of different modeling constraints}
\begin{figure}[H]%
\centering
\includegraphics[width=0.9\textwidth]{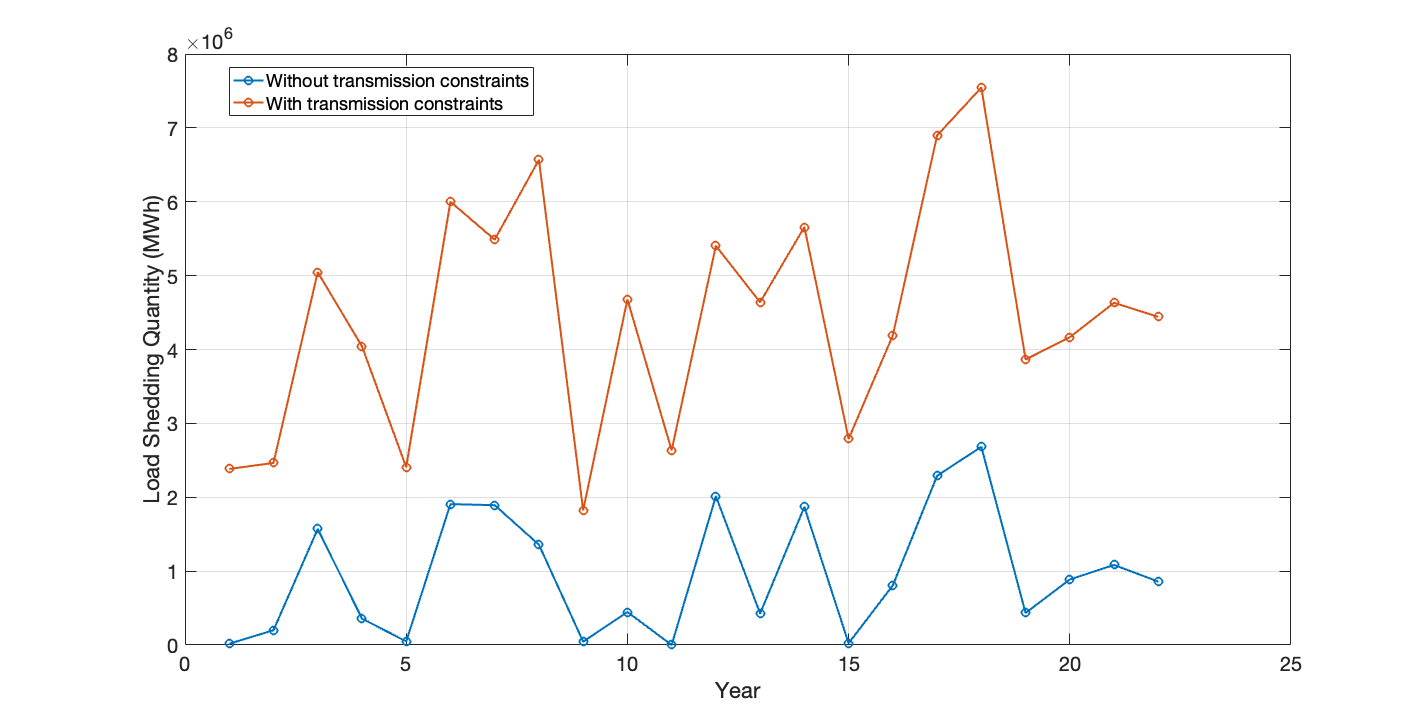}
\caption[Comparison for load shedding quantity over 22 years under different model constraints]%
{\textbf{Comparison for load shedding quantity over 22 years under different model constraints} }\label{fig: yearlylsq}
\end{figure}

\begin{figure}[H]%
\centering
\includegraphics[width=0.9\textwidth]{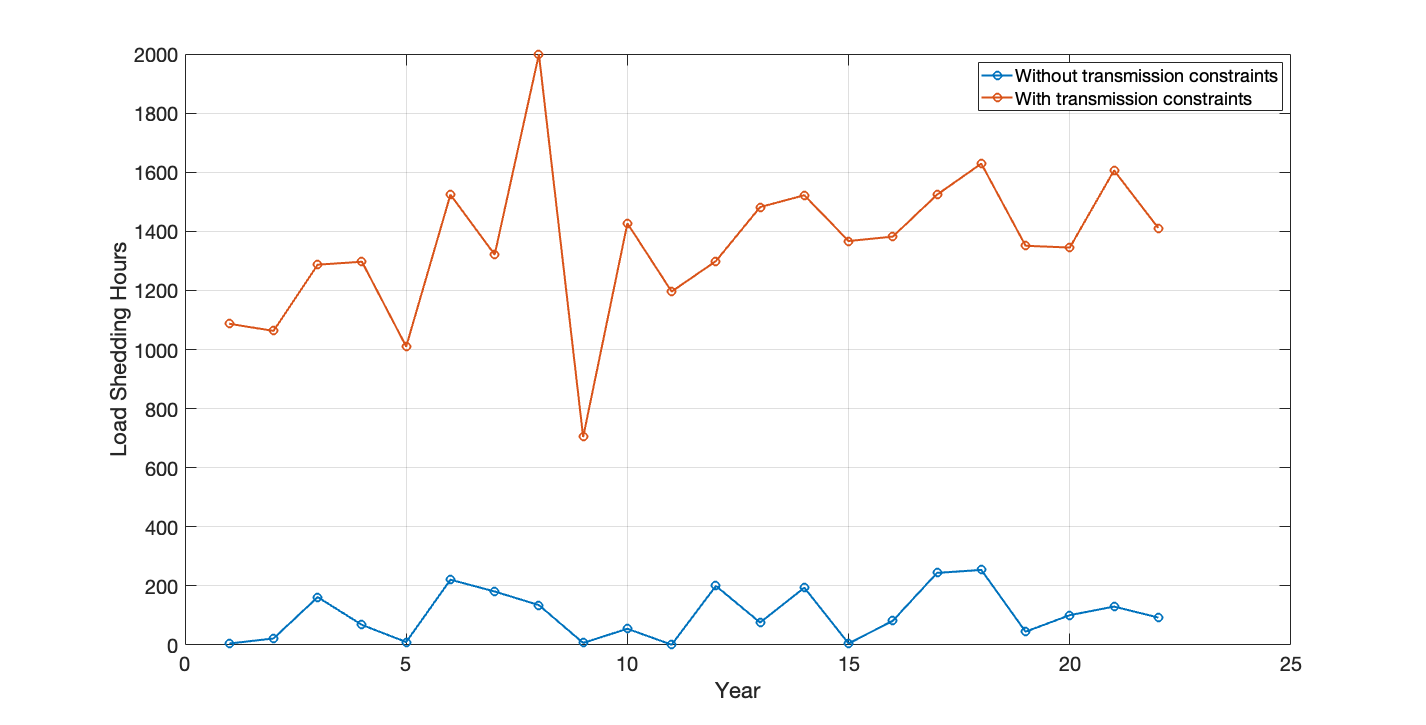}
\caption[Comparison for load shedding hours over 22 years under different model constraints]%
{\textbf{Comparison for load shedding hours over 22 years under different model constraints} }\label{fig: yearlylsf}
\end{figure}

\begin{figure}[H]%
\centering
\includegraphics[width=0.9\textwidth]{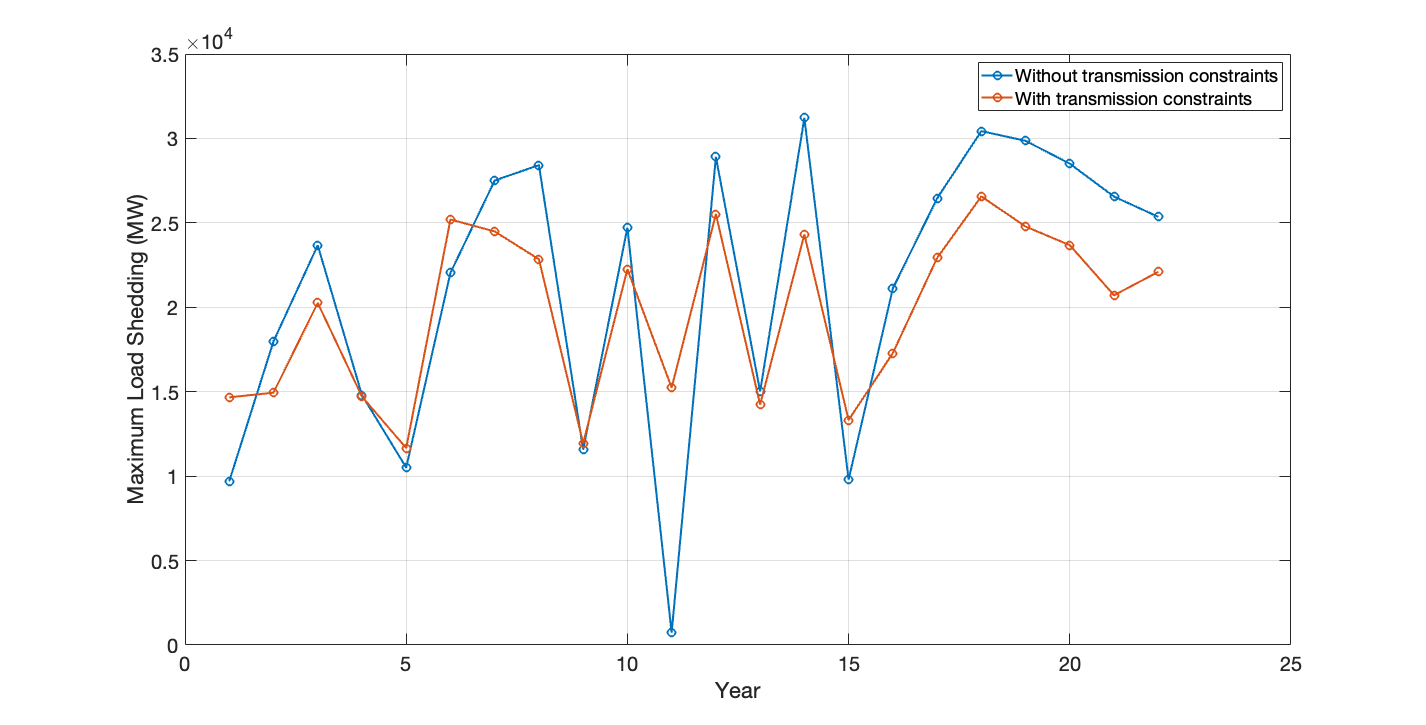}
\caption[Comparison for maximum load shedding in an hour over 22 years under different model constraints]%
{\textbf{Comparison for maximum load shedding in an hour over 22 years under different model constraints} }\label{fig: yearlylsm}
\end{figure}

In Section~\ref{sec: dcopf}, we introduced extended DC-OPF constraints that capture the complexities of real-world scenarios more accurately. In this section, we aim to examine the impact of transmission constraints, specifically phase angle constraints and transmission capacity limits. This analysis is crucial as it highlights the impacts of including system operation constraints in multi-year, multi-scenario studies, which is one of the key contributions of this research. To evaluate the role of these constraints, we compare two cases in terms of load shedding quantity, load shedding hours, and maximum load shedding, as depicted in Figures~\ref{fig: yearlylsq},~\ref{fig: yearlylsf}, and~\ref{fig: yearlylsm}, respectively:

\begin{itemize}
    \item Case 1: Standard DC-OPF formulation without transmission constraints. (Namely, the formulation in Section~\ref{sec: dcopf}  without ~\ref{eq: eq6},~\ref{eq: eq7},~\ref{eq: eq8}, ~\ref{eq: eq14}. constraint~\ref{eq: eq13} has a constant upper and lower bound.)
    \item Case 2: Standard DC-OPF formulation (with transmission constraints. (Namely, the formulation in Section~\ref{sec: dcopf}  without constraint~\ref{eq: eq14}. Constraint~\ref{eq: eq13} has a constant upper and lower bound.)
\end{itemize}

The findings highlight the importance of considering system operation constraints. Neglecting these constraints can lead to a significant underestimation of system vulnerabilities, particularly for evaluating load shedding quantity and load shedding hours. This aligns with the expected outcome, as disregarding transmission line constraints and the power flow relationship with phase angle essentially eliminates the grid's topological structure, allowing power to be delivered anywhere in the grid.

However, the analysis of the maximum load shedding quantity in an hour reveals a mixture of scenarios. In the absence of transmission constraints, certain years exhibit higher maximum load shedding during a specific hour. Initially, this may seem counter-intuitive, but it can be attributed to the underlying assumption of "no transmission constraints" modeling. By assuming that power can be transmitted to any location at any time, the reliance on energy storage batteries diminishes, resulting in less energy being stored. Consequently, during periods of significant renewable shortage and negligible stored energy, the maximum load shedding amount can be substantial.

\begin{figure}[H]%
\centering
\includegraphics[width=0.9\textwidth]{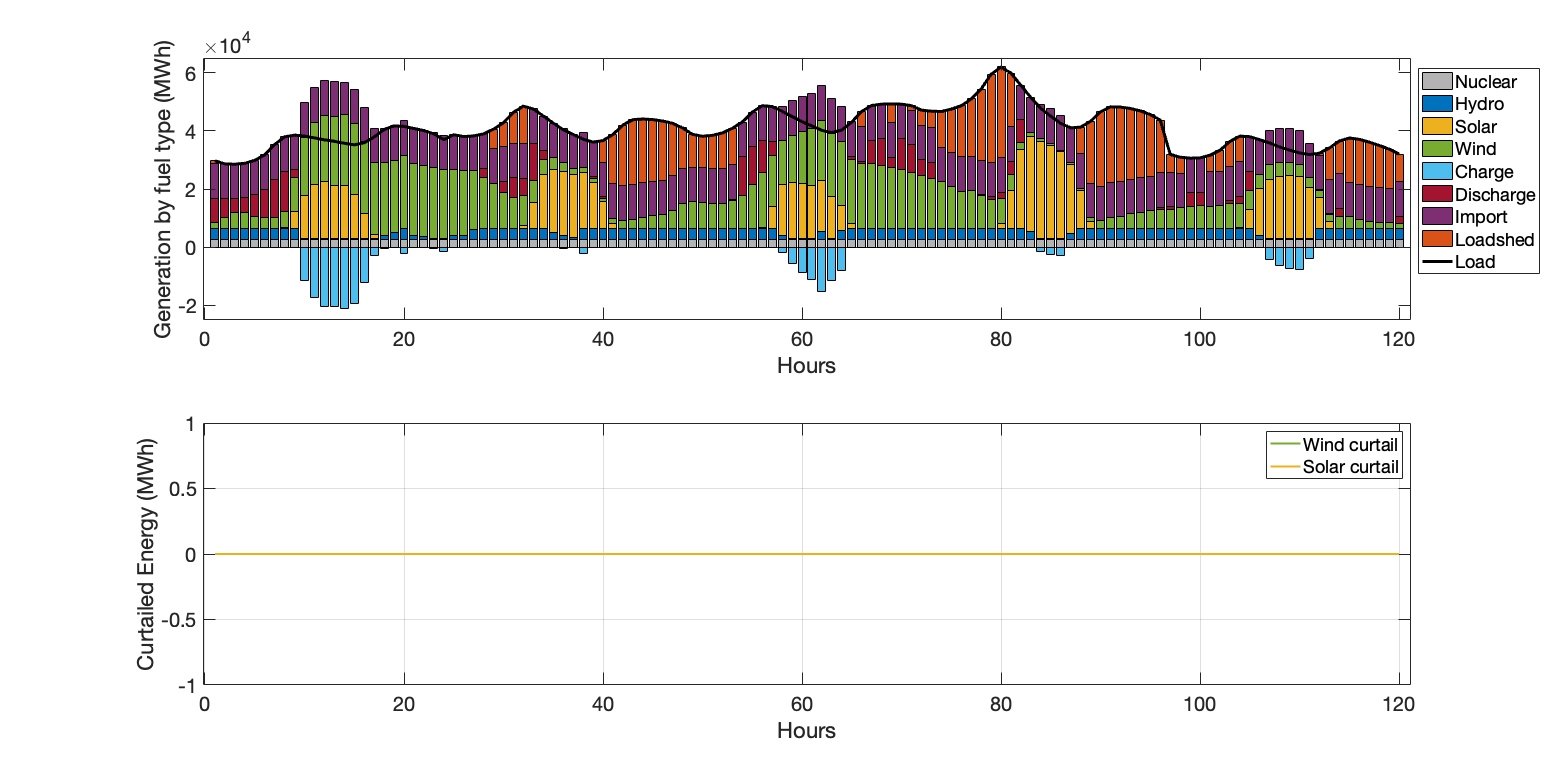}
\caption[Comparison for maximum load shedding in an hour over 22 years without network constraints]%
{\textbf{Comparison for maximum load shedding in an hour over 22 years without network constraints} }\label{fig: wonet}
\end{figure}

\begin{figure}[H]%
\centering
\includegraphics[width=0.9\textwidth]{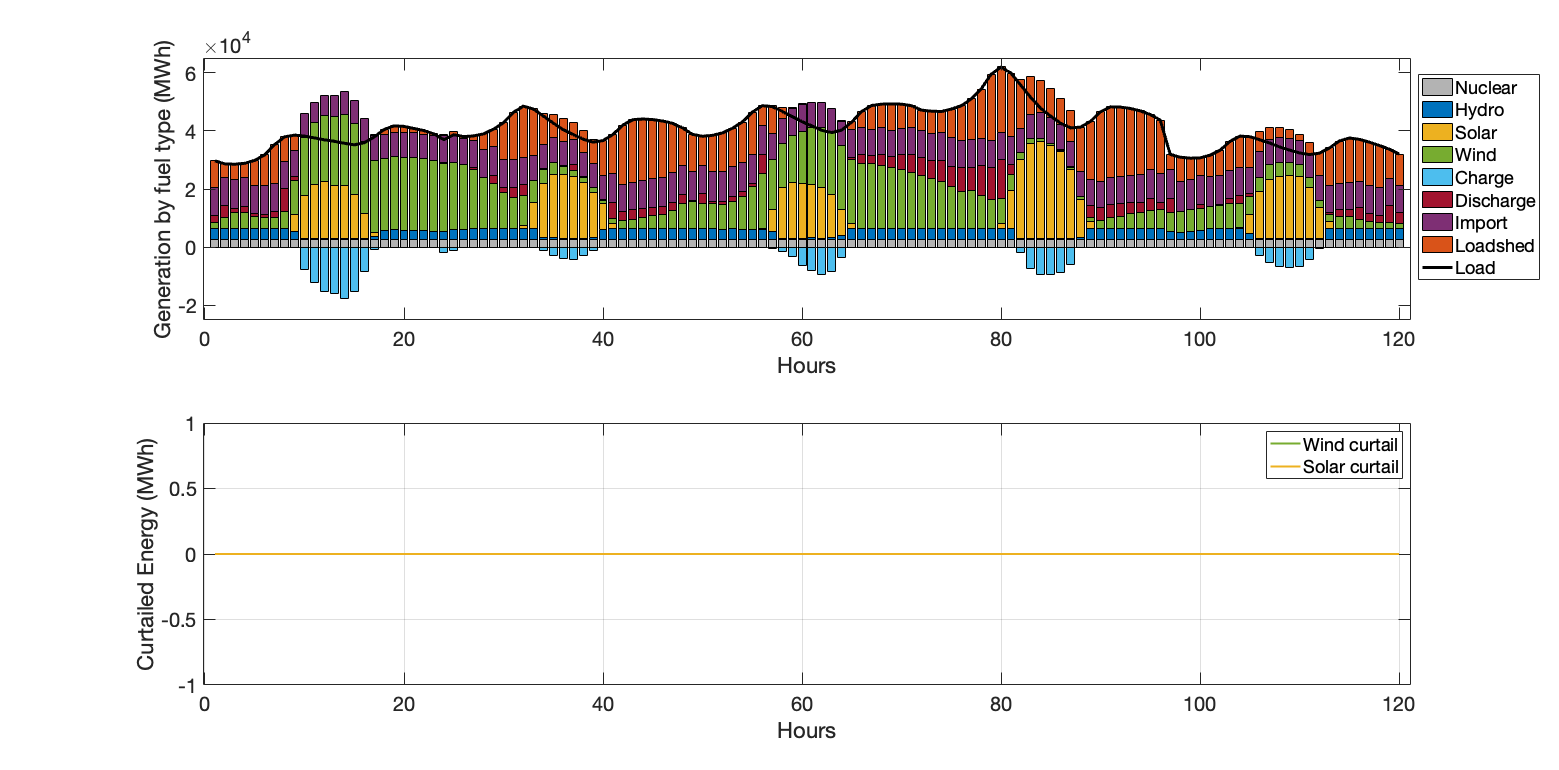}
\caption[Comparison for maximum load shedding in an hour over 22 years with network constraints]%
{\textbf{Comparison for maximum load shedding in an hour over 22 years with network constraints} }\label{fig: withnet}
\end{figure}

Figure~\ref{fig: wonet} illustrates five consecutive days, which includes the hour with the maximum observed load shedding (hour 80) in year 14 without transmission constraints. In contrast, Figure~\ref{fig: withnet} portrays the same five days, but with transmission constraints considered. As depicted in Figure~\ref{fig: wonet}, there is less battery charging during hours 35-40 and more battery discharging during hours 67-73, resulting in a lower energy storage level to cover the load peak around hour 80. Consequently, the load shedding quantity during hours 78-80 is significantly greater compared to the scenario with transmission constraints (Figure~\ref{fig: withnet}).

\section{Load profile before and after electrification} \label{sec: electrificationcomparison}
\begin{figure}[H]%
\centering
\includegraphics[width=0.9\textwidth]{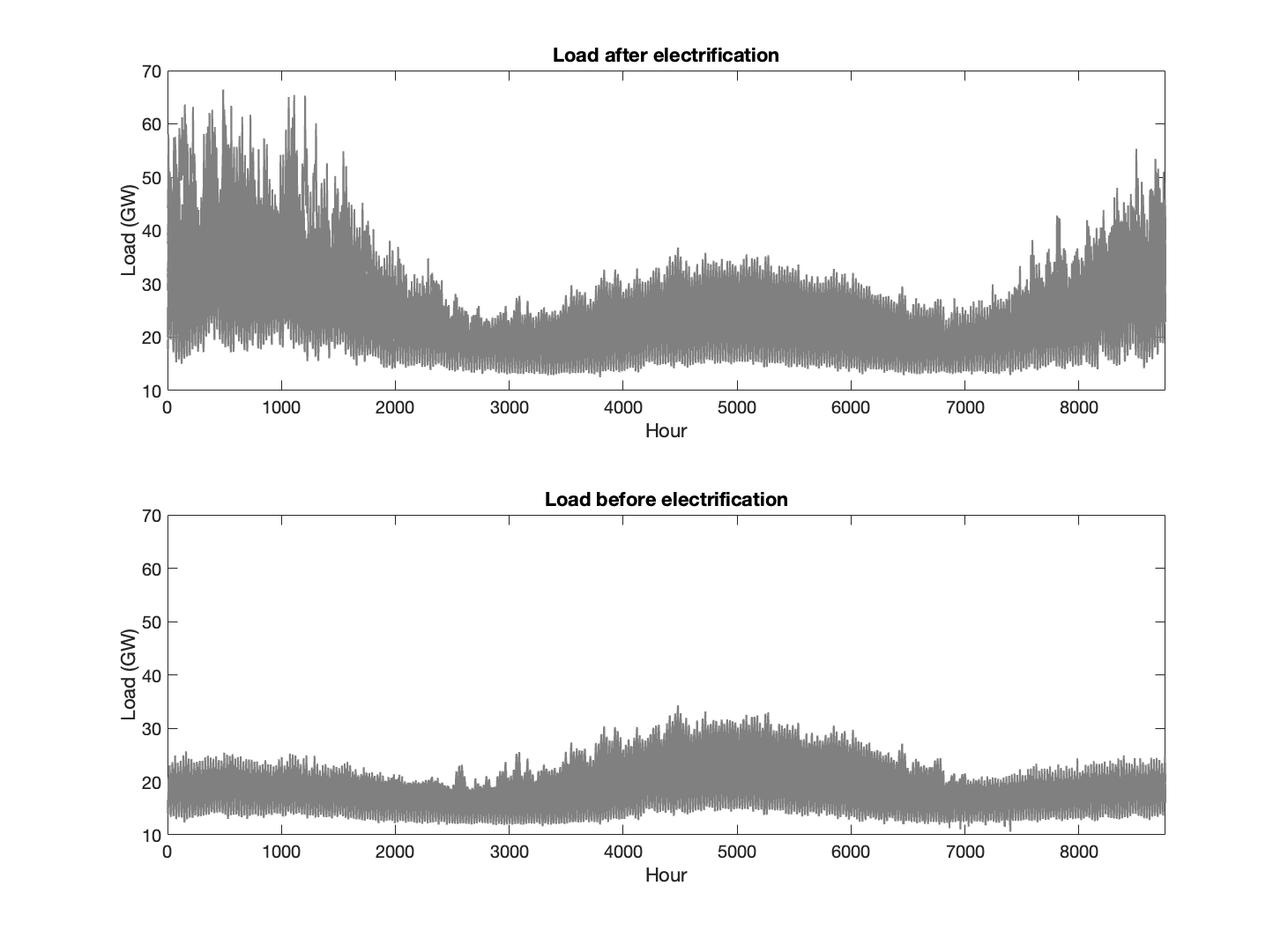}
\caption[Comparison for load profile of 22 years before and after electrification]%
{\textbf{Comparison for load profile of 22 years before and after electrification} }\label{fig: Loadeletrification}
\end{figure}

The load profile before and after electrification is shown in Figure~\ref{fig: Loadeletrification}. Our results aligned with the projection in~\cite{NYISOphaseI2019} that by 2050, the peak demand shifts from summer to winter. 

\section{Daily downstate curtailment in summer and winter, baseline analysis}
\begin{figure}[H]%
\centering
\includegraphics[width=0.9\textwidth]{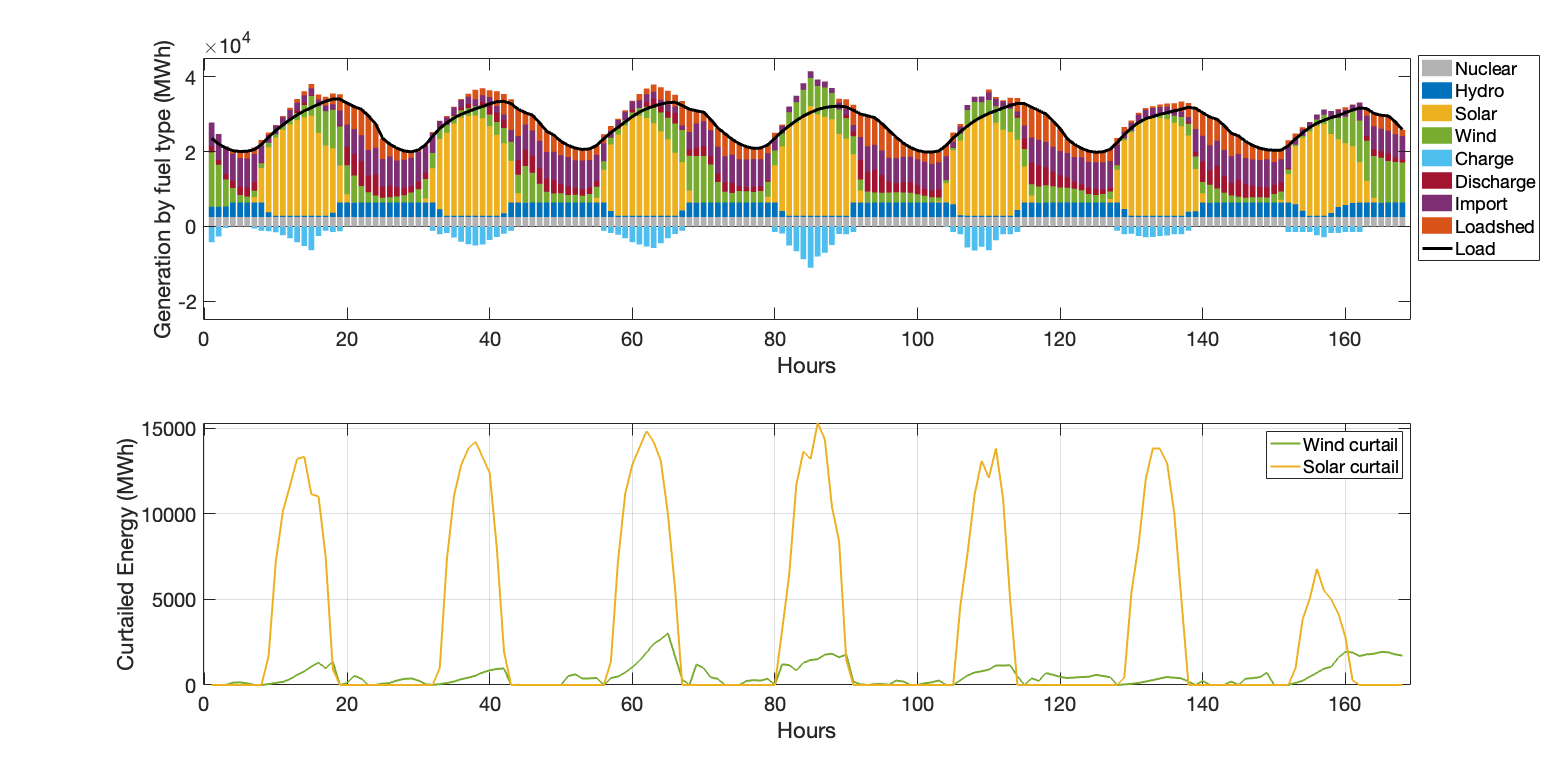}
\caption[Energy composition and renewable curtailment for a typical summer week with load shedding]%
{\textbf{Energy composition and renewable curtailment for a typical summer week with load shedding} }\label{fig: allstatelssummer}
\end{figure}

\begin{figure}[H]%
\centering
\includegraphics[width=0.9\textwidth]{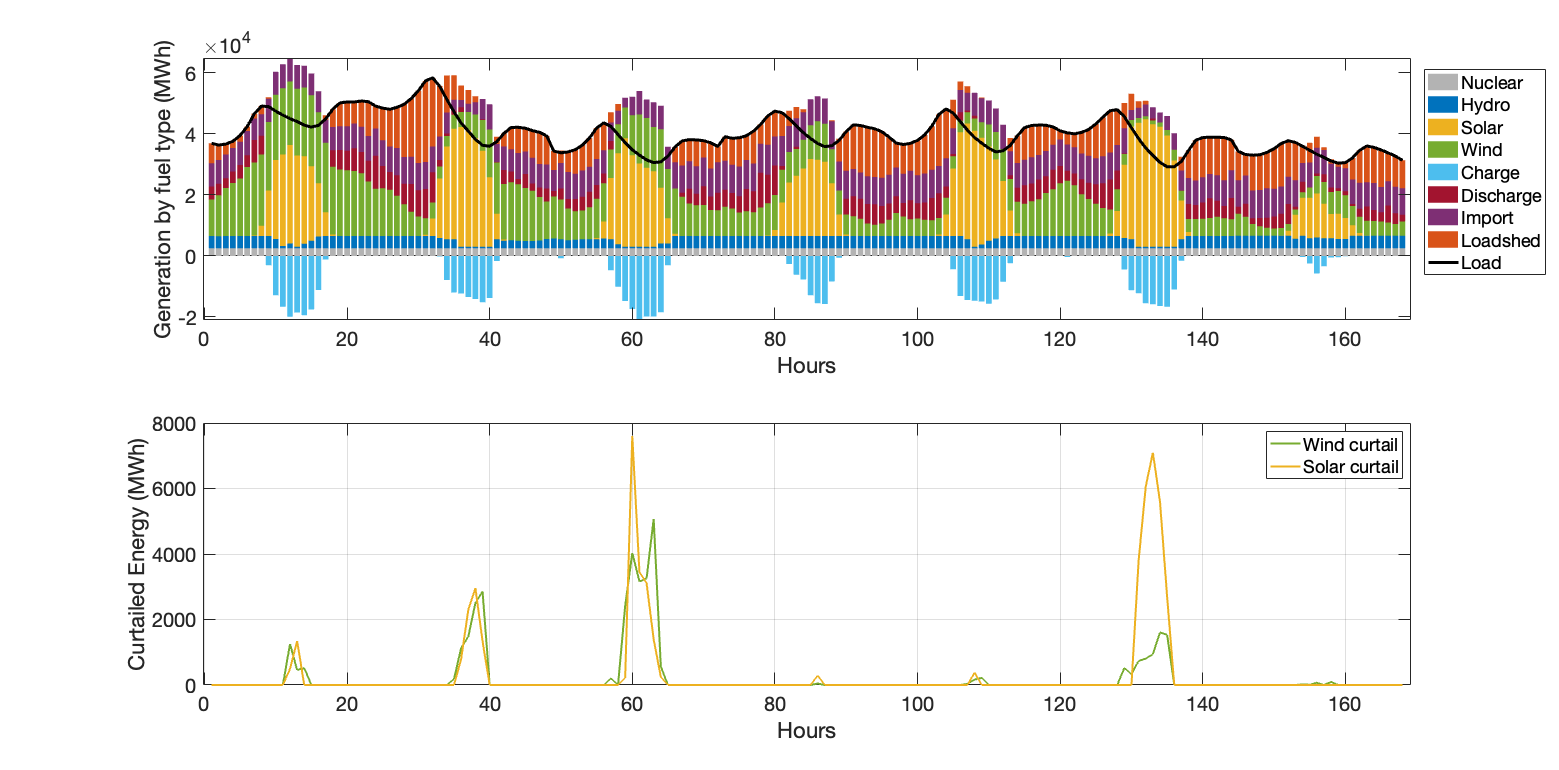}
\caption[Energy composition and renewable curtailment for a typical winter week with load shedding]%
{\textbf{Energy composition and renewable curtailment for a typical winter week with load shedding} }\label{fig: allstatelswinter}
\end{figure}

\begin{figure}[H]%
\centering
\includegraphics[width=0.9\textwidth]{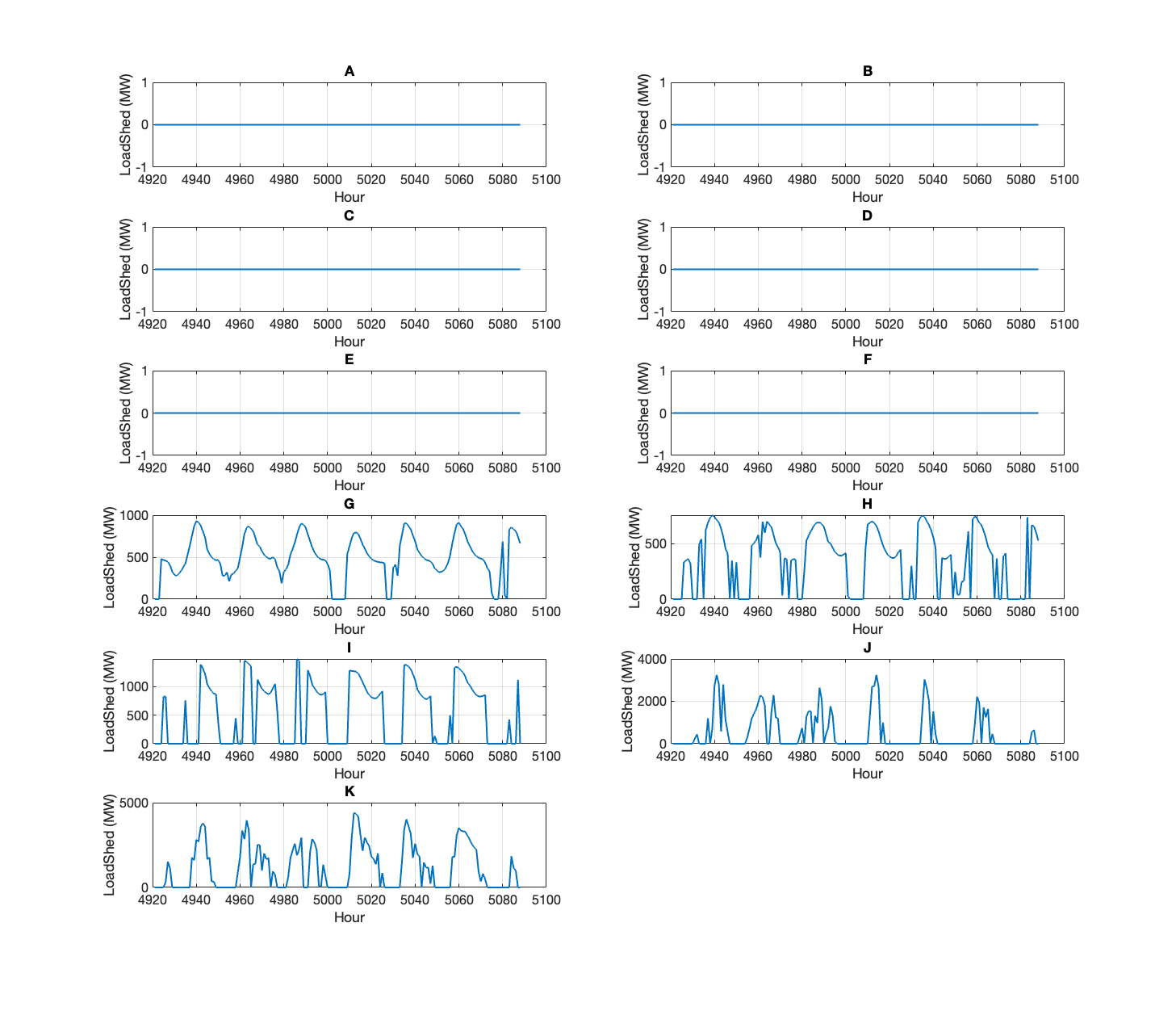}
\caption[Zonal load shedding for a typical week in summer]%
{\textbf{Zonal load shedding for a typical week in summer} }\label{fig: zonallssummer}
\end{figure}

As explained in the main text, winter and summer exhibit distinct patterns. Figures~\ref{fig: allstatelssummer} and~\ref{fig: allstatelswinter} illustrate a typical summer week and a typical winter week with load shedding, respectively. During the summer week, there is significantly more solar availability in terms of intensity and duration. However, wind availability remains low throughout the week. Consequently, load shedding primarily occurs after sunset. It is worth noting that curtailment of wind and solar energy is observed during midday, indicating that transmission line congestion hampers the efficient utilization of renewable energy. A closer examination of the zonal load shedding pattern depicted in Figure~\ref{fig: zonallssummer} reveals that load shedding exclusively transpires in the downstate zones (G-K). This outcome aligns with expectations since downstate zones heavily depend on wind availability and thus face greater vulnerability during summer nights under prolonged wind droughts.

\begin{figure}[H]%
\centering
\includegraphics[width=0.9\textwidth]{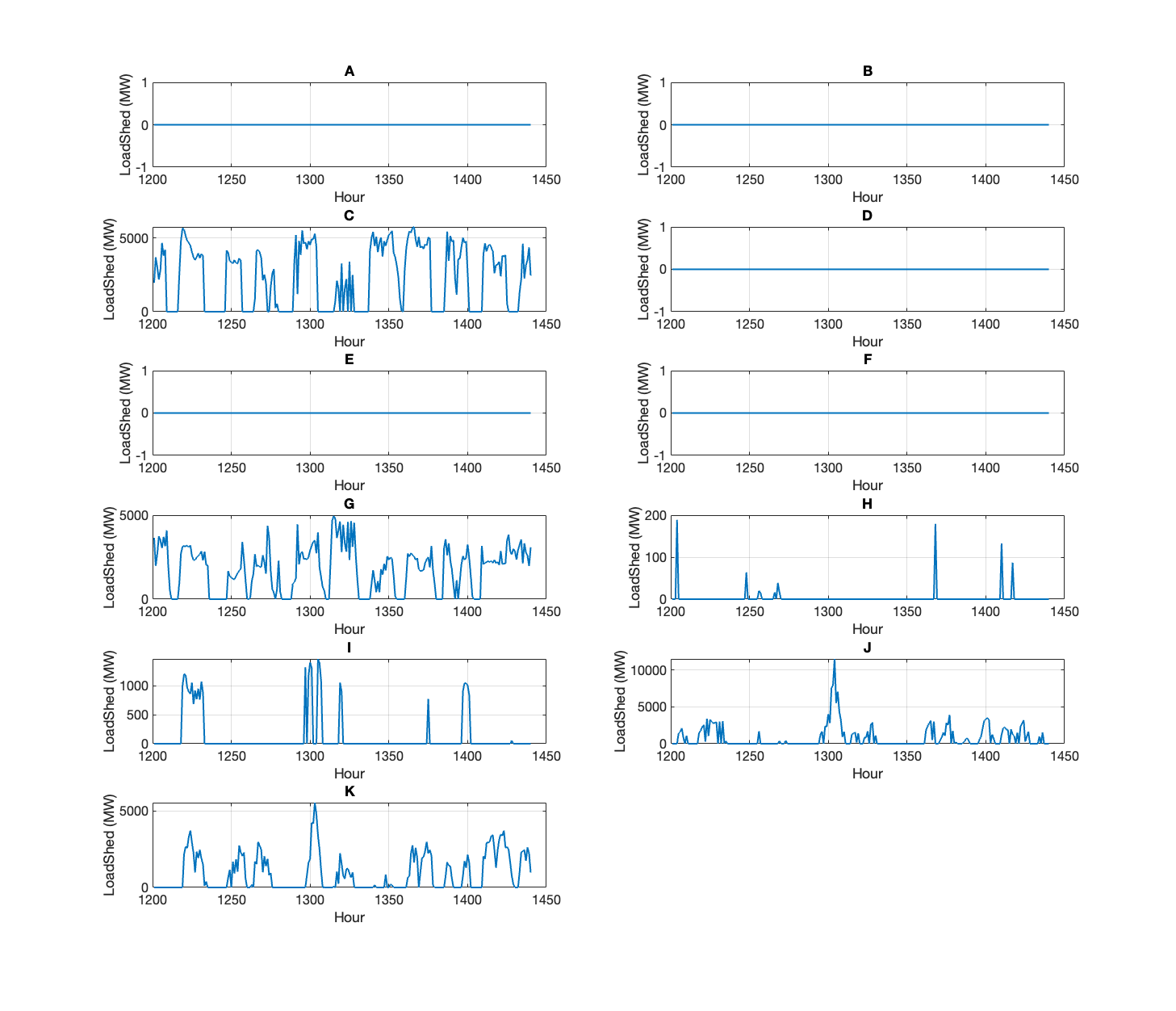}
\caption[Zonal load shedding for a typical week in winter]%
{\textbf{Zonal load shedding for a typical week in winter} }\label{fig: zonallswinter}
\end{figure}
In contrast to summer, winter exhibits a different vulnerability pattern. With the electrification of the heating load (refer to Section~\ref{sec: electrificationcomparison}), the overall load profile substantially increases. Consequently, load shedding occurs when there is high demand coupled with wind drought, even during the middle of the day when solar power is available. Load shedding is not limited to downstate zones but also affects upstate zones due to lower temperatures during cold waves, as shown in Figure~\ref{fig: zonallswinter}.

The amount of load shedding is the result of both load shedding intensity and duration, generally driven by high heating demand co-occurring with low renewable availability. The primary conditions are spatially different in intensity and duration:
\begin{itemize}
    \item Load shedding intensity:
    \begin{itemize}
        \item Low \emph{minimum} temperature accompanied by low wind power availability in load centers, or
        \item Extremely low \emph{minimum} temperature across the whole state
    \end{itemize}
    \item Load shedding duration:
    \begin{itemize}
        \item Extremely low \emph{average} temperature in load centers
        \item Low \emph{average} temperature accompanied by low \emph{maximum} solar power in load centers, or
        \item Low \emph{average} temperature in load center accompanied by wind droughts across the whole state
    \end{itemize}
\end{itemize}

\section{Summer load shedding comparison: baseline vs high temperature}

\begin{figure}[H]%
\centering
\includegraphics[width=0.9\textwidth]{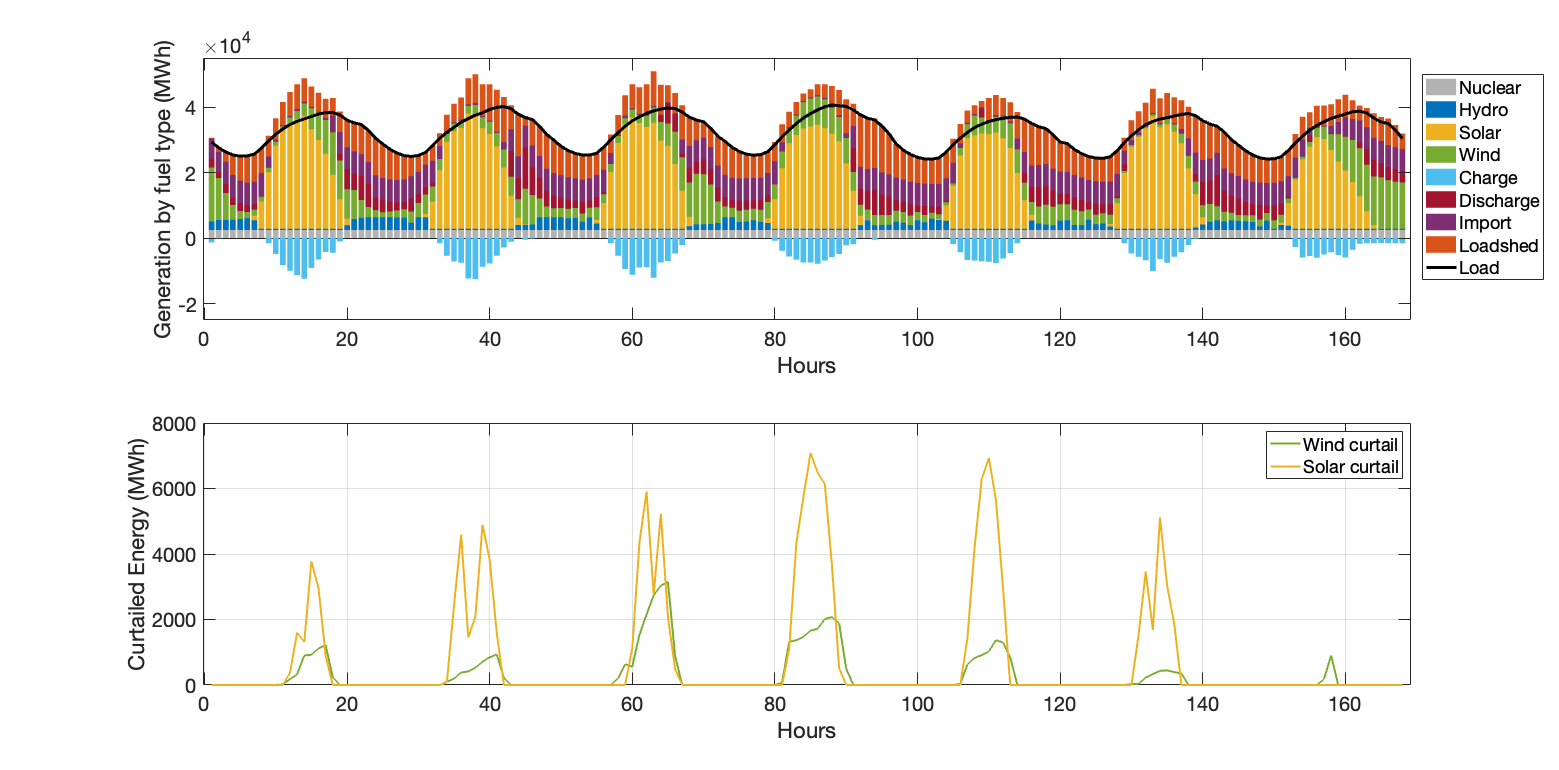}
\caption[Energy composition and renewable curtailment for a typical summer week with load shedding under a severe temperature increase scenario]%
{\textbf{Energy composition and renewable curtailment for a typical summer week with load shedding under a severe temperature increase scenario} }\label{fig: S300summerweek}
\end{figure}

Figure~\ref{fig: S300summerweek} presents the same typical summer week as depicted in Figure~\ref{fig: allstatelssummer} but with an extreme temperature rise of 5.64 \textdegree C. The overall demand profile experiences an increase due to the elevated temperatures. As a result, Load shedding occurs not only during nighttime but also throughout the daytime with solar power curtailment, which indicates heavily congested transmission interfaces. In comparison to the intense maximum hourly load shedding observed in the baseline winter scenario, the summer load shedding in the severe temperature increase scenario is characterized by mild intensity. However, the load shedding duration is prolonged especially during the wind drought week.

\section{Analysis of the dynamic rating modeling of transmission lines}

\begin{figure}[H]%
\centering
\includegraphics[width=0.7\textwidth]{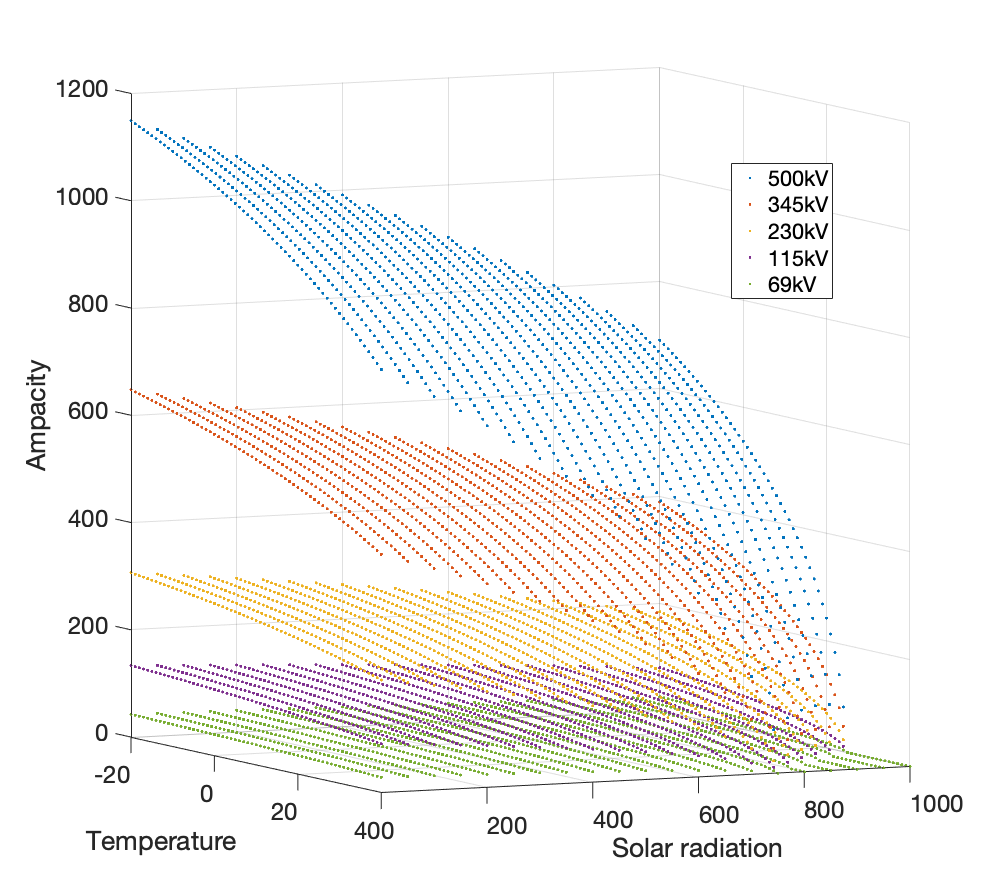}
\caption[Sensitivity of the Ampacity with respect to temperature and solar radiation]%
{\textbf{Sensitivity of the Ampacity with respect to temperature and solar radiation} }\label{fig: tempsolar}
\end{figure}

\begin{figure}[H]%
\centering
\includegraphics[width=0.7\textwidth]{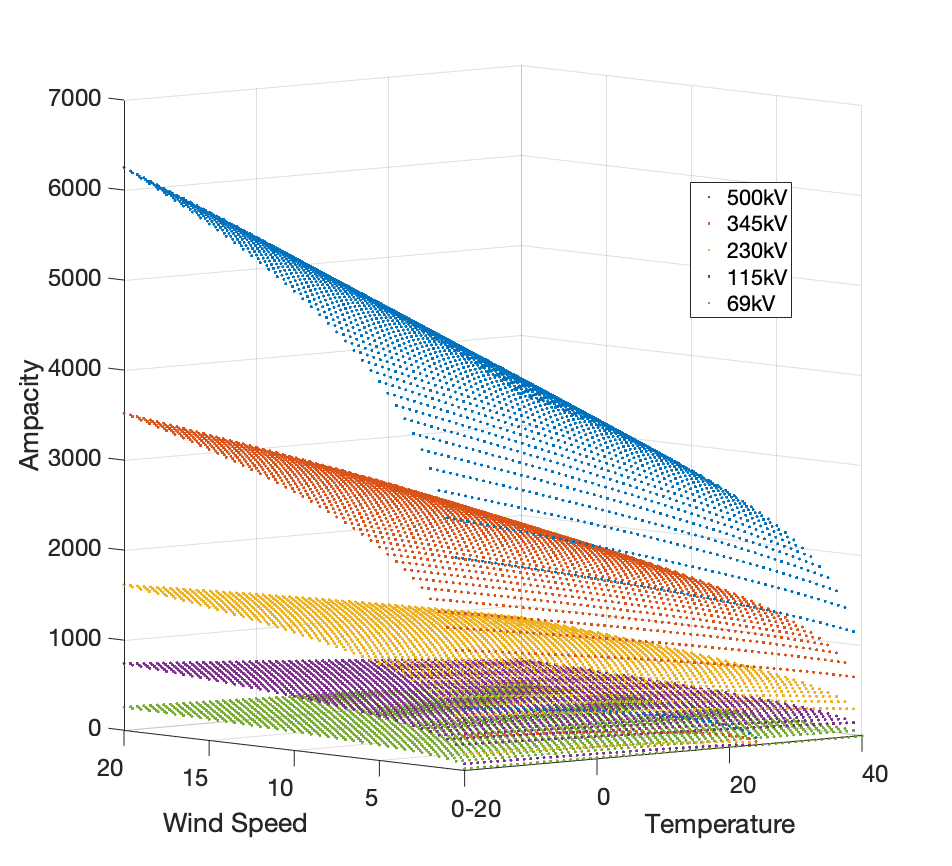}
\caption[Sensitivity of the Ampacity with respect to wind speed and temperature]%
{\textbf{Sensitivity of the Ampacity with respect to wind speed and temperature} }\label{fig: windtemp}
\end{figure}

As detailed in Section~\ref{sec: dynamicrating}, the rating of transmission lines is influenced by various weather variables: temperature, wind speed, and solar radiation. Our sensitivity analysis, consistent with previous findings~\cite{Dyresult}, reveals that the rating is most sensitive to wind speed, followed by solar radiation and temperature, as illustrated in Figures~\ref{fig: tempsolar} and~\ref{fig: windtemp}.

\begin{figure}[H]%
\centering
\includegraphics[width=0.9\textwidth]{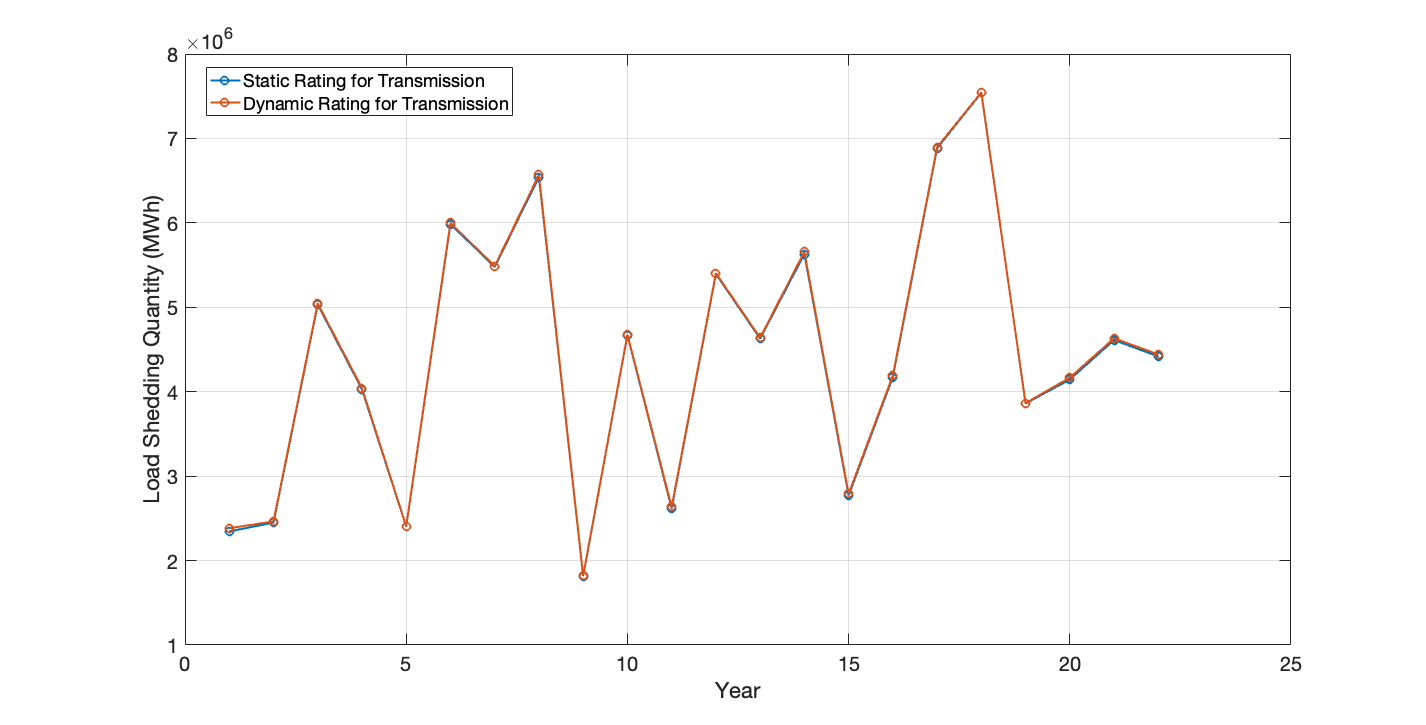}
\caption[Load shedding quantity with static and dynamic transmission line rating]%
{\textbf{Load shedding quantity with static and dynamic transmission line rating} }\label{fig: DyLSQ}
\end{figure}

\begin{figure}[H]%
\centering
\includegraphics[width=0.9\textwidth]{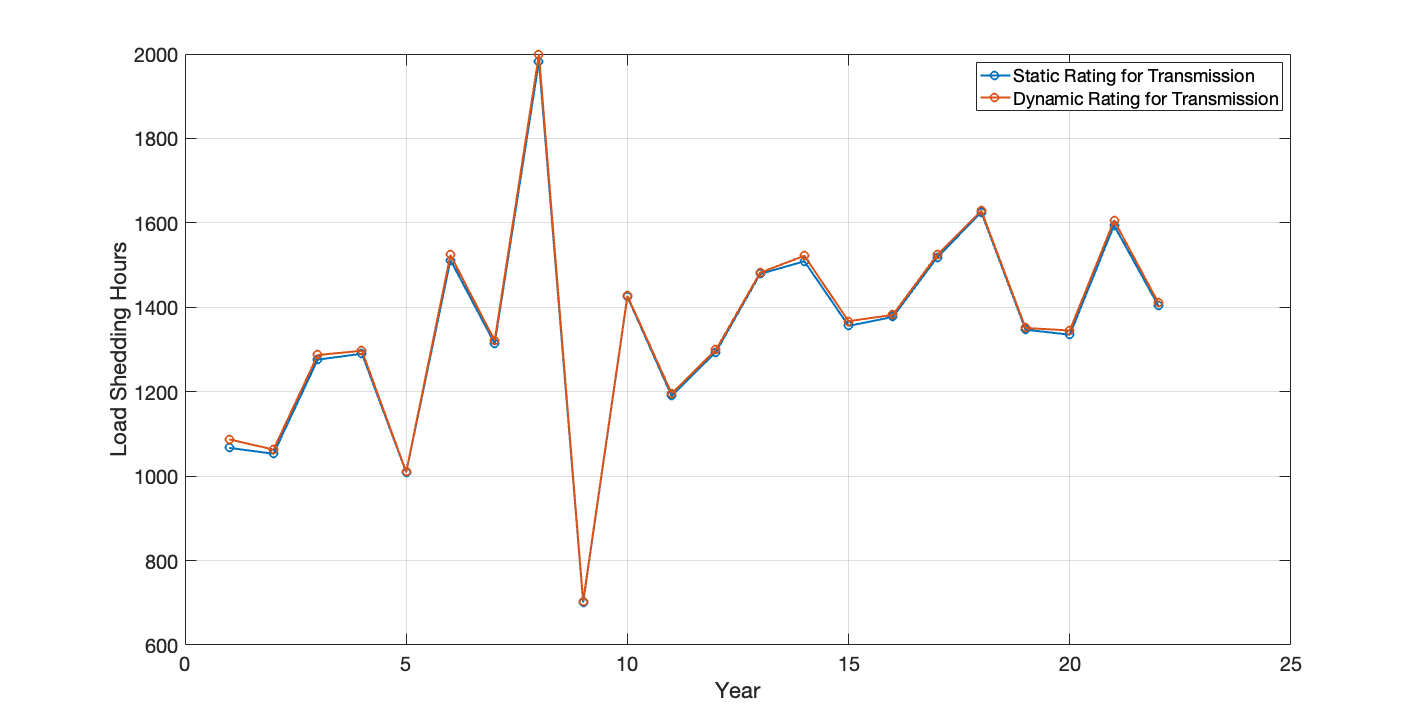}
\caption[Load shedding duration with static and dynamic transmission line rating]%
{\textbf{Load shedding duration with static and dynamic transmission line rating} }\label{fig: DyLSF}
\end{figure}

\begin{figure}[H]%
\centering
\includegraphics[width=0.9\textwidth]{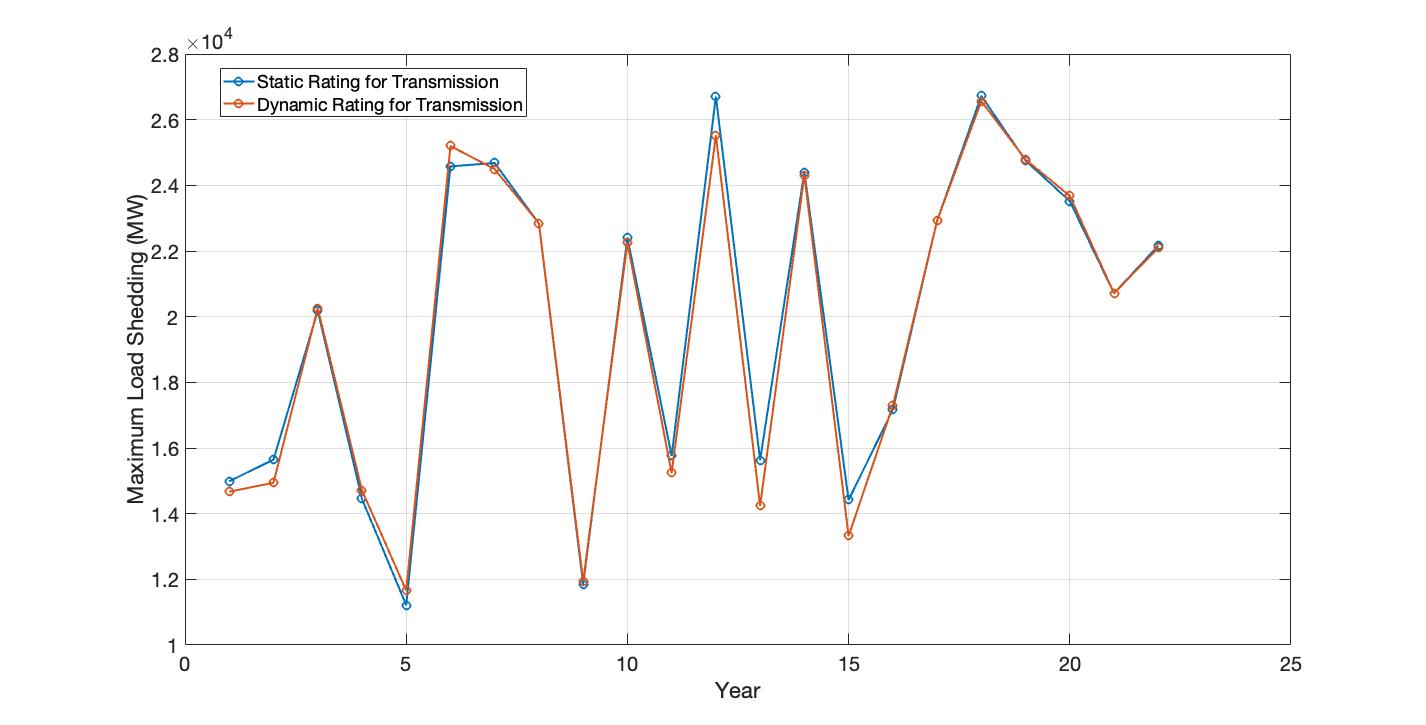}
\caption[Load shedding intensity with static and dynamic transmission line rating]%
{\textbf{Load shedding intensity with static and dynamic transmission line rating} }\label{fig: DyLSM}
\end{figure}

\begin{figure}[H]%
\centering
\includegraphics[width=0.9\textwidth]{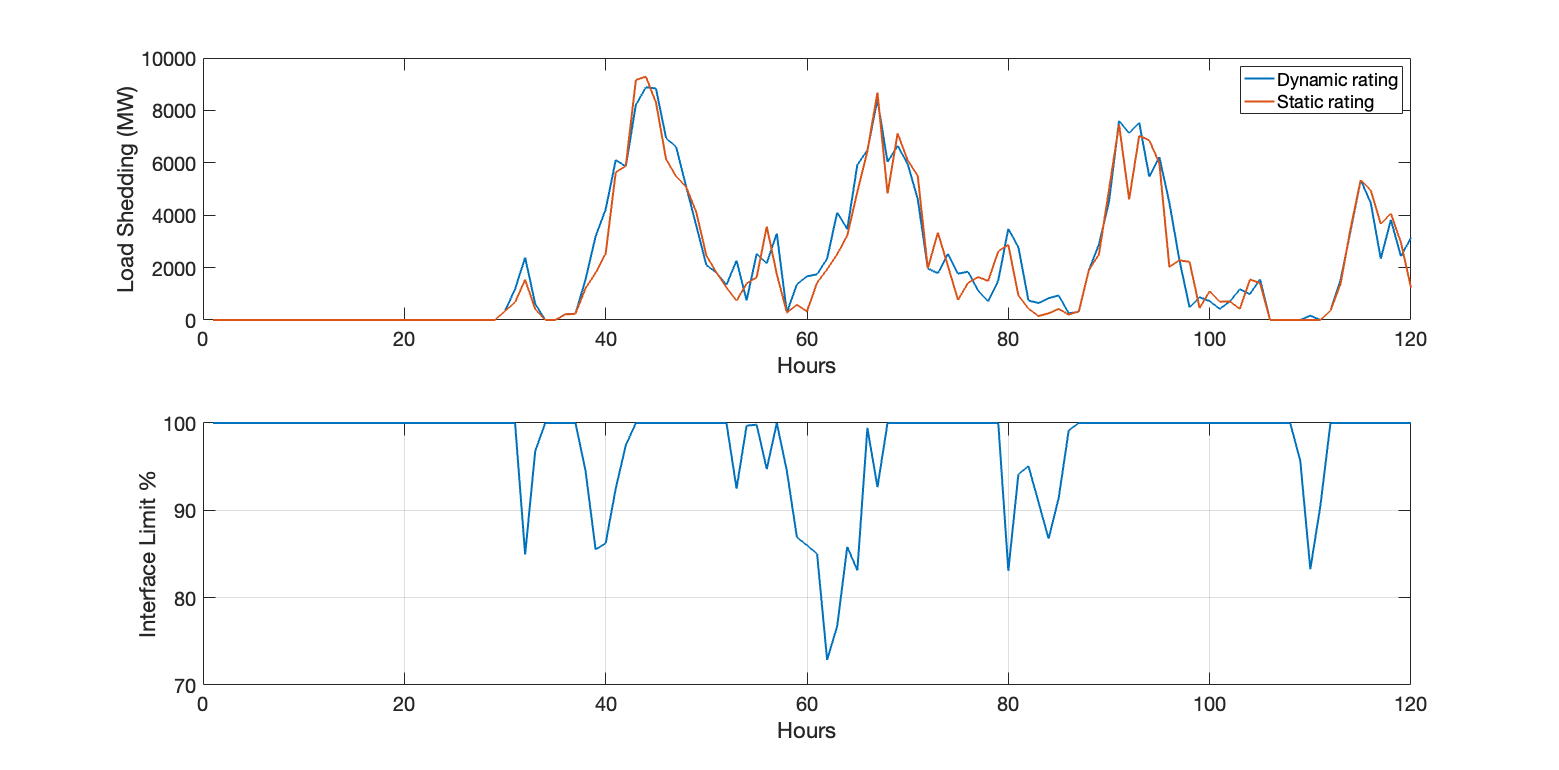}
\caption[Comparison between dynamic and static rating]%
{\textbf{Comparison between dynamic and static rating} }\label{fig: st_vs_dy}
\end{figure}

\begin{figure}[H]%
\centering
\includegraphics[width=0.9\textwidth]{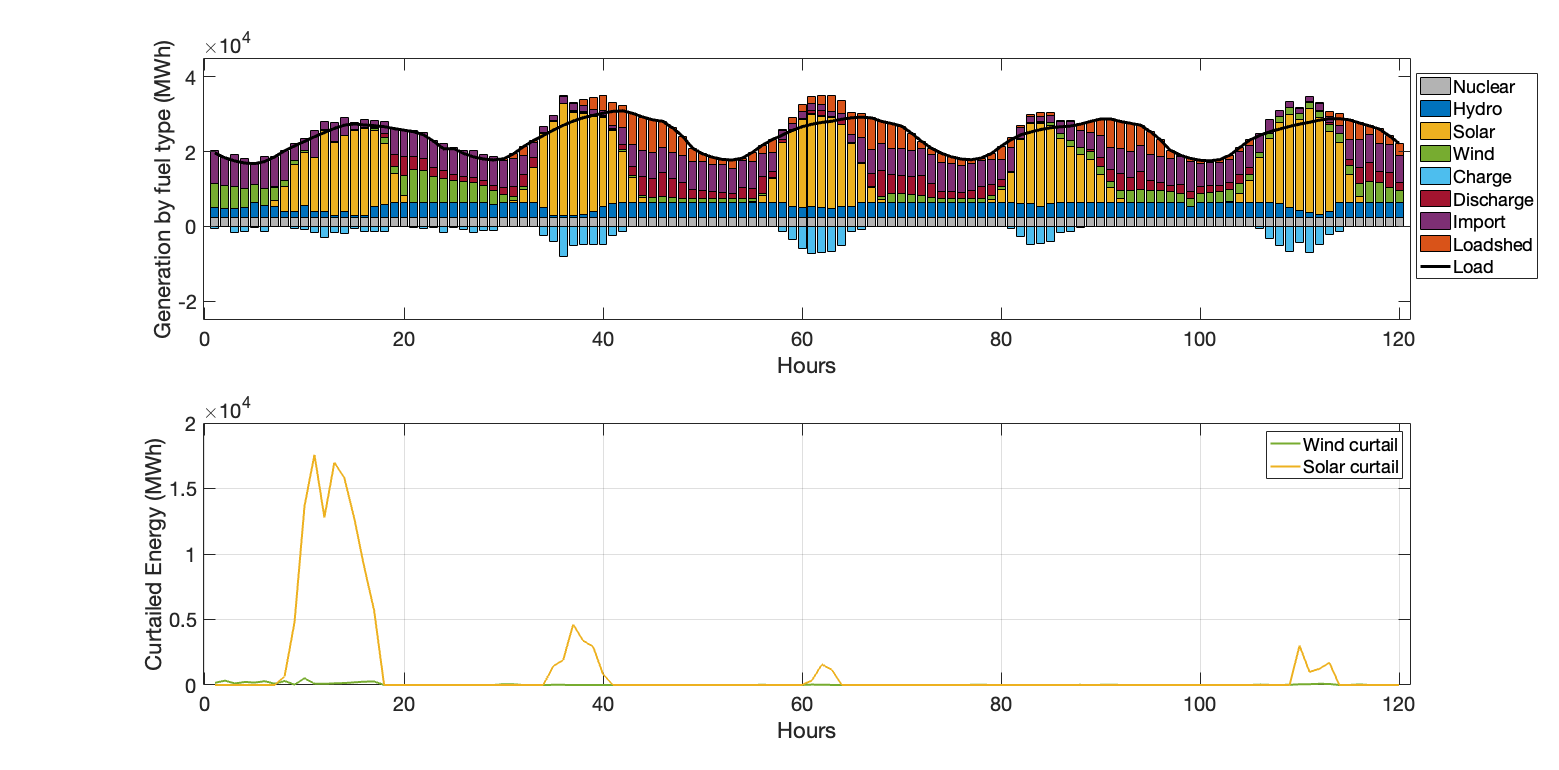}
\caption[Overview of the system condition over five consecutive days]%
{\textbf{Overview of the system condition over five consecutive days} }\label{fig: dy5dayoverview}
\end{figure}

To comprehensively gauge the impact of dynamic rating modeling on system vulnerabilities, we compare the baseline case, spanning 22 years of data, with and without dynamic rating modeling. This comparison is summarized in Figures~\ref{fig: DyLSQ},~\ref{fig: DyLSF} and~\ref{fig: DyLSM} for the three evaluation metrics. Overall, the dynamic rating has a negligible effect on load shedding quantity and duration because the probability of encountering extremely hot, sunny, and windless hours is low. The maximum load shedding does exhibit a slightly larger difference. Unlike load shedding quantity, which is averaged over time, load shedding intensity quantifies instantaneous load shedding. Dynamic modeling of transmission line ratings alters quite a few optimization problem parameters, resulting in potentially different outcomes not only during hours when the parameters change. For instance, Figure~\ref{fig: st_vs_dy} depicts load shedding over five consecutive days with dynamic and static ratings in the upper panel, and transmission line ratings in the lower panel. Instantaneous load shedding results vary slightly with and without dynamic rating, but overall quantities remain similar. A notable difference in load shedding occurs when dynamic ratings lead to a significant drop in transmission capacity (for example, around hour 60 and hour 80), indicating heightened system vulnerability due to lowered transmission capacity. Figure~\ref{fig: dy5dayoverview} provides an overview of the system's condition, revealing that significant drops in transmission capacity occur during daytime with minimal wind availability. Note that these occur not necessarily during the hottest part of the day (hours 80-85 represent 8 am to 1 pm). As wind speed increases in the afternoon after 1 pm, transmission ratings return to normal and no increased load shedding is observed during that period. This analysis further validates our earlier findings that low wind availability plays a pivotal role in system vulnerabilities.

\begin{figure}[H]%
\centering
\includegraphics[width=0.7\textwidth]{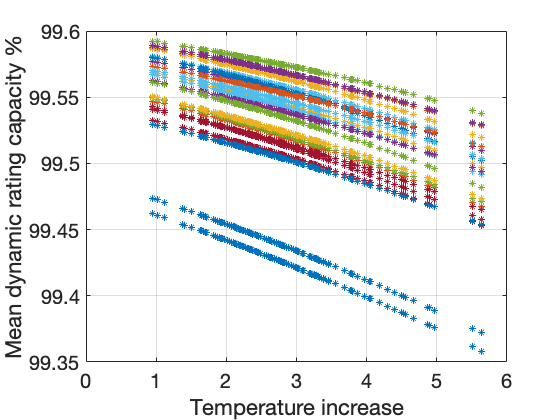}
\caption[Average dynamic rating vs temperature increase]%
{\textbf{Average dynamic rating vs temperature increase} }\label{fig: MeanRating}
\end{figure}

\begin{figure}[H]%
\centering
\includegraphics[width=0.7\textwidth]{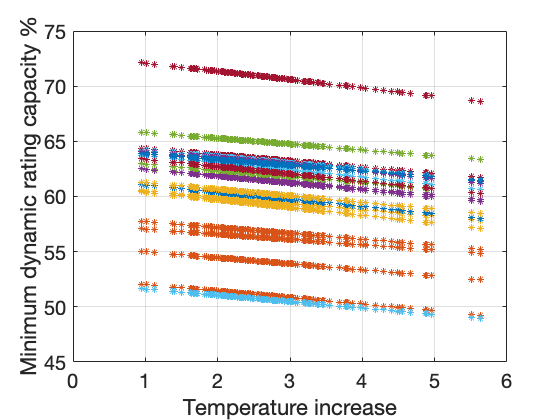}
\caption[Minimum dynamic rating vs temperature increase]%
{\textbf{Minimum dynamic rating vs temperature increase} }\label{fig: MinRating}
\end{figure}

Considering that temperature has a relatively smaller impact on dynamic ratings, a temperature increase of 1-5.64 \degree C exerts a limited influence on overall ratings. Figures~\ref{fig: MeanRating} and~\ref{fig: MinRating} illustrate the average and minimum capacity changes in rating with increasing temperatures, with each line representing a year. The average ratings remain above 99\% over the 22-year period, with only a slight decline as temperatures rise. The minimum rating exhibits some variability across years, but the maximum decrease remains below 3.7\% as temperatures increase. In summary, temperature increases, especially during summer, may potentially constrain power delivery from generation centers to load centers and exacerbate system vulnerabilities. However, this impact is relatively minor compared to the load increase resulting from high temperatures and reduced wind power availability during wind droughts.

\bibliography{sample}
